\documentclass[twocolumn]{aastex6}

\usepackage{setspace}
\usepackage{ulem}
\usepackage{color,ulem}
\usepackage{multirow}
\usepackage{amsmath,amsfonts,amsthm,bm}
\usepackage{booktabs}
\usepackage{natbib}
\usepackage{bm}
\include{epsf}
\include{epstopdf}
\DeclareGraphicsExtensions{.png,.jpg,.pdf,.eps}

\newcommand{\beq}{\begin{equation}}
\newcommand{\eeq}{\end{equation}}

\newcommand{\Bug}{{\bf U}_g}
\newcommand{\Bup}{{\bf U}_p}
\newcommand{\bug}{{\bf u}_g}
\newcommand{\bup}{{\bf u}_p}
\newcommand{\bupi}{{\bf u}_{p{\rm i}}}
\newcommand{\mstar}{M_{\star}}

\newcommand{\meanrhop}{\left<\rho_p\right>_{xy}}

\def\la{\hbox{\raise.35ex\rlap{$<$}\lower.6ex\hbox{$\sim$}\ }}
\def\ga{\hbox{\raise.35ex\rlap{$>$}\lower.6ex\hbox{$\sim$}\ }}

\def\Fjortoft{Fj\o rtoft\ }

\def\St{{\rm{St}} }
\def\Rieff{{\rm Ri}_{{\rm eff}}}
\def\Riphi{\Ri_{{\phi}}}
\def\Riphinot{\Ri_{{\phi,0}}}
\def\Rir{\Ri_{{r}}}
\def\Ri{{\rm Ri}}
\def\Ro{{\rm Ro}}
\def\rmi{{\rm i}}
\def\rmj{{\rm j}}
\def\beq{\begin{equation}}
\def\eeq{\end{equation}}
\def\beqa{\begin{eqnarray}}
\def\eeqa{\end{eqnarray}}

\def\sub#1{_{_{#1}}}
\def\order#1{{\cal O}\left({#1}\right)}

\def\mc#1{\mathcal{#1}}
\newcommand{\REV}[1]{{\bf{{\color{purple} #1}}}}

\newcommand{\sfrac}[2]{\small \mbox{$\frac{#1}{#2}$}}

\shorttitle{Turbulence in sheared particle layers of protoplanetary disks}
\shortauthors{Sengupta \& Umurhan}

\begin{document}

\title{Turbulence in particle laden midplane layers of planet forming disks}

\author{Debanjan Sengupta\altaffilmark{1,2,3,4}}
\author{Orkan M. Umurhan\altaffilmark{1,5,6}}
\altaffiltext{1}{NASA Ames Research Center; Mail Stop 245-3, Moffett Field, CA 94035, USA}
\altaffiltext{2}{Universities Space Research Association; 7178 Columbia Gateway Drive, Columbia, MD 21046, USA}
\altaffiltext{3}{Department of Astronomy, New Mexico State University, Las Cruces, NM 88003-8001 (debanjan@nmsu.edu)}
\altaffiltext{4}{NASA Postdoctoral Program (NPP) Fellow}
\altaffiltext{5}{SETI Institute, 389 Bernardo Way, Mountain View, CA 94043, U.S.A. (oumurhan@seti.org)}
\altaffiltext{6}{Cornell Center for Astrophysics and Planetary Sciences, Cornell University, Ithaca, NY, USA}
\altaffiltext{3}{Department of Astronomy, New Mexico State University, Las Cruces, NM 88003-8001 (debanjan@nmsu.edu)}


\begin{abstract}

We examine the settled particle layers of planet forming disks in which the streaming instability (SI) is thought to be either weak or inactive.  A suite of low-to-moderate resolution three-dimensional simulations in a $0.2H$ sized box, where $H$ is the pressure scale height, are performed using PENCIL for two Stokes numbers, \St$=0.04$ and $0.2$, at 1\% disk metallicity.   We find a complex of Ekman-layer jet-flows emerge subject to three co-acting linearly growing processes:   (1) the Kelvin-Helmholtz instability (KHI), (2) the planet-forming disk analog of the baroclinic Symmetric Instability (SymI), and (3) a later-time weakly acting secondary transition process, possibly a manifestation of the SI, producing a radially propagating pattern state. For \St$=0.2$, KHI is dominant and manifests as off-midplane axisymmetric rolls, while for \St$=0.04$ the axisymmetric SymI mainly drives turbulence. SymI is analytically developed in a model disk flow, predicting that it becomes strongly active when the Richardson number (Ri) of the particle-gas midplane layer transitions below 1, exhibiting growth rates $\le\sqrt{2/\Ri - 2}\cdot\Omega$, where $\Omega$ is local disk rotation rate. For fairly general situations absent external sources of turbulence it is conjectured that the SI, when and if initiated, emerges out of a turbulent state primarily driven and shaped by at least SymI and/or KHI. We also find that turbulence produced in $256^3$ resolution simulations are not statistically converged and that corresponding $512^3$ simulations may be converged for \St$=0.2$. Furthermore, we report that our numerical simulations significantly dissipate turbulent kinetic energy on scales less than 6-8 grid points.

\end{abstract}


\section{Introduction} \label{sec:intro}
An understanding of how the basic building blocks of planets form
remains elusive.  In the standard picture, the nascent solar nebula is populated with sub-$\mu$m grains that, through collisional sticking, grow until they reach mm-cm scales; however, various dynamical growth barriers prevent further incremental growth en route to the eventual formation of these $50-100$~km sized planetesimals \citep[for a deeper discussion see][]{Estrada_etal_2016, Drazkowska_2022}. Overcoming the so-called cm-barrier has been the subject of intense research for up to two decades now.  Several proposed routes that can circumvent this barrier and produce overdensities that are gravitationally bound have been considered of late,  including (but not limited to) particle concentration by giant vortices \citep[see recent work by][]{Lyra_etal_2018,Raettig_etal_2021} and particle density enhancements resulting from turbulent concentration \citep[e.g.,][]{Chambers_2010,Hartlep_Cuzzi_2020}. The leading candidate process, having received the most attention, is the Streaming Instability \citep[][SI, hereafter]{Youdin_Goodman_2005,Johansen_etal_2007}, which can routinely produce gravitationally bound overdensities \citep[e.g.,][]{Simon_etal_2017,Abod_etal_2019}. The SI -- which produces high density clumps through a strong resonance between two counterflowing streams \citep{Squire_Hopkins_2018a} -- is promising for several reasons including the correspondence between the observed angular momentum orientation distribution of cold classical Kuiper Belt objects and that of gravitationally bound overdensities produced in high resolution SI simulations \citep{Nesvorny_etal_2019}.  On the other hand, if planetesimal forming disk regions experience some kind of hydrodynamic or magneto-hydrodynamic  turbulence \citep[e.g., see review of][]{Lyra_Umurhan_2019}, the efficacy of the SI at producing gravitationally bound overdensities remains uncertain and subject to ongoing debate \citep{Chen_Lin_2020,Umurhan_etal_2020,Gole_etal_2020, Schaefer_etal_2020}.
\par

For what protoplanetary disk conditions then should the SI be expected to lead to clumps dense enough to trigger gravitational collapse?  Assuming that the disk is not subject to some sort of external turbulence source and the disk's particle size distribution is monodisperse, this question has been rephrased by asking what combination of disk metallicity ($Z$) and particle Stokes number (St) leads to SI activity strong enough to produce gravitationally bound overdensities \citep{Carrera_etal_2015,Yang_etal_2017,Li_Youdin_2021}?  Based on a survey of 3D-axisymmetric and full 3D particle-gas simulations, these studies have sought to determine a critical St number dependent metallicity, $Z_c(\St)$, for which values of $Z>Z_c$ are likely to lead to gravitationally bound clumps.  Up until recently, $Z_c$ appeared to be parabolic-like in $\St$, with a minimum value $Z_{c,{{\rm min}}} \approx 0.015$  occurring roughly at around $\St \approx 0.04$.  However, during the preparation stage of this manuscript the study by \citet{Li_Youdin_2021} was released suggesting that this minimum $Z_c$ value may go well below $Z=0.01$, occurring at $\St \approx 0.3$ instead, and that $Z_c$ shows a strong upward jump in value for values of $\St \lessapprox 0.015$.  The reasons for the discrepencies between these various investigations has yet to be clarified. A further important clue was identified by \citet{Sekiya_Onishi_2018} in which they, based on an independent parameter study of the SI, conjecture that the outcome of particle-gas disk simulations is actually a function of St and the ratio $Z/\Pi$, where $\Pi$ is the nondimensionalization of a disk's local background radial pressure gradient.\par

In almost all cases considered, midplane-settled particle layers go through a nominally turbulent pre-clump phase before strong clumping manifests; this is especially true for input values of $\St \lessapprox 0.1$ where this turbulent phase can last up to several dozens of orbit times.  For values of $Z<Z_c$ this turbulent state appears to persist relatively unabated \citep[e.g., see the corresponding simulations of][]{Sekiya_Onishi_2018}.  
\par
It is generally assumed that the SI simultaneously coexists and/or emerges out of a shear driven turbulent state.  This shear state, originally envisioned by \citet{Weidenschilling_1980} to be central to particle-disk scenarios, and leading to the Kelvin-Helmholtz instability (``KHI" hereafter) and roll-up (``KH-roll-up" hereafter), should also develop Ekman type flow structure owing to the presence of strong rotation \citep{Cuzzi_etal_1993,Dobrovolskis_etal_1999}. In the recent study of full 3D particle-disk simulations by \citet{Gerbig_etal_2020} it was shown that for input parameters $Z$ and $\St$ that should {\it not} lead to strong SI activity, the Richardson numbers (Ri) of the turbulent state seem to routinely exceed the classical limiting value of $1/4$ expected for non-rotating stratified flow setups \citep{Miles_1961,Howard_1961}. Indeed, there have been a series of antecedent studies considering the problem of KH-roll-up with strong rotation in either a restricted non-axisymmetric two-dimensional geometry \citep[i.e., dynamics restricted to the azimuthal-vertical plane of the disk, most notably][]{Gomez_Ostriker_2005,Johansen_etal_2006,Barranco_2009} or considered in full 3D via a facsimile single fluid model with an imposed composition gradient \citep{Barranco_2009,Lee_etal_2010a,Lee_etal_2010b}.  All of these studies indicate that activity may persist for values of $\Ri > 1/4$ and likely less than $\Ri < 1$, and conclude that rotation somehow pushes the boundary of stability away from the traditional value of $1/4$; exactly how far this boundary extends is not settled under the relevant conditions.

With these considerations in mind, we set out to better understand how midplane settled protoplanetary disk particle layers behave when the SI is either weak or effectively extinguished.  In this study we are focused on disk models with no external sources of turbulence.  One set of specific aims here is to characterize the shear flow that manifests within the streaming layer; to witness its transformation into a non-steady (and likely turbulent) state; and to identify the mechanism(s) that drive this transition.  Could the insights gained as a result of this exercise lead to better understanding of the Ri $>1/4$ findings of \citet{Gerbig_etal_2020}?  
\par
The study by \citet{Sekiya_Onishi_2018} offers some preliminary glimpses. These authors conducted a suite of low-to-medium resolution simulations (that include parameter inputs that do not lead to strong density clumping) in which they showcase vertically integrated particle density that manifests azimuthally oriented banded structure.  Presumably the rotationally modified KHI or some other fluid dynamical process(es), possibly including a very weak operation of the SI, sculpts these phenomena.  In this regard the unpublished study of \citet{Ishitsu_etal_2009} offers further insights wherein they investigated the purely 3D axisymmetric development of a settled particle-gas midplane layer finding relatively pronounced fluid dynamical development in 3-5 orbit times for low St ( = 0.001) with correspondingly weak and/or dispersed particle clumping \citep[in particular see Fig. 3 of][]{Ishitsu_etal_2009}. Understanding the flow structure underpinning this effect when particle clumping is weak and St is low therefore deserves further scrutiny: what about the underlying flow state thwarts the SI's emergence?
\par

Another one of our broader aims is to characterize the turbulent kinetic energy spectra during various stages of the layer's development in order to help assess the kind of turbulence that might be emerging.  Beyond very recent investigations reported in the geophysical fluid dynamics literature, little is known about the character and nature of the turbulent kinetic energy spectrum in flows that are simultaneously subject to strong rotation and stratification \citep[][]{Alexakis_Biferale_2018}. Moreover, beyond brief glimpses reported in \citet{Li_etal_2018}, to our knowledge there seems to be no published insights in the matter for protoplanetary disk scenarios like considered here.

We approach these questions by conducting a limited series of 3D axisymmetric and full 3D particle-gas shearing box numerical simulations employing the widely used numerical platform PENCIL. We follow the approach taken by numerous previous investigators in our initial setup by adopting a monodisperse distribution of particles characterized by a single \St and positioned along a Gaussian distribution with respect to the disk midplane.  There are no external sources of turbulence. The experiment is then monitored as the particles collapse and drive dynamical activity.  Our simulations do not have particle self-gravity turned on at any stage.  We consider two values of St, $= 0.04, 0.2$, with a metallicity of $Z=0.01$, as parameter inputs that ought not lead to active SI and/or putatively Roche-density exceeding overdensities -- e.g., as based on Fig. 8 of \citet{Carrera_etal_2015} and  Fig. 9 of \citet{Yang_etal_2017}\footnote{However, as noted earlier, \citet{Li_Youdin_2021} report that the SI ought to be active for both sets of model parameters we have adopted here.  We keep this in mind throughout this discussion.}. In this sense, our parameter inputs might be considered analogous to the subset of those examined by \citet{Sekiya_Onishi_2018} that lead to weak clump production.  We wish to better understand the emergent flow state under these weakly clumping conditions in order to extend the insights made by \citet{Sekiya_Onishi_2018} in this regard. As such, we are primarily concerned with the particle-gas dynamical state right on up to the point where either the SI emerges, in some possibly weak incarnation, or the flow exhibits a patterned state. \par 

This study is organized as follows. In section \ref{sec:model} we present the numerical model and simulation setup with the publicly available PENCIL code. The results of these hydrodynamic simulations with particles and gas, specially the system's transition to a turbulent state is discussed in section \ref{sec:turbulence_transition}. In section \ref{sec:statistice}, the turbulence statistics from the simulations are analyzed, which include a calibration of PENCIL. A selected set of linear theory analyses for the dynamics of the shear driven midplane settled particle layer is presented in section \ref{sec:linear_analysis} using tools independent of PENCIL. We discuss our findings and their implication in the context of several previous studies in section \ref{sec:discussion}. Given the substantial content of this paper, readers are encouraged first to skip to section \ref{sec:summary} where we, in brief, summarize the main findings of this work.

\section{Analytical \& Numerical Model}\label{sec:model}

The nascent planet forming environment is a complex system containing gas with dust as the solid counterpart. The formal modeling of such systems is generally formulated with the Euler's equation for the gaseous component, along with the solids treated as a pressureless fluid. The dynamics of the gas and the dust are coupled via a drag force experienced by the dust, arising from a headwind due the pressure supported gas that slightly reduces the radial velocity of the solids. The continuity and momentum conservation equations for the disk gas, in cylindrical coordinate $(\hat{R},\hat{\phi},\hat{z})$ with unit vector $\hat{\bf r}=R\hat{\bf R}+\phi\hat{\bf{\phi}}+z\hat{\bf z}$, can respectively be written as:

\begin{equation}\label{eqn:gascontinuity}
\frac{\partial \rho_g}{\partial t}+\nabla\cdot\left(\rho_g \Bug \right)=0;
\end{equation}
\begin{equation}\label{eqn:gasmomentum}
\frac{\partial \Bug}{\partial t} +\left(\Bug\cdot\nabla\right)\bug = -\Omega^2{\bf \hat{r}} +\frac{\rho_p}{\rho_g}\frac{\Bup - \Bug}{t_f}-\frac{1}{\rho_g}\nabla P,
\end{equation}

\noindent where $\Bug$ and $\Bup$ are the total gas and particle velocities, $P$ is the gas pressure, $\Omega=\sqrt{G\mstar/R^3}$ is the local orbital frequency with $G$ and $\mstar$ being the universal gravitational constant and the stellar mass, respectively. With this, the local keplerian velocity can be expressed as $V_k=R\Omega$. The corresponding equations for the particles treated as a fluid (hereafter we often refer to it as the {\it{particle-fluid}}) read

\begin{equation}\label{eqn:dustcontinuity}
\frac{\partial \rho_p}{\partial t}+\nabla\cdot \left(\rho_p\Bup\right)=0;
\end{equation}
\begin{equation}\label{eqn:dustmomentum}
\frac{\partial\Bup}{\partial t}+\left(\Bup\cdot \nabla\right)\Bup=-\Omega^2 {\bf \hat{r}} -\frac{\Bup -\Bug}{t_f}.
\end{equation}

\noindent The second term on the RHS of each of Eqs. \ref{eqn:gasmomentum} and \ref{eqn:dustmomentum} represent the drag between the gas and the dust components, which is proportional to their relative velocities, normalized by $t_f$, a mechanical relaxation timescale also known as the friction time. Particles, being a pressure-less fluid, move with the local Keplerian velocity $U_{p,K}=R\Omega$, whereas the gas feels the radial pressure gradient, the $\nabla P$ term in Eq. (\ref{eqn:gasmomentum}), which makes their motion slightly sub-Keplerian. The reduction in gas speed is quantified by the parameter $\eta$ given by

\begin{equation}\label{eqn:eta}
\eta=-\frac{1}{2}h^2\frac{\partial \ln\rho_g}{\partial \ln R}=\frac{1}{2}\beta h > 0,
\end{equation}

\noindent where $h=H_g/R$ is the disk aspect ratio and 

\begin{equation}\label{eqn:beta}
\beta=-h\frac{\partial \ln \rho_g}{\partial \ln R}.
\end{equation}

In modeling the system, the parameter $\beta$ is often a representation of the global radial pressure gradient in the system. In systems such as described here, the ratio of the reduction in local Keplerian speed $\eta V_k$ and the sound speed $c_s$, a measure of the dynamical compressibility of the system, is designated by  $\Pi=\eta V_k/c_s$. With Eqs. (\ref{eqn:eta}-\ref{eqn:beta}), $\Pi$ can be expressed in terms of $\beta$ as 

\begin{equation}
    \Pi=-\frac{1}{2}\beta.
\end{equation}

\noindent In all our simulations, the value of $\Pi$ is chosen as $0.05$.

\begin{deluxetable*}{cl}
\tabletypesize{\footnotesize}
\tablecolumns{2}
\tablewidth{0pt}
\tablecaption{Variables used in theoretical modeling}\label{tbl:vartheory}
\tablehead{
\colhead{Variable} \vspace{-0.2cm} & \colhead{Meaning}\\
&  }
\vspace{-0.2cm}
\startdata
\vspace{-0.2cm}\\
$H,H_g$ & Gas scale height (appearing interchangeably)\\
$H_p$ & Particle scale height\\
$\Omega$, $\Omega_0$ & Keplerian frequency (appearing interchangeably) \\
$R$ & orbital distance from central star \\
$V_K$& Keplerian velocity \\
$\Bug$ & Total gas velocity vector\\
$\bug$ & Perturbation gas velocity vector\\
$u_g,v_g,w_g$ & $3$ components of gas velocity\\
$\left<u_g\right>_y$,
$\left<v_g\right>_y$,
$\left<w_g\right>_y$
& azimuthal average of gas velocities\\
$\left<u_g\right>_{xy}$,
$\left<v_g\right>_{xy}$,
$\left<w_g\right>_{xy}$
& radial-azimuthal average of gas velocities\\
$\Bup$ & Total particle-fluid velocity vector\\
$\bup$ & Perturbation particle-fluid velocity vector\\
$u_p,v_p,w_p$ & $3$ components of particle-fluid velocity\\
$\left<u_p\right>_y$,
$\left<v_p\right>_y$,
$\left<w_p\right>_y$
& azimuthal average of particle-fluid velocities\\
$\left<u_p\right>_{xy}$,
$\left<v_p\right>_{xy}$,
$\left<w_p\right>_{xy}$
& radial-azimuthal average of particle-fluid velocities\\
$\vec{x}_{{\rm i}}$ & Position vector for particle $i$\\
$\bupi$ & Lagrangian velocity vector for particle $i$\\
$x_{\rm i},y_{\rm i},z_{\rm i}$ & $3$ components of particle i's position\\
$u_{p{\rm i}},v_{p{\rm i}},w_{p{\rm i}}$ & $3$ components of particle i's Lagrangian velocity\\
$c_s$ & local isothermal sound speed \\
$\alpha$ & turbulence strength \\
$\rho_g$ & gas volume density \\
$\rho_p$ & particle volume density \\
$\rho_m$ & Box-averaged mean solid density \\
$\epsilon$ & dust to gas mass ratio \\
$\rho_{g,0}$ & azimuthally averaged midplane $\rho_g$ \\
$\left<\rho_p\right>_{y,0}$ & azimuthally averaged midplane $\rho_p$ \\
$\left<\rho_p\right>_{xy}$ & radial-azimuthal average of particle fluid field \\
$t_{f}$ & friction/stopping time \\
$\St$ & Stokes number \\
$\Rir$ & Richardson number based on radial velocity \\
$\Riphi$ & Richardson number based on azimuthal velocity \\
$\Ri_{\phi,0}$ & Midplane estimate for $\Riphi$ \\
Re & Reynolds number \\
\vspace{-0.2cm}\\
\enddata
\end{deluxetable*}

 %
 %
 %
 %

\subsection{Numerical Setup}\label{setup}

For numerical solutions of Eqs. (\ref{eqn:gascontinuity}-\ref{eqn:dustmomentum}), we use the PENCIL  code\footnote{\url {http://pencil-code.nordita.org/ }}, which is sixth order in space and third order in time. The hydrodynamic equations are solved in a shearing box setup \citep{Goldreich_Lynden-Bell_1965,Umurhan_Regev_2004,Latter_Papaloizou_2017} which is a small box in the Cartesian coordinate system corotating with local $\Omega$, corresponding to a distance $R$ from the central star. The shearing box approximation assumes the radial $(x)$ and the azimuthal $(y)$ dimensions of the box $(L_x,L_y)$ are small compared to $R$, whereas the vertical dimension $(L_z)$ is not constrained by the shearing box approximation. The unperturbed azimuthal gas velocity in the corotating frame can be written as $V\sub{K}=-q\Omega x$, where

\begin{equation}\label{eqn:q}
q=-\frac{\,d \ln\Omega}{\,d \ln R},
\end{equation}

\noindent which is $-3/2$ for a Keplerian disk. Here, $q$ is the measure of the linear shear the simulation box is subjected to. We will assume $q=-3/2$ throughout. With the shearing box setup, we solve Eqs. (\ref{eqn:gascontinuity}-\ref{eqn:gasmomentum}) in the isothermal approximation with equation of state $P=\rho c_s^2$ with $c_s$ being the local isothermal sound speed. We write the total velocity components as a sum of a perturbation field plus Keplerian flow, i.e., $\Bug = V_k{\bf {\hat y}} + \bug$ and $\Bup = V_k{\bf {\hat y}} + \bup$ for gas and particles (respectively), resulting in the form,

\begin{equation}\label{eqn:pencilgascontinuity}
\frac{\partial \rho_g}{\partial t}+\nabla\cdot \left(\rho_g\bug\right)+V\sub{K}\frac{\partial\rho_g}{\partial y}=f_D\left(\rho_g,\bug\right),
\end{equation}
\begin{equation}\label{eqn:pencilgasmomentum}
\begin{split}
& \frac{\partial \bug}{\partial t}+\left(\bug\cdot\nabla\right)+V\sub K\frac{\partial \bug}{\partial y}=-c_s^2\nabla\ln\rho_g+h\beta\Omega\hat{x}+ \\
& \left( 2\Omega v \hat{x}-\frac{1}{2}\Omega u \hat{y}-\Omega^2 z \hat{z}  \right)+\frac{\rho_p}{\rho_g}\frac{\bug-\bup}{t_f}+f_{\nu}\left(\bug ,\rho_g\right).
\end{split}
\end{equation}

\noindent The perturbation gas velocity and its respective Cartesian components are written with $\bug \equiv (u_g, v_g, w_g)$, and similarly for the perturbation particle-fluid velocity and its components as $\bup \equiv (u_p, v_p, w_p)$.  The third term on the LHS of Eq. (\ref{eqn:pencilgasmomentum}) is the advection of gas due to the shear. The terms in the parenthesis on the RHS denotes the combined effects of the centrifugal force, Coriolis force and stellar gravity. The pressure term of Eq. (\ref{eqn:gasmomentum}) is decomposed into two components: a local and a global pressure gradient, represented by the first and second terms on the RHS of Eq. (\ref{eqn:pencilgasmomentum}).  The local particle mass volume density is $\rho_p$. Note that, the large scale pressure gradient present in a typical protoplanetary disk is modeled as a constant forcing represented by the term $h\beta\Omega$, and is unresponsive to the gas dynamics.\par

In all the simulations, a periodic and a shear-periodic boundary condition has been used in azimuthal and radial directions respectively. In the vertical direction a reflective boundary condition has been used. It is important to remark here that \cite{Li_etal_2018} made a detailed study on the effect of different choices of vertical boundary conditions and they found that the thickness of the dust layer changes with different choices. In particular, the thickness of the settled dust layer is smaller when a periodic boundary condition is used. In this work we have not explored the effects of the different setup and stick to the reflective one in all our simulations.

For these simulations we choose values of $h\beta$ such that in the absence of particles $u_g = w_g = 0$ and $v_g = -0.05 c_s$ everywhere, indicating a weakly pressure supported Keplerian steady state. Eqs. (\ref{eqn:pencilgascontinuity}-\ref{eqn:pencilgasmomentum}) are solved on an Eulerian grid $(x_j,y_j,z_j)$. In order to stabilize the code in cases where steep gradients appear in the solutions, PENCIL uses sixth-order hyper-viscosity and hyper-diffusivity which are represented by $f_D$ in Eq. (\ref{eqn:pencilgascontinuity}) and $f_{\nu}$ in Eq. (\ref{eqn:pencilgasmomentum}), respectively. These two terms allow the fields to dissipate their energy near the smallest scale while preserving the power spectra at the large scales. For more details on these schemes, the reader is referred to section \ref{sec:hyperdiffusion}.

The use of hyperdissipation over the normal (second-order) dissipation scheme greatly improves the bandwidth of the inertial range obtained from the simulations.  In simulations with $\sim 2500$ grids per $H_g$, a bandwidth of more than a decade is obtained using the hyperdissipation scheme (see Fig. \ref{Calibration_Spectra_3D}) which is impossible to obtain  with a normal second-order viscosity prescription. However, even with this scheme, a considerable part of the simulation domain is lost in dissipation with roughly one-third of the Nyquist frequency (corresponding to $\pi/dx$ where $dx$ is grid size, i.e., a 2$dx$ wave) not giving anything meaningful as far as the gas and particle dynamics are concerned.  This issue is examined in more detail in section \ref{sec:spectra}.
\par

The equations for the solid component are implemented in the form of Lagrangian super-particles \citep{Johansen_etal_2007} . The simulation box is seeded with $N_{par}$ super-particles, each labeled by i, with position vector $\vec{x}_{\rmi} \equiv (x_{\rmi},y_{\rmi},z_{\rmi})$  randomly chosen from a Gaussian distribution with scale-height $H_{p,0}$.  Each particle's corresponding perturbation velocity vector $\bupi$ is similarly chosen to be random such that $|\bupi| < 0.002c_s$. The evolution set Eqs. (\ref{eqn:dustcontinuity} - \ref{eqn:dustmomentum}) for each solid's position and velocity are solved in the form

\begin{equation}
\frac{d \vec{x}_\rmi}{d t}=-q\Omega x_\rmi \hat{y} +\bupi,
\end{equation}

\noindent and

\beq
\frac{d \bupi}{d t}=\left(2\Omega v_{pi}\hat{x} -\frac{1}{2}\Omega u_{p\rmi} \hat{y} -\Omega^2 z_\rmi \hat{z}\right)  -\frac{\bug(x_\rmi,y_\rmi,z_\rmi) -\bupi}{t_f}.
\label{particle_evolution}
\eeq

For simulations with mono-disperse solids, particles are chosen as a swarm of identical particles with a single Stokes number \St and a predetermined disk metallicity $Z$, interacting with the gas collectively through the drag force. In order to achieve a smooth solution for the super-particle properties, a triangular shaped cloud (TSC) scheme \citep{Youdin_Johansen_2007, Hockney_Eastwood_1981} is adopted (also see PENCIL CODE Manual), which uses a second order interpolation and assignment method,  by a quadratic spline or quadratic polynomial. This scheme provides an interpolated estimate for $\bug(x_\rmi,y_\rmi,z_\rmi)$ based on the gas velocities values that are known on the fixed Eulerian grid set $(x_\rmj,y_\rmj,z_\rmj)$ (indexed by $\rmj$), and given $\vec{x}_\rmi$ and $\bupi$, constructing an estimate for $\bup$ on the Eulerian grid for ultimate use in Eq. (\ref{eqn:pencilgasmomentum}). For more details of this scheme, the reader is referred to \citet[][Appendix A]{Youdin_Johansen_2007}.
\par
The properties of super-particles are determined based on the parameters used for the simulation box. The surface density $\Sigma$ of the box, with a midplane gas density $\rho_0$, is $\sqrt{2\pi}\rho_0 H$ where $H=c_s/\Omega$ is the gas scale height. With this, the mean gas density in the box becomes $\rho_m=\Sigma/L_z$. The representative density of each super-particle thus reads

\begin{equation}
\rho_{\rm par,swarm}=\frac{Z\rho_m}{N_{\rm par}/(N_xN_yN_z)},
\label{swarm_definition}
\end{equation}
\noindent where $N_{par}$ is the total number of super-particles introduced in the box with number of grids $N_x$, $N_y$ and $N_z$ in the $x$, $y$ and $z$ directions, respectively. Similarly, the total mass represented by each is given by

\begin{equation}
m_{\rm par,swarm}=\frac{Z\rho_m V_{\rm box}}{N_{\rm par}},
\end{equation}
\noindent where $V_{box}=L_xL_yL_z$ is the volume of the simulation box.\par

For post analysis purposes, in order to construct an effective particle-fluid velocity field on the gas fluid's Eulerian simulation grid we do the following: (1) for each Eulerian grid cube with coordinate $(x_\rmj,y_\rmj,z_\rmj)$ and side $\Delta$ we find the set of all particles $\{\rmi \}$ that lie within the cube, (2) the total number of particles in the cube are added and a value of $\rho_p(x_\rmj,y_\rmj,z_\rmj)$ is assigned after multiplication by Eq. (\ref{swarm_definition}),  followed by (3) taking the average of all $\bupi$ contained in the same grid box and assigning its value to the particle-fluid's Eulerian velocity, i.e., $\bup(x_\rmj,y_\rmj,z_\rmj)$. A value of  $\bup = 0, \rho_p = 0$ is assigned when there are no particles in the grid box. As a matter of course, when there are 2 or more particles found within the grid box we calculate a standard deviation and assign it to the vector field $\delta\bup$.





\subsection{Initial Conditions \& Simulation Sets}\label{initial_conditions}

In all our simulations, the gas is assumed to follow the isothermal equation of state $P=\rho_g c_s^2$, where in code units, we assign $c_s=1$, $\rho_g=1$ along with $\Omega=1$. This choice of initial conditions translates to a gas scale height $H_g=c_s/\Omega=1$.
\footnote{However, despite these simplifications, we explicitly quote all quantities in terms of their physical units throughout this study.}
The initial metallicity is assumed to be $Z=0.01$ which sets the initial mass of the solids in the box. Given that the main objective of this work is a thorough investigation of the turbulence generation mechanism in the settled dust layer, we choose combinations of \St and $Z$ that are not expected to readily lead to SI in the simulations  \citep[see][for acceptable parameters leading to SI]{Carrera_etal_2015}. For this work, we choose \St$=0.04$ and $0.2$, values which are thought to to lead to weak SI growth when combined with $Z=0.01$, although as mentioned earlier there is uncertainty in this expectation \citep[e.g.,][]{Li_Youdin_2021}. 

The size of the simulation box is set as $(L_x,L_y,L_z) \equiv (0.2,0.2,0.2)H_g$ and the initial positions of the super-particles are assigned randomly following a Gaussian distribution, constrained by a predetermined initial scale-height $H_p=0.1 H_g$. For the 3D simulations, we choose resolutions of $128^3$ (low) and $256^3$ (medium) for both the \St values chosen. $N_{\rm par}$ is set accordingly in order to achieve one particle per grid. We also present a high-resolution simulation with $512$ grid points in each direction for \St$=0.2$ with a lower number of particles to save computation time. In table \ref{tbl:simulations}, we present the list of simulations along with the relevant parameters.

In terms of diagnostics, the evolving scale heights and velocity fields of the particles are calculated dynamically by the numerical code. In order to compute the scale height $H_p$ of the particles representative of the full domain, the simulation box is divided into $N_x$ slices in the radial direction. $H_p$ is then calculated first for each individual slice following the rms of the particle vertical distances from the midplane,

\begin{equation}\label{eqn:particlescaleheight}
    H_{p,k}=\sqrt{\frac{1}{N_{par,k}}\sum \left|z_\rmi-\langle z\rangle\right|^2},
\end{equation}
where $H_{p,k}$ is the scale-height for the $k^{th}$ slice, $z_\rmi$ denotes the position of the super-particle i contained in that slice, $N_{{\rm par},k}$ is the total number of particles and $\langle z\rangle$ is the average vertical position of all $N_{{\rm par},k}$ particles belonging to the $k^{th}$ slice. The final scale-height $H_p$ is calculated by taking the weighted average of all $H_{p,k}$ from Eq.\ref{eqn:particlescaleheight} over all 2D slices:

\begin{equation}
    H_p=\frac{1}{N_x}\sum H_{p,k}.
    \label{Hp_def}
\end{equation}



%
%

\begin{deluxetable*}{c c c c c c c} 
\label{tbl:simulation}
\tabletypesize{\scriptsize}
\tablewidth{0pt} 
\tablecaption{List of simulations and relevant parameters. \label{tbl:simulations}} 
\tablehead{ 
\colhead{Simulation} \vspace{-0.2cm} & \colhead{Domain\tablenotemark{a}} & \colhead{$N_{grid}$} & \colhead{} & \colhead{} & \colhead{}  & \colhead{}\\
 \vspace{-0.2cm} 
 & &  & \colhead{$N_{par}$} & \colhead{St} & \colhead{$Z$} & \colhead{$N_{par}/$ grid}\\ 
\colhead{Name} & \colhead{$L_x\times L_y \times L_z$} & \colhead{$N_x\times N_y\times N_z$} & & & &}
\startdata 
\vspace{-0.2cm}\\
A2D-04H & $0.2\times 0.2\times 0.2$ & $512\times 1\times 512$ & $2^{18}$ & $0.04$ & $0.01$ & $1$ \\
A2D-04M & $0.2\times 0.2\times 0.2$ & $1024\times 1\times 1024$ & $2^{20}$ & $0.04$ & $0.01$ & $1$ \\
A2D-04SH & $0.2\times 0.2\times 0.2$ & $2048\times 1\times 2048$ & $2^{22}$ & $0.04$ & $0.01$ & $1$ \\
\vspace{-0.2cm}\\
\hline
\vspace{-0.2cm}\\
 A2D-2H & $0.2\times 0.2\times 0.2$ & $512\times 1\times 512$ & $2^{18}$ & $0.2$ & $0.01$ & $1$ \\
A2D-2SH & $0.2\times 0.2\times 0.2$ & $2048\times 1\times 2048$ & $2^{22}$ & $0.2$ & $0.01$ & $1$ \\
\vspace{-0.2cm}\\
\hline
\vspace{-0.2cm}\\
B3D-04L & $0.2\times 0.2\times 0.2$ & $128\times 128\times 128$ & $2^{21}$ & $0.04$ & $0.01$ & $1$ \\
B3D-04M & $0.2\times 0.2\times 0.2$ & $256\times 256\times 256$ & $2^{24}$ & $0.04$ & $0.01$ & $1$ \\
\vspace{-0.2cm}\\
\hline
\vspace{-0.2cm}\\
B3D-2L & $0.2\times 0.2\times 0.2$ & $128\times 128\times 128$ & $2^{21}$ & $0.2$ & $0.01$ & $1$ \\
B3D-2M & $0.2\times 0.2\times 0.2$ & $256\times 256\times 256$ & $2^{24}$ & $0.2$ & $0.01$ & $1$\\
B3D-2H & $0.2\times 0.2\times 0.2$ & $512\times 512\times 512$ & $2^{24}$ & $0.2$ & $0.01$ & $0.125$\\
 \vspace{-0.2cm}\\
 \hline 
 \vspace{-0.2cm}
 F3D-512\tablenotemark{b} & $2\pi \times 2\pi \times 2\pi$ & $512\times 512\times 512$ & \nodata & \nodata & \nodata & \nodata \\
 \vspace{-0.2cm}\\
 \enddata 
  \tablenotetext{$\rm{a}$}{All lengths in units of $H$.}
 \tablenotetext{$\rm{b}$}{$512^3$ forced simulation used to identify useful sub-domain}
\tablecomments{Simulation Sets presented in this paper. The y-dimension in 3D axisymmetric runs are arbitrarily noted as $0.2$.}
\end{deluxetable*}

\begin{figure}
\begin{center}
\leavevmode
\includegraphics[width=9.5cm]{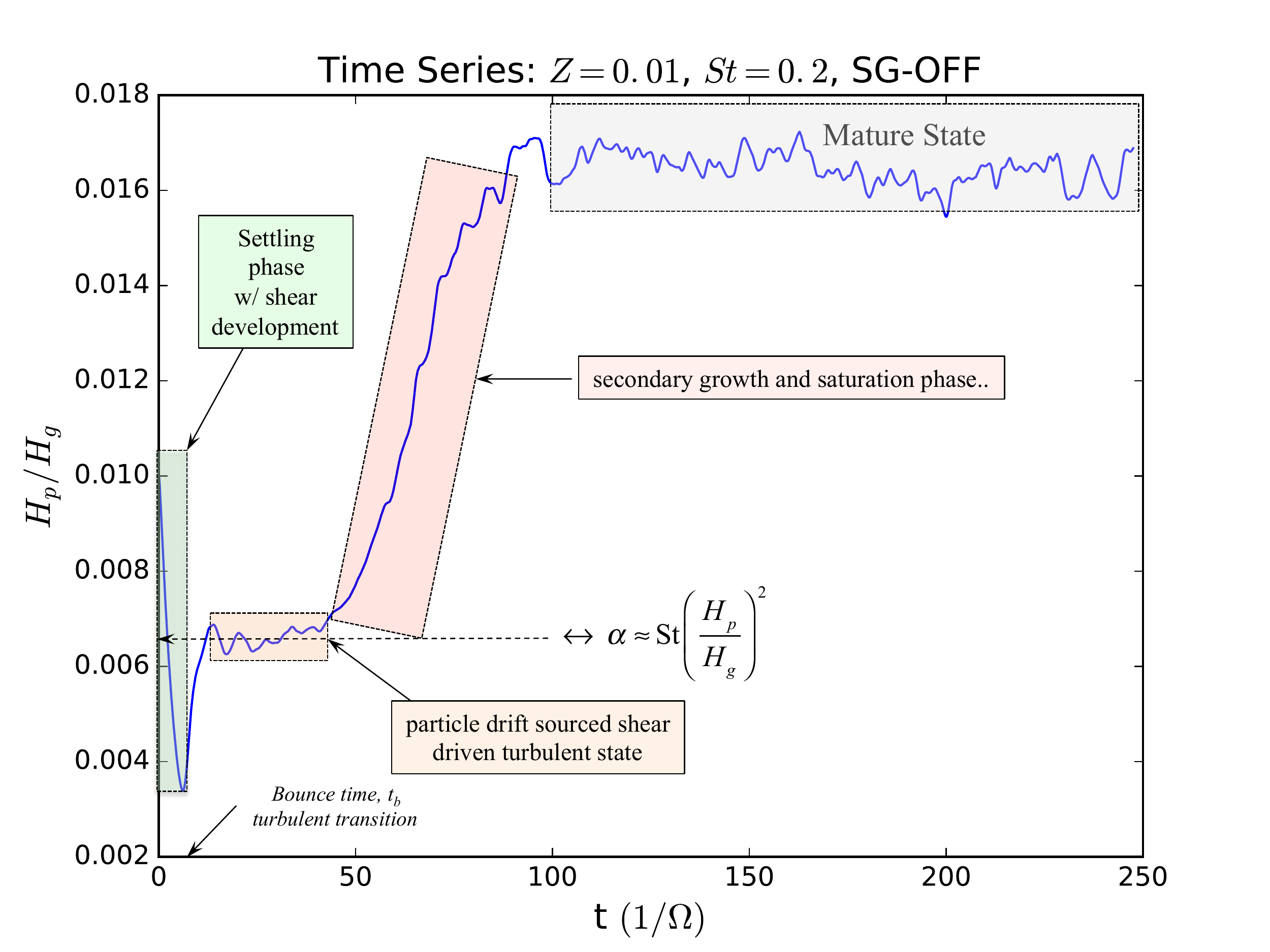}
\par
\end{center}
\caption{Time series of particle scale height for 3D simulation B3D-2L with $Z=0.01$ and St$=0.2$.
The various stages of development are labeled. }
\label{Stages_of_Development}
\end{figure}

\section{Transition to Turbulent State}\label{sec:turbulence_transition}
\subsection{Stages of Development}
Fig. \ref{Stages_of_Development} summarizes
several shared characteristic stages exhibited by simulations during their development over time. This sequence of phases are also generally typical of SI simulations reported in the literature. We describe the stages as: (1) the dust {\it settling phase} in which the settling and drifting dust generates strong velocity shears in both the gas and dusty components, particularly in the radial and perturbation azimuthal component velocity fields; (2) {\it the bounce} out of which the fluid state is sufficiently dynamically unstable so that the midplane trajectory of the settling dust particles is reversed (at some time $t_b$) and the layer starts to thicken some; and (3) a {\it particle drift driven shear turbulent state} in which the shear turbulence is maintained and a quasi-steady turbulent state emerges where the particle layer settles onto a corresponding steady scale height from which we infer an effective measure of the turbulent state via $\alpha \approx {\rm{St}} \Big(H_p/H\Big)^2$ \citep[e.g., a la][]{Cuzzi_etal_1993,Dubrulle_etal_1995}. All simulations reported here further exhibit some type of: (4), a {\it secondary growth phase} followed by (5) a drifting {\it pattern state}. These latter two stages may or may not be an instance of the SI. We further describe the details of these stages in what follows.
\begin{figure*}
\begin{center}
\leavevmode
\includegraphics[width=14cm]{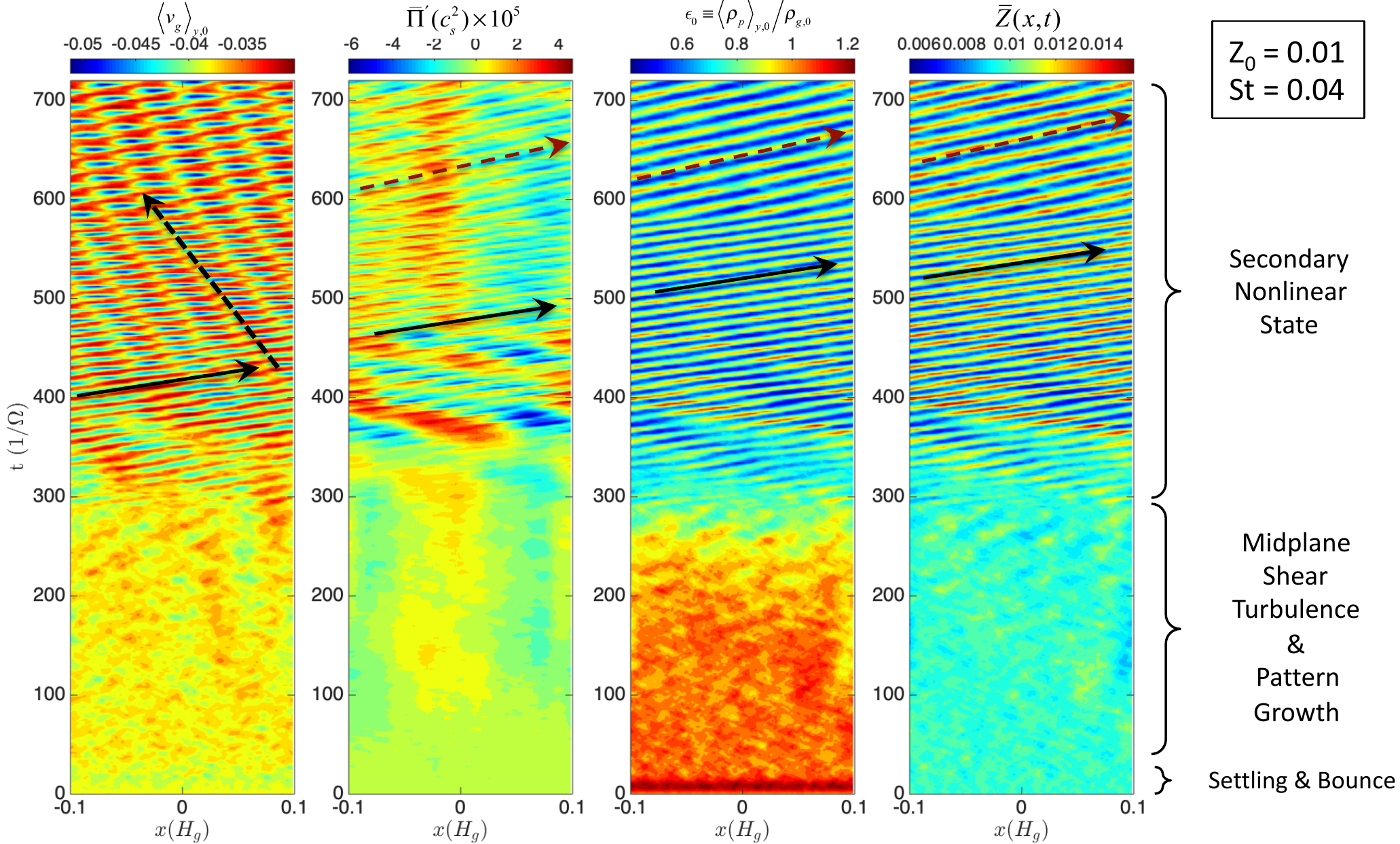}
\par
\end{center}
\caption{Composite figure showing spacetime diagrams for several quantities with St = 0.04 (simulation B3D-04L).  }
\label{spacetime-st-0.04}
\end{figure*}

\subsection{SpaceTime Plots and Observed Pattern Drift}

\begin{figure*}
\begin{center}
\leavevmode
\includegraphics[width=14cm]{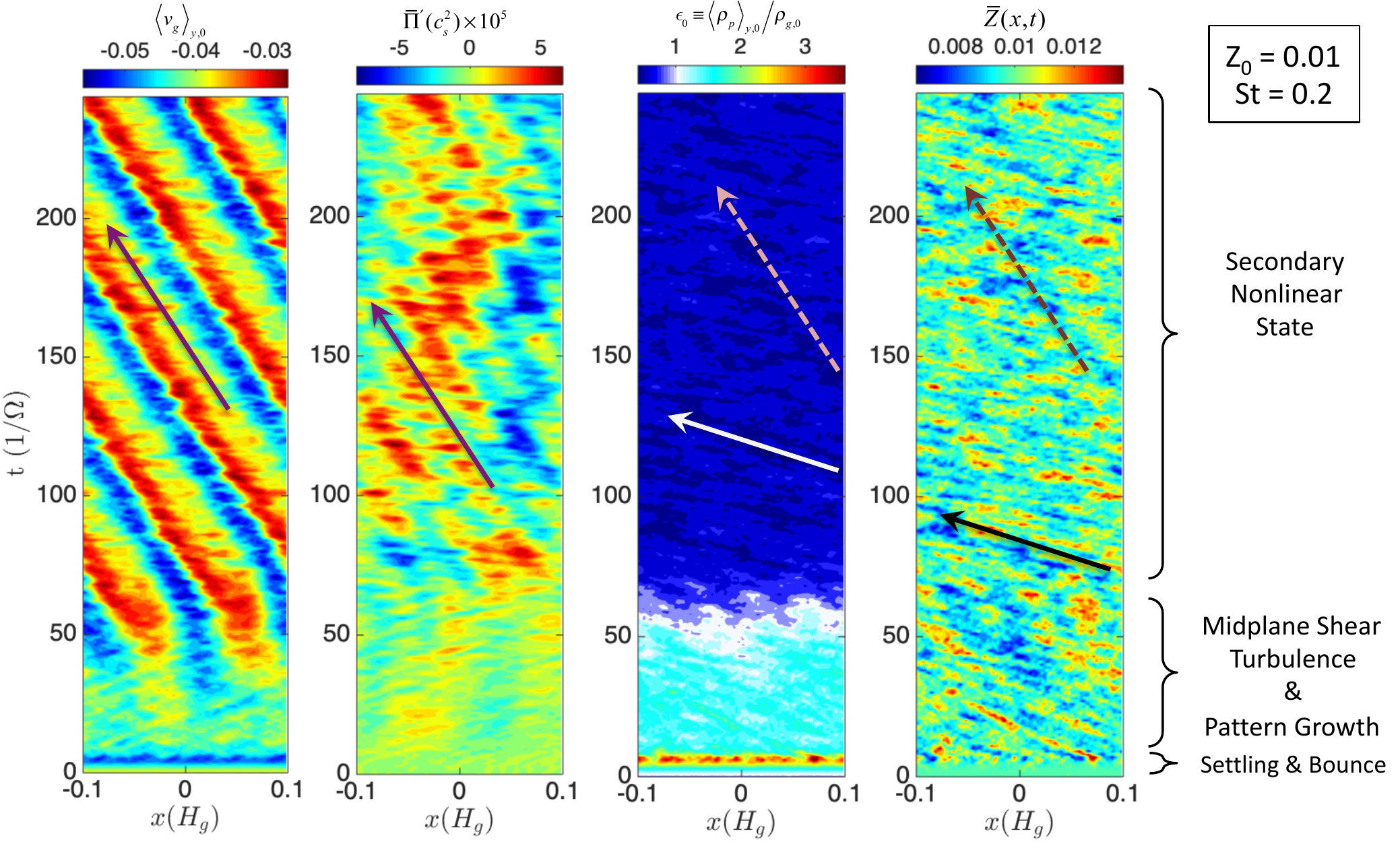}
\par
\end{center}
\caption{Like Fig. \ref{spacetime-st-0.04} except with St = 0.2 (simulation B3D-2L).  }.
\label{spacetime-st-0.2}
\end{figure*}

In Figs. (\ref{spacetime-st-0.04}-\ref{spacetime-st-0.2}) we show the space-time plots based on the low resolution $128^3$ simulations (B3D-04L, B3D-2L). As a function of radius $x$ and time $t$, each figure displays: (i) midplane azimuthal gas velocity, $\left<v_g\right>_{y,0}$ averaged over $y$ direction, (ii) the midplane gas pressure perturbation per unit gas density, $\Pi'(x,t)$, (iii) the ratio of the azimuthally averaged midplane particle density, i.e., $\left<\rho_p\right>_{y,0}$, to the midplane gas density, i.e.,, $\left<\rho_g\right>_{y,0}$, where, in other words,

\beq
\epsilon_0(x,t) \equiv \frac{\left<\rho_p\right>_{y,0}(x,t)}{\left<\rho_g\right>_{y,0}(x,t)};
\eeq

\noindent and (iv) the azimuthally averaged metallicity, $\bar Z(x,t)$. Because the particle layers are close to the midplane and given that the box sizes considered here are small, the gas densities throughout the domain are nearly constant. This allows us to replace $\left<\rho_g\right>_{y,0}$ instead with the global average $\rho_{g,0}$. The exception is when we analyze a perturbation pressure quantity defined by

\beq
\Pi' \equiv c_s^2 \Big(\ln \left<\rho_g\right>_{y,0}(x,t) - \ln \rho_{g,0}\Big).  
\eeq

\noindent Inspection of the Figs. (\ref{spacetime-st-0.04}-\ref{spacetime-st-0.2}) readily shows that the density/pressure fluctuations are indeed weak, effectively rendering these dynamics nearly incompressible.  The radial metallicity is defined as

\beq
\bar Z(x,t) \equiv { \displaystyle \int_{-\infty}^\infty \left<\rho_p\right>_y(x,z,t)dz}\Big/{\sqrt{2\pi} H \rho_{g,0}},
\eeq

\noindent where $\left<\rho_p\right>_y$ is the azimuthally averaged particle density.\par

\begin{figure*}
\begin{center}
\leavevmode
\includegraphics[width=17cm]{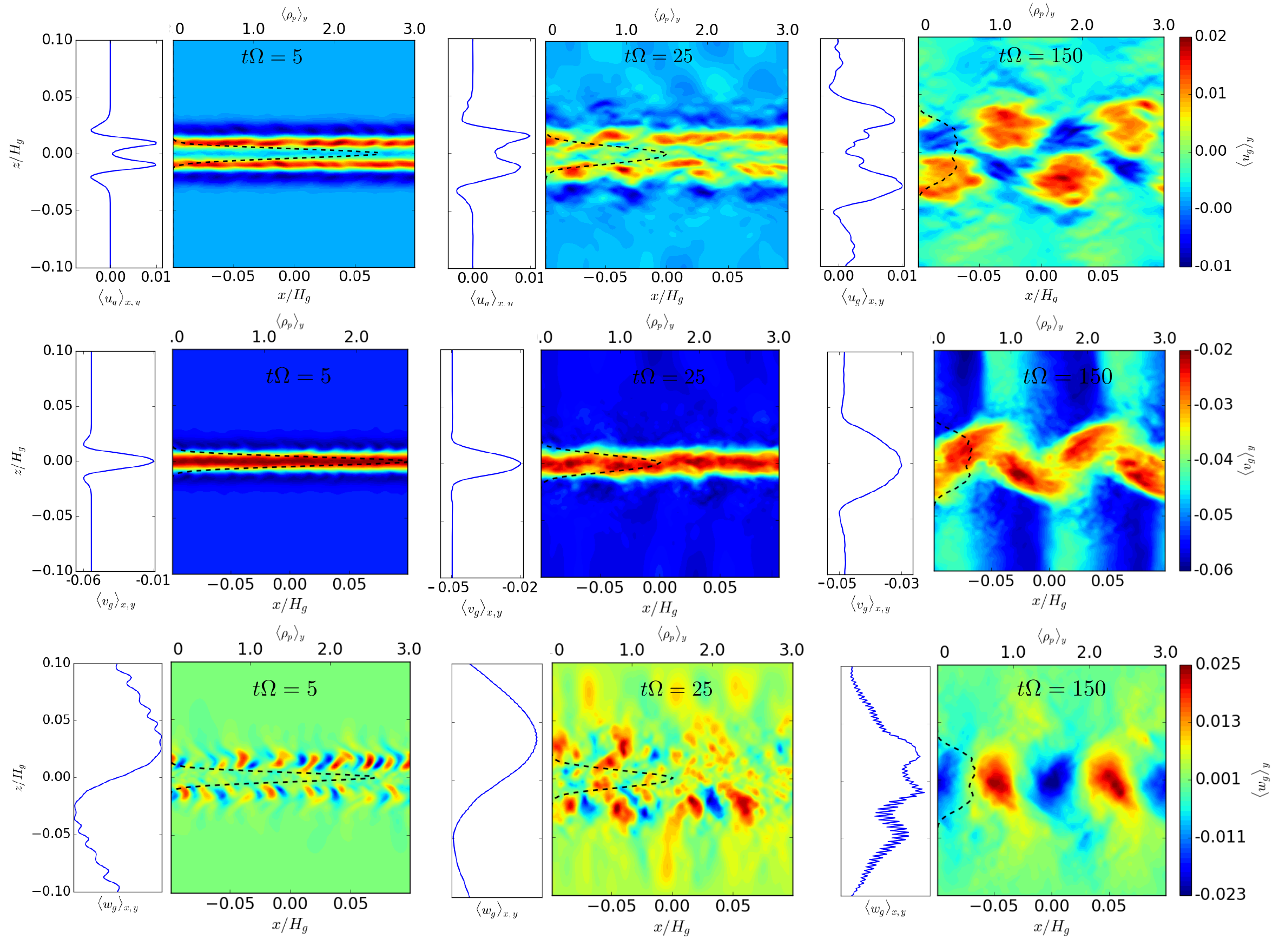}
\par
\end{center}
\caption{Flow structure development for the azimuthally averaged gas velocities by component: $u_g$ (top row, $\left<u_g\right>_y$), $v_g$ (middle row, $\left<v_g\right>_y$) and $w_g$ (bottom row, $\left<w_g\right>_y$).  The figure shows the results for simulation B3D-2L with \St$=0.2$ at $t\Omega=5$ (Bounce state, left column), $25$ (shear driven turbulence state, middle column) and $150$ (mature state, right column). Each figure has two components: the sub-figures on the left show the combined azimuthally and radially averaged vertical profiles of the corresponding flow velocity components ($\left<u_g\right>_{xy}$,$\left<v_g\right>_{xy}$,$\left<w_g\right>_{xy}$). The color plot in each case shows the azimuthally averaged velocity components in the simulation domain. The line plot, superimposed on the 2-D color plot, with the axis on the top, shows the corresponding vertical profile of the solid density ($\left<\rho_p\right>_y$) averaged over $x$ and $y$.}.
\label{fig:velocity-profile-st-0.2}
\end{figure*}

Fig. \ref{spacetime-st-0.04} shows the development for St = 0.04.  The settling and bounce phase, which occurs within $t\Omega=12$,is clearly evident in the $\epsilon_0$ quantity (3$^{\rm rd}$ panel). This initial stage is followed by a relatively long period of time ($t\Omega \sim 250$) in which the fluid appears to be in a turbulent state.  By  $t\Omega=300$ the flow transitions into a symmetry breaking patterned state, in which all quantities exhibit an outwardly propagating traveling wave with approximate wave speed $\approx 7\times 10^{-3} c_s$ (solid black lines in Fig. \ref{spacetime-st-0.04}).  The patterned state appears to fill 2.5 wavelengths on the simulation's radial domain. $\left<v_g\right>_{y,0}$ also exhibits an inwardly propagating secondary pattern with a longer approximated pattern speed $\approx 4\times 10^{-4} c_s$ (hatched black line in left panel of Fig. \ref{spacetime-st-0.04}). This same inwardly propagating pattern is also weakly visibly in the $\bar\Pi'$ field, for which we also note its extremely low amplitudes, $< \order{10^{-4} c_s^2}$, which is consistent with the dynamics here being largely incompressible.\par

During the midplane shear turbulence phase $\left<v_g\right>_{y,0}$ shows weak fluctuations about a mean perturbation velocity $\approx -0.037 c_s$ (i.e., sub-Keplerian).  After transition into the secondary nonlinear state,  $\left<v_g\right>_{y,0}$ increases its oscillation amplitude exhibiting relatively steady fluctuations above this mean value, indicated by the red colored contours in the first panel of Fig. \ref{spacetime-st-0.04}, together with more pulsed fluctuations below this mean value, shown by the blue contours of the same.  The weaker left propagating pattern is only weakly visible in  the midplane particle density and metallicity plot (right two panels of Fig. \ref{spacetime-st-0.04}). A close inspection of these two quantities at about $t\Omega=570$ shows that there is an abrupt downshifting of the outwardly propagating pattern speed to $\approx 4\times 10^{-3} c_s$, slightly more than 40 percent of what it was earlier (hatched magenta lines in the right panels of Fig. \ref{spacetime-st-0.04}).  While the midplane particle densities hover between 1.1 and 1.2 $\rho_{g,0}$ , after the transition into the patterned state $\left<\rho_p\right>_{y,0}$ falls well below $\rho_{g,0}$ as the ratio $\epsilon_0$ generally drops down into the $0.5< \epsilon_0 < 0.8$ range. There are only narrow spatial extents where $\epsilon_0$ only slightly exceeds 1.  We also observe that the metallicity lies in the range $0.006 < Z < 0.014$.
 \par
 
 \begin{figure*}
\begin{center}
\leavevmode
\includegraphics[width=17cm]{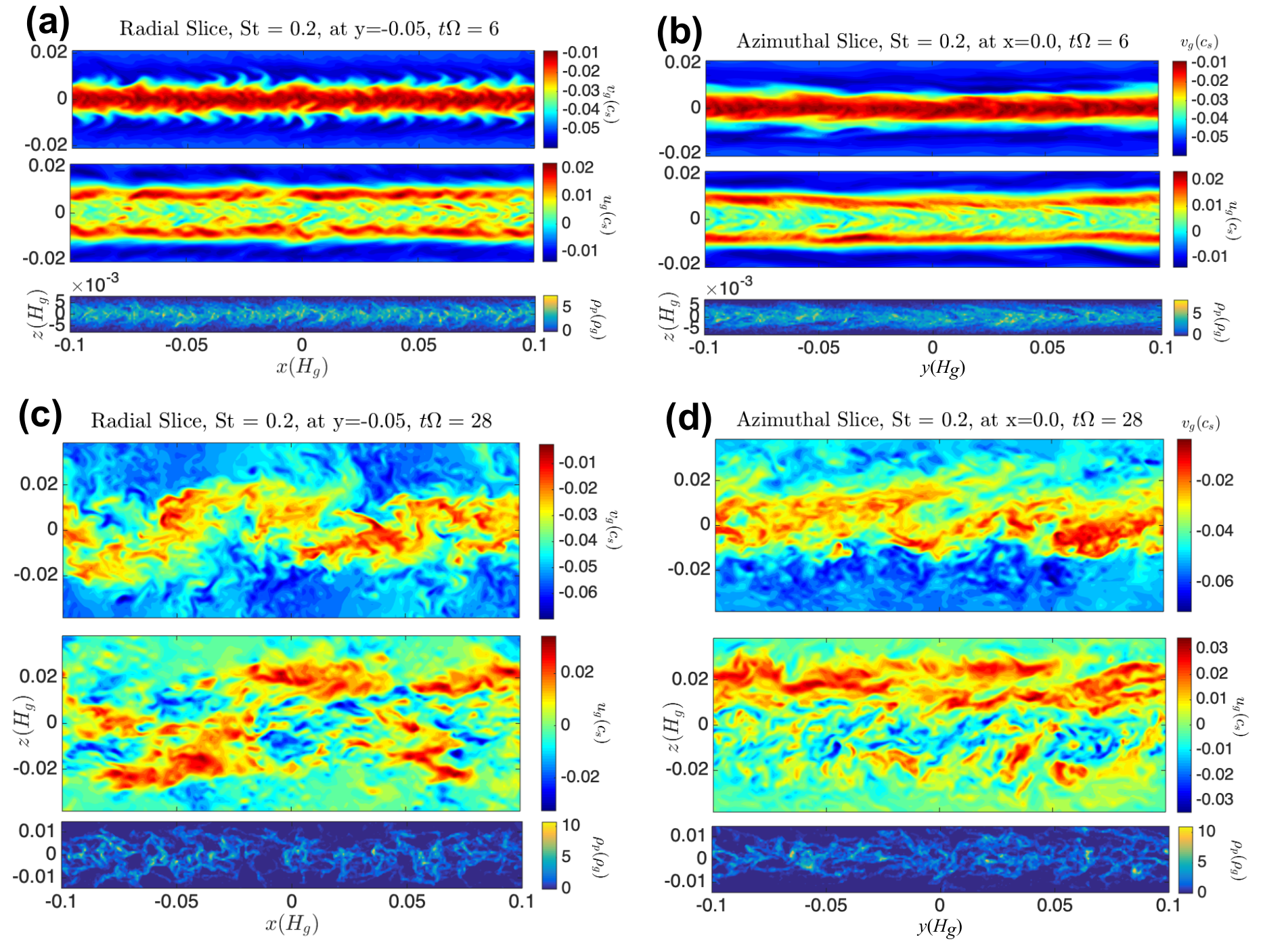}
\par
\end{center}
\caption{Four sets of flow visualizations as slices for St = 0.2 at two different times representing early turbulence development (bounce) phase and the nominally shear turbulent phase.  Each set consists of three rows depicting $u_g$, $u_g$ and $\rho_p$ (respectively): (a) radial slice at $y = -0.05 H$ and $t\Omega = 6$, (b) an azimuthal slice at $x = 0$ and $t\Omega = 6$, (c) radial slice at $y = -0.05 H$ and $t\Omega = 28$, and (d) azimuthal slice at $x = 0$ and $t\Omega = 28$.  These depict the B3D-2M simulation set.}
\label{fig:slices-st-0.2}
\end{figure*}

 Fig. \ref{spacetime-st-0.2} shows the analogous evolution for St = 0.2.  The development sequence is similar to the St = 0.04 case, with a settling/bounce phase ($t\Omega <10$), followed by a turbulent state up to about $t\Omega \approx 40 $, finally leading into a secondary nonlinear patterned state exhibiting about two wavelengths in the radial domain.  However, here the pattern propagation in the secondary state is opposite than what it is in the St = 0.04 case: $\left<v_g\right>_{y,0}$ and $\bar\Pi'$ show inwardly propagating pattern speeds $\approx 1.7\times 10^{-3} c_s$ (solid magenta lines of Fig. \ref{spacetime-st-0.2}'s two left panels), while correspondingly less discernible in the particle fields $\epsilon_0$ and $\bar Z$ (hatched lines of Fig. \ref{spacetime-st-0.2}'s two right panels).  During the bounce phase $\left<v_g\right>_{y,0}$ shows a strong burst (also examined further in the next section), followed by a slow growth of a period-two non-propagating pattern during the midplane turbulent phase (i.e., $10 \lessapprox t\Omega \lessapprox 40$).  The transition into the patterned state becomes manifest ($t\Omega \approx 70$) with an amplitude variation in $\left<v_g\right>_{y,0}$ about a nominal equilibrium value of around $\approx -0.04 c_s$ with extremes between $\pm 0.01 c_s$, painting the picture of an emergent jet flow.
  \par
 Interestingly, $\bar Z$ appears to show a fast moving radial streak pattern (solid black line of Fig. \ref{spacetime-st-0.2}'s far right panel) with a pattern speed $\approx 9\times 10^{-3} c_s$.  $\epsilon_0$ shows a deep spike at the extreme bounce phase (with $\epsilon_0 \approx 2.7$) followed by settling into a quasi-steady turbulent value with $\epsilon_0 \approx 1.5$ before transitioning into the patterned state with a typical value of $\epsilon_0 \approx 0.8$.  Aside from the possibly weak expression of the fast inward drifting pattern, the metallicity shows no particular organization with its values remaining well in the range of 0.0075 and 1.2.

\begin{figure*}
\begin{center}
\leavevmode
\includegraphics[width=17cm]{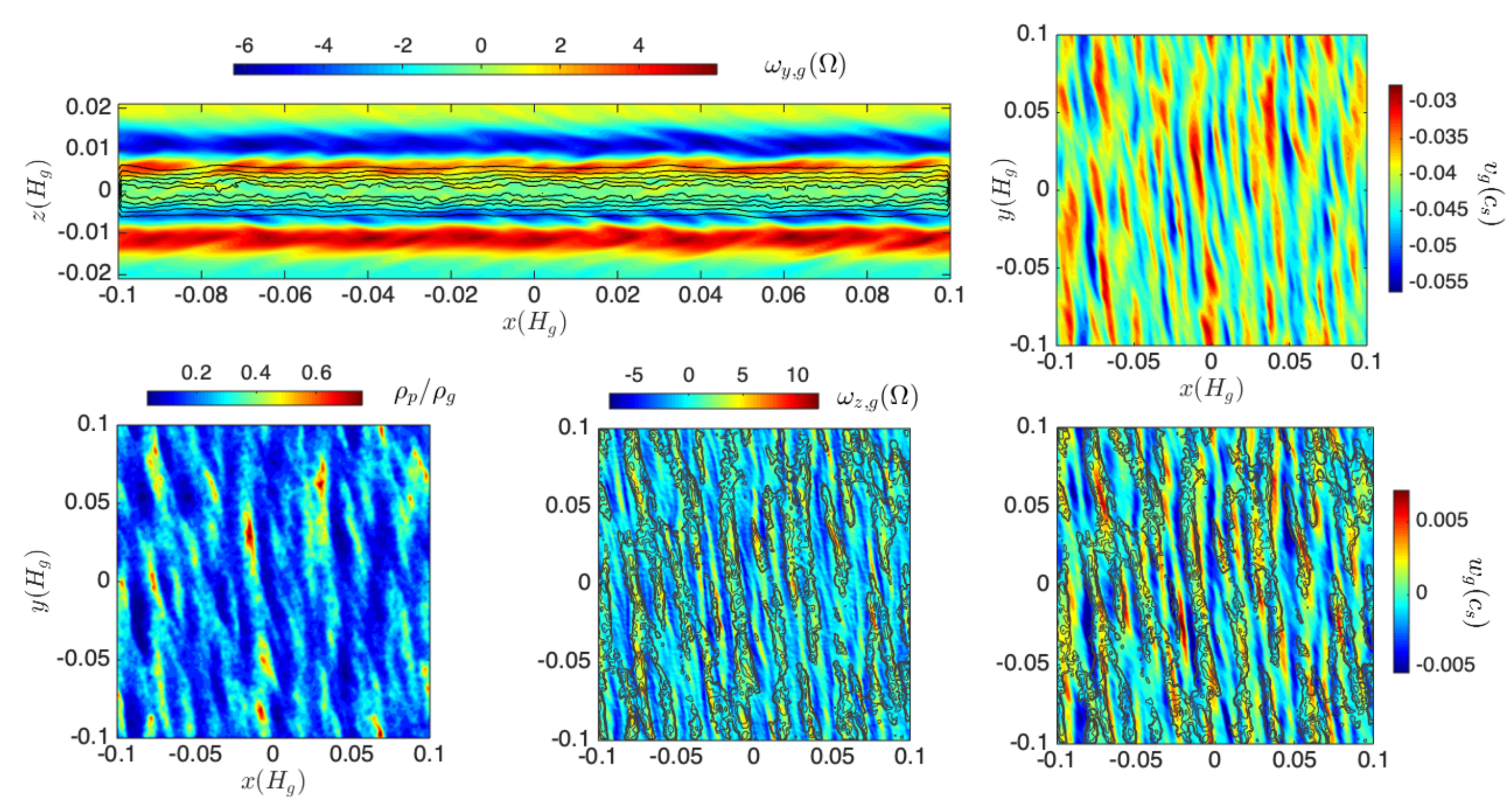}
\par
\end{center}
\caption{More flow visualization for St = 0.2 utilizing B3D-2M simulations at $t\Omega = 6$ (bounce phase).  Top row panels: (left) the azimuthal gas vorticity ($\omega_{y,g}$) formed from the azimuthally averaged gas velocities (see text). The off-midplane counterflowing jet layers ($0.005 < |z/H_g| < 0.015$) show the beginnings of dipolar vortex pair formation, with approximately 14-16 cat's eyes pairs in evidence each above and below the midplane. Overlain in black contours are corresponding isolevels of the azimuthally averaged $\rho_p/\rho_g$. As a function of radial and azimuthal coordinate, the right panel shows the vertical average of $v_g$ across a narrow layer ranging from $0.0055 \le z/H_g \le 0.0095$, corresponding to 5 grid points and nominally centered on the positive $\omega_{g,y}$ vorticity anomaly above the midplane (the red layer) of the left panel. Bottom row panels: vertical averages across the same narrow layer for (left) $\rho_p/\rho_{g}$, (middle) the vertical gas vorticity ($\omega_{z,g}$), and (right) the vertical gas velocity.  Across the positive $\omega_{g,y}$ vorticity anomaly layer the average of $\rho_p/\rho_g$ is $\approx 0.22$. The latter two figures show $\rho_p/\rho_g$ in dark contour lines for values (0.22, 0.25, 0.4, 0.5), all denoting lines of density enhancement compared to the layer's mean.  Positive $w_g$ appears correlated with $\rho_p/\rho_g > 0.22$}.
\label{fig:Flow_Sandwich_Plotter_t6}
\end{figure*}

\subsection{$3$D Simulations}\label{simulation_results_description}
We now present several views of the simulation results and describe their notable characteristics.  We focus our discussion on the early bounce phase and the shear turbulent states of development. 

\subsubsection{St = 0.2}
Fig. \ref{fig:velocity-profile-st-0.2} shows azimuthal and azimuthal-vertical averages of the three gas velocity components, $\left<u_g\right>_y, \left<v_g\right>_y,\left<w_g\right>_y$ and $\left<u_g\right>_{xy}, \left<v_g\right>_{xy}, \left<w_g\right>_{xy}$, respectively. The plot shows three time snapshots nominally representing the three stages of development shown in Fig. \ref{Stages_of_Development}. Each $x$-$z$ slice shows with black dashed lines the corresponding averaged particle densities as a function of disk height, denoted by $\left<\rho_p\right>_y$. The bounce phase is deepest at between $t\Omega = 5$-6, where the emergence of a pair of counterflowing radial jets in $u_g$ can be seen contained in 2 midplane symmetrically placed layers $0.005 < |z/H_g| < 0.02$. The position of these jets are also highlighted in Fig. \ref{fig:Richardson_Number_Plots} (The shaded region in the bottom left figure) in the context of a discussion on the Richardson's Number of the system (see Section \ref{sec:Richardson_Numbers}). Most importantly, the particle layer with $H_p = 0.003H_g$ is localized well away from the off-midplane jet layers.   Moreover, the jet layers shows signs of developing cat's eyes in $u_g$ indicating the ongoing emergence of a dominantly axisymmetric dynamic, which can also be seen in the azimuthally averaged $w_g$ field.  The quantity $\langle v_g\rangle_y$ exhibits a strong azimuthally directed jet mostly coinciding with the extent of the particle layer.  $v_g$ asymptotes to the predicted particle-free pressure balanced limiting value ($ \rightarrow -0.05 c_s$) far from the particle layer.
\par
Fig. \ref{fig:slices-st-0.2}a-b displays slices through the flow field at $t\Omega = 6$.  The first of these displays the gas quantities as a function of radius at a nominal azimuthal position (here $y=-0.05 H_g$).  The vertical extent of the particle layer is well within 0.005$H$ of the midplane (we note that for this snapshot $H_p \approx 0.003 H_g$). The particle field is diffusely filamentary exhibiting outward directed chevron patterning, which is also weakly apparent in the radial velocity $u_g$ field in the same region.  $\rho_p$ typically falls in the values of 1-3 $\rho_g$, with extreme events as high as $5\rho_g$.  Away from the midplane  $u_g$ exhibits the strong counterflowing structure together with the aforementioned signs of roll-up.  The azimuthal velocity has an imprint of the activity seen in $u_g$ in the counterflowing layers above the midplane. There is clear evidence of dynamical activity in the midplane layer as well, but its severity is muted in comparison to what manifests in the counterflowing layers. \par

\begin{figure*}
\begin{center}
\leavevmode
\includegraphics[width=15cm]{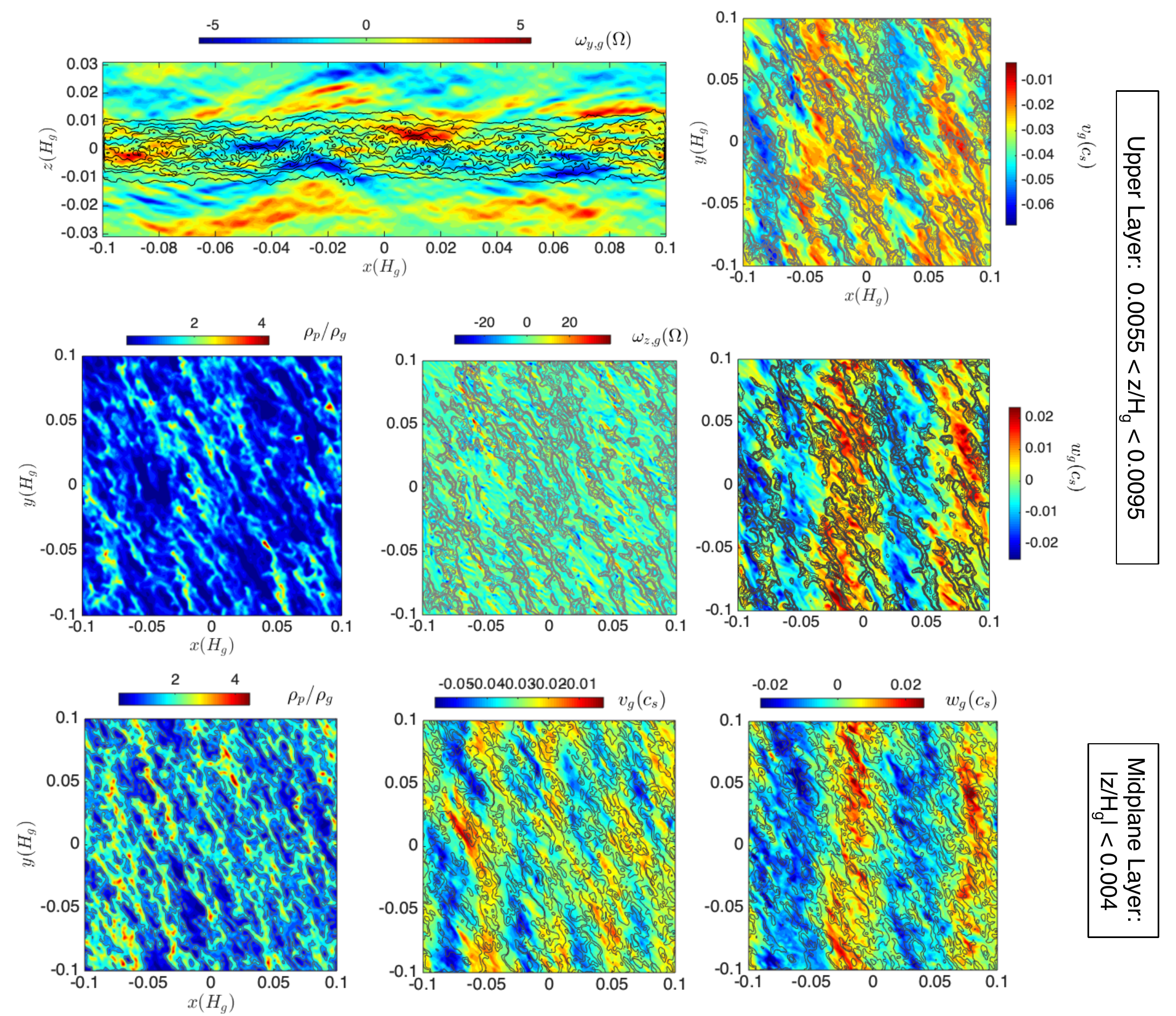}
\par
\end{center}
\caption{Similar to Fig. \ref{fig:Flow_Sandwich_Plotter_t6}, but at $t\Omega = 28$ (shear turbulent phase).  Top row panels: (left) azimuthal average of $\omega_{y,g}$, (right) vertical average of $v_g$ of same off midplane layer of Fig. \ref{fig:Flow_Sandwich_Plotter_t6}.  Middle row panels: 
similar off midplane layer averages of (left) $\rho_p/\rho_g$,
(middle panel) $\omega_{z,g}$, and (right) $w_g$.  The overlain contours,
$\rho_p/\rho_g : (0.81, 1, 1.25, 1.5 , 2 ,3)$,
all represent enhanchements in $\rho_p/\rho_g$ over and above the layer average of 0.81.  
Note the appearance of broadly zonal flow in $v_g$ exhibiting finer scale wispy substructure that is found in other quantities.  
Density contours shown for  in which the first is the layer's mean.
Bottom row panels: shows analagous midplane centered averages ($|z/H_g| \le 0.004, \leftrightarrow $ 9 grid points): (left) $\rho_p/\rho_g$, (middle), $v_g$, and (right) $w_g$. Density contours shown for $\rho_p/\rho_g = 1, 1.54$ only, where the latter corresponds the layer's mean.}
\label{fig:Flow_Sandwich_Plotter_t28}
\end{figure*}

Fig. \ref{fig:slices-st-0.2}b shows an azimuthal slice at the radial position $x=0$.  The particle density field similarly displays filamemtary chevron patterning directed toward the increasing azimuthal direction, conforming with the mean vertical structure of $v_g$ which is greater near the midplane than further away. Fig. \ref{fig:Flow_Sandwich_Plotter_t6} displays the azimuthal vorticity defined as

\beq
\omega_{y,g} \equiv \frac{\partial\left<u_g\right>_y}{\partial z} - \frac{\partial\left<w_g\right>_y}{\partial x}.
\eeq

\noindent Overlain are contours of constant azimuthally averaged particle density, once again depicting that those regions remain far away from the active layers above and below the midplane.  The figure shows a radial-azimuthal planar plot of $v_g$ vertically averaged across a layer containing the positive $\omega_{y,g}$ vorticity anomaly above the midplane, i.e., for $0.0055 H_g \lessapprox z \lessapprox 0.0095 H_g$.  The imprint of the strong developing axisymmetric dynamic is evident with the emergence of zonal-flow like structure with radial periodicity of $\sim 0.02H$.  For the same layer the figure also shows an average of the particle velocity field exhibiting fluctuations with the same pattern.  Similarly, the layer restricted vertical average of $w_g$ is shown with contours of the vertical layer average of $\rho_p$, showing a strong correlation between positive vertical velocity and positive density anomaly indicating that the emergent roll-up dynamics vertically advects the settled particle layer below.
\par
In the shear turbulent phase all quantities show the signs of turbulent motions, but with some retention of basic counterflowing jet flow that led to instability. As the second column of Fig. \ref{fig:velocity-profile-st-0.2} shows ($t\Omega = 25$), the jet layer structure in $\left<u_g\right>_y$ has fragmented while still retaining some discernible axisymmetric structure.  Structure in  $\left<u_g\right>_y$ shows  vertical spread (up to $\pm 0.05 H_g$). A similar vertical spread is also seen in $\meanrhop$ (with $H_p\approx 0.05 H_g$). Midplane asymmetry has developed in  $\left<u_g\right>_{xy}$; but,  $\left<v_g\right>_y$ remains largely intact with clear evidence of the emergence of some organized axisymmetric structure near the midplane. The averaged vertical velocity field $\left<w_g\right>_y$ has fragmented into small scale structures that extend as far as those structures observed for $\left<u_g\right>_y$.

However, the radial and azimuthal slices at this shear turbulent stage, shown in Fig. \ref{fig:slices-st-0.2}c-d, tell a story that is lost if one focuses purely on the azimuthal averages.  Fig. \ref{fig:slices-st-0.2}c displays a large scale radial sinusoidal pattern appearing in $v_g$, with wavelength about half of the box size.  Imprinted on that pattern are small scale unsteady turbulent motions.  $\left<u_g\right>_y$ also shows a pattern of strong positive value following the sinusoidal structure observed in $v_g$, with the regions in between interspersed with regions of negative velocity.  Moreover, the spatial distribution of the particles appear restricted to with $\pm 0.01H$, but now shows more dramatic filamentary structure with densities in places as large as $10 \rho_g$.  The filaments appear comparably oriented with the midplane as with the vertical. Fig. \ref{fig:slices-st-0.2}d, which depicts an azimuthal slice at the middle of the box $x=0$, shows similar disordered turbulent quality imprinted on broad segregated zones of positive or negative mean values of $u_g$ and $v_g$.  Similarly, the azimuthal slice of $\rho_p$ shows filamentary structure like seen in the radial slice with the only difference possibly being that the filaments are more aligned parallel with the midplane than with the vertical. 

Fig. \ref{fig:Flow_Sandwich_Plotter_t28} shows how $\omega_{y,g}$ is developing strong coherence conforming to the period-2 radial wave structure mentioned above.  The particle layer, while still mainly contained around the midplane, also expresses the period-2 wave structure.  Moreover, vertical averages of $\rho_p$ and $v_g$ across the same off-midplane layer discussed in Fig. \ref{fig:Flow_Sandwich_Plotter_t6} show that the azimuthally aligned structures start to fragment  with a tilt $\approx45^\circ$ from the upper left toward the lower right.  In this orientation $\rho_p$ shows wispy high density structure that is reminiscent of filamentary density structures characteristic of simulations in which the SI is known to be operative \citep[e.g., see Figure 1 of][and several others]{Simon_etal_2017}.  The layer average of $v_g$ exhibits a period-2 axisymmetrically banded zonal flow structure with similarly finely layered 45$^\circ$ oriented wisps seen in $\rho_p$. However, there does not appear to be any correlation between high density filaments with the relative departures of $v_g$ with respect to its layer mean: high density filaments appear together with both high and low amplitude values of $v_g$, only the relative gross orientation of the finer scale structures seem to correlate.  A similar correlated pattern is seen between $\rho_p$ and the layer average of $w_g$. Aside from streak orientation, there is even less correlation between high values of $\rho_p$ and the corresponding layer average of $\omega_{z,g}$, where the gas vertical vorticity is defined as

\beq
\omega_{z,g} \equiv
\frac{\partial v_g}{\partial x} - 
\frac{\partial u_g}{\partial y} .
\eeq

\begin{figure}
\begin{center}
\leavevmode
\includegraphics[width=8.5cm]{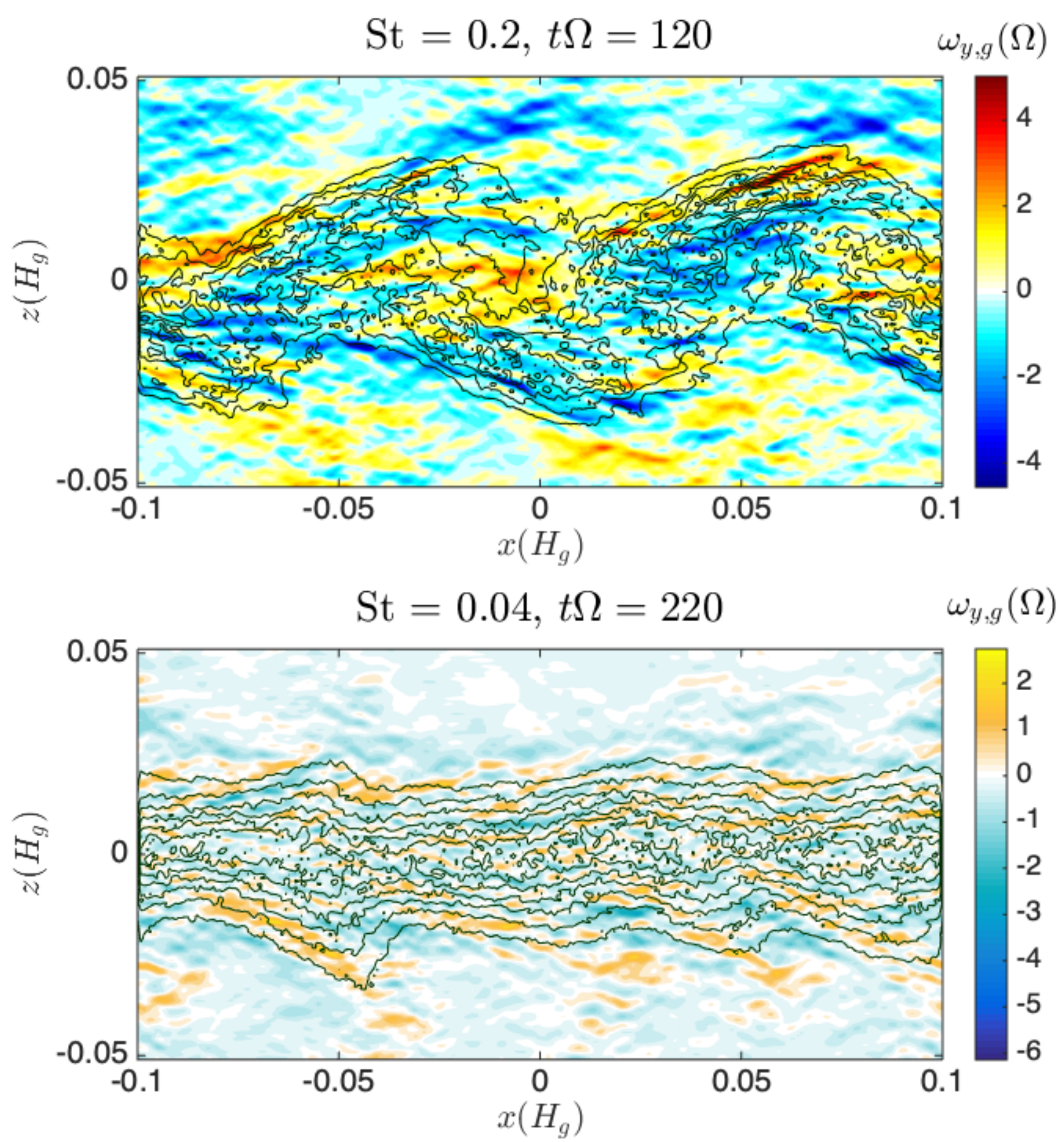}
\par
\end{center}
\caption{Radial slice depictions of $\omega_{y,g}$ and $\left<\rho_p\right>_y$ during late stage development after secondary transition. (top) St = 0.2 at $t\Omega = 120$ with contours on $\left<\rho_p\right>_y$ ranging from 0 to 0.8 $\rho_{g,0}$ (simulation B3D-2M) 
(bottom) St = 0.04 at $t\Omega = 220$ 
with $\left<\rho_p\right>_y$ contours
ranging from 0 to 1.2$\rho_{g,0}$
(simulation B3D-04M).
}
\label{fig:late_stage_vorticities}
\end{figure}

Finally, in the secondary pattern state, the layer expresses a strong period-2 sinusoidal disturbance in all quantities.  This midplane layer undulation phenomenon has been observed in several simulations in which the SI is the primary dynamical driver \citep[e.g.][and others]{Yang_etal_2017,Yang_etal_2018,Li_etal_2018,Gerbig_etal_2020}.  The final column of Fig. \ref{fig:velocity-profile-st-0.2} most unambiguously illustrates this state of affairs.  The interlaced but steadily disintegrating off-midplane configuration of  $\left<u_g\right>_y$ found during the shear turbulence phase has transitioned into a coherent midplane crossing zig-zagging oscillatory pattern.  Interestingly,  while $\left<v_g\right>_y$ shows similar period-2 oscillatory character but where the near midplane azimuthal jet profile now appears crenellated, far from the midplane the azimuthal gas velocity field shows an alternating vertically oriented radial pattern where the far field value of $\left<v_g\right>_y$ now oscillates around its particle-free limiting value.  Likewise, $\left<w_g\right>_y$ shows radial oscillation indicating that the particle layer is similarly sinusoidally undulating.  This is borne out in the top panel of Fig. \ref{fig:late_stage_vorticities}, where the particle layer has entered into an organized sinusoidal configuration.  A detailed examination of the nature of this stage is reserved for a future publication.

\begin{figure*}
\begin{center}
\leavevmode
\includegraphics[width=17cm]{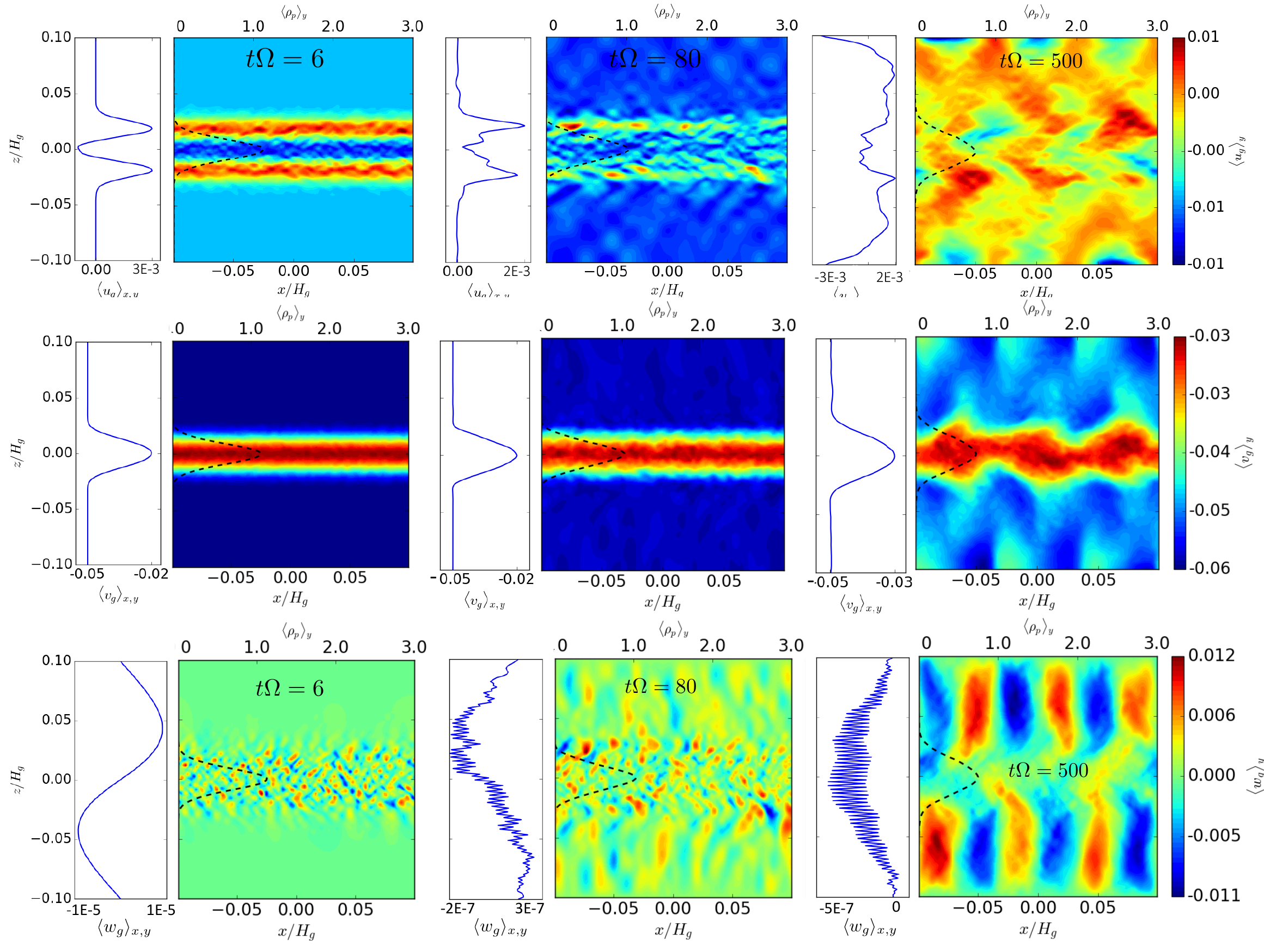}
\par
\end{center}
\caption{Representation of the development of velocity profiles, similar to figure \ref{fig:velocity-profile-st-0.2}, for \St$=0.04$ (simulation B3D-04L): snapshots at $t\Omega=6$ (bounce phase, left column), $t\Omega=80$ (shear driven turbulence phase, middle column) and $t\Omega=500$ (mature state, right column).  Note the period-3 banded structure in all three fields in the mature state. }
\label{fig:velocity-profile-st-0.04}
\end{figure*}

\begin{figure*}
\begin{center}
\leavevmode
\includegraphics[width=17cm]{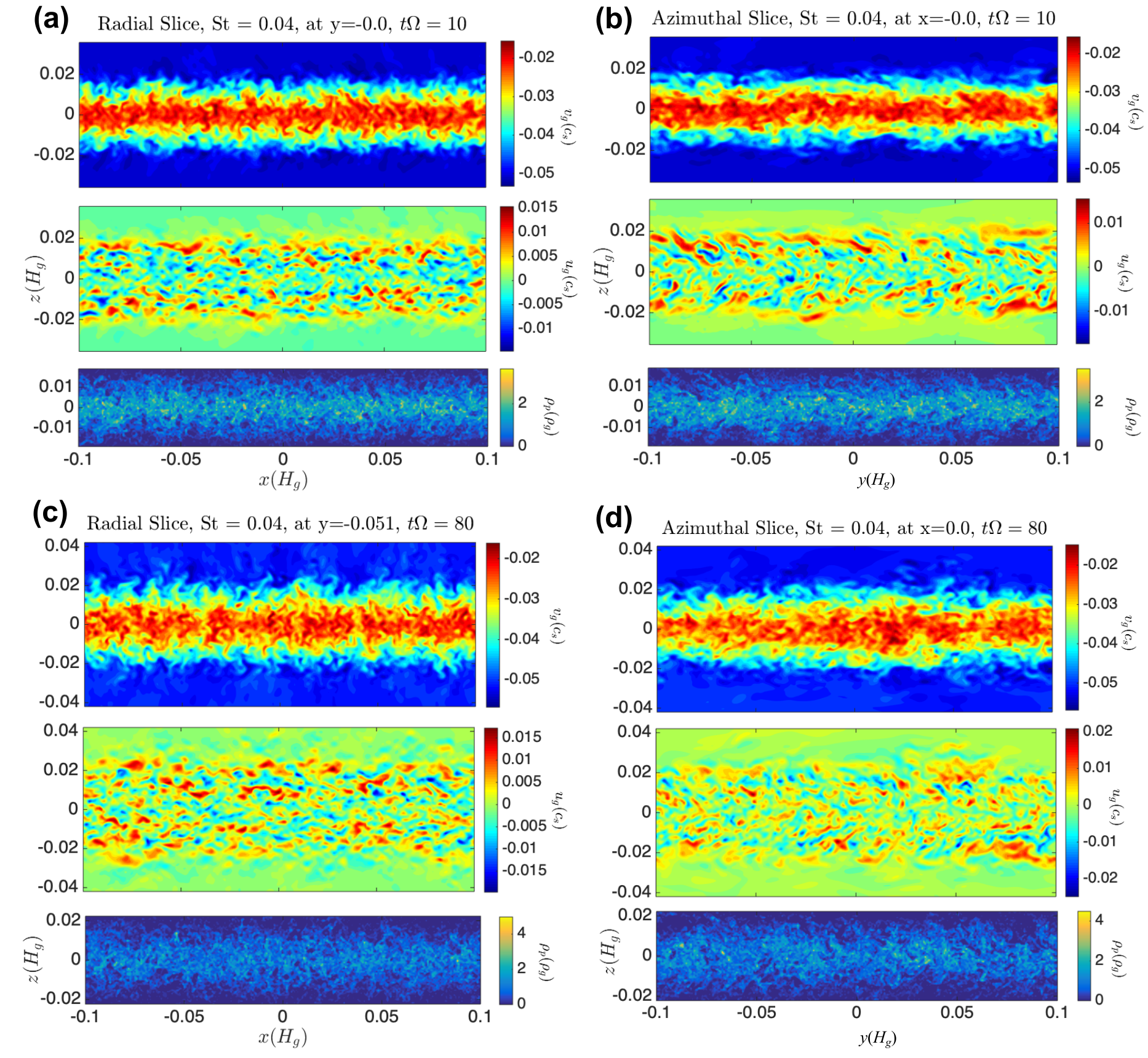}
\par
\end{center}
\caption{ Flow visualizations analogous to Fig. \ref{fig:slices-st-0.2} except for St = 0.04:
(a) radial slice at $y = -0.0 H$ and $t\Omega = 10$, (b) an azimuthal slice at $x = 0$ and $t\Omega = 10$, (c) radial slice at $y = -0.05 H$ and $t\Omega = 80$, and (d) azimuthal slice at $x = 0$ and $t\Omega = 80$.  These depict the B3D-04M simulation results.}
\label{fig:slices-st-0.04}
\end{figure*}

\subsubsection{St = 0.04}
Analogous to Fig. \ref{fig:velocity-profile-st-0.2}, 
Fig. \ref{fig:velocity-profile-st-0.04} depicts azimuthally averaged flow fields during the three stages of development.  During the early developing bounce phase (see figure's first column)
$\left<u_g\right>_y$ develops a jet-like structure above and below the midplane just like for \St$ = 0.2$ case, but its amplitude is weaker by a factor of $10$ as $\left<u_g\right>_{xy}$ clearly illustrates.  There are no obvious development of Kelvin's cat's eyes unlike the \St$ = 0.2$ case. There is no discernible structure in $\left<v_g\right>_y$ aside from weak perturbations atop the dominant midplane jet structure.  Similarly, $\left<v_g\right>_y$ shows perturbations that are of very small scale and amplitude and confined to within the layer containing the bulk of the particles, whose scale height is around $H_p\sim 0.01H_g$.  
\par 
Once again, the radial/azimuthal slice images (Fig. \ref{fig:slices-st-0.04}a-b) demonstrate that in fact the layer is strongly active during the bounce phase.  The radial slice of $v_g$ shows that this layer is undergoing significant dynamical activity with the appearance of plumes up through to where the particle layer effectively terminates ($\sim 2.5H_p$).  The plumes' lengthscales are between $0.005H_g$ and $ 0.01H_g$. The $u_g$ field shows activity restricted to within the dust layer; in contrast to the \St$ = 0.2$ case where dynamics in $u_g$ extends far beyond the dust layer. $u_g$ also shows structure on the scale of the plumes, but whose horizontal scales are anywhere from 2 to 3 times the vertical scales. Structure in  $u_g$ also appears to be larger in size up past one to two particle scale heights and, moreover, shows no obvious organization like there seen during the bounce phase of \St$ = 0.2$ (c.f., Fig. \ref{fig:slices-st-0.2}a).  The radial slice of $\rho_p$ shows that the filaments  are far more diffuse and seem to follow the textures seen in $v_g$; overall the $\rho_p$ field is far more nondescript compared to the St = 0.2 case. The azimuthal slices shown in Fig. \ref{fig:slices-st-0.04}b follow the general tenor of the qualities exhibited in the radial slice case with perhaps the only real difference being that the $v_g$ and $u_g$ fields are slightly more azimuthally elongate especially at heights about 1-2 $H_p$ from the midplane.
\par
For St = 0.04 the turbulence phase takes root by $t\Omega = 40$.  The second column of Fig. \ref{fig:velocity-profile-st-0.04} shows that the weak jet structure that emerged during the bounce phase has fragmented somewhat and that its overall structure has significant asymmetries.  The vertical extent of the particle layer has expanded some and $\left<v_g\right>_y$ is now showing signs of dynamical unsteadiness.  The $\left<w_g\right>_y$ too shows that there is a qualitative transition with structures growing in size and extending vertically across the domain, with the appearance of some amount of diffuse vertical alignment in the field.
\par
Remarkably, the radial/azimuthal slices (Fig. \ref{fig:slices-st-0.04}c-d) during this turbulent phase seem to show that the overall qualitative character of the unsteady motions emerging during the bounce phase characterise the turbulent flow as well.  Aside from stretching its vertical extent a bit, the character of $v_g$ and $u_g$, in both of their  azimuthal and radial slices, look very much like what they look like during the bounce phase: unsteady motions with plumes in $v_g$ at 1-2$H_p$ distance from the midplane, with small scale structures in $u_g$ on similar scales.  Perhaps the only significant difference is that filaments in $\rho_p$ are somewhat finer, where higher values of $\rho_p$ are achieved compared to the early development.  Nonetheless, $\rho_p$ is generally diffuse especially when compared to the situation encountered in the corresponding St = 0.2 case (c.f., see the $\rho_p$ fields of Fig. \ref{fig:slices-st-0.2}c-d).
\par
\begin{figure}
\begin{center}
\leavevmode
\includegraphics[width=9cm]{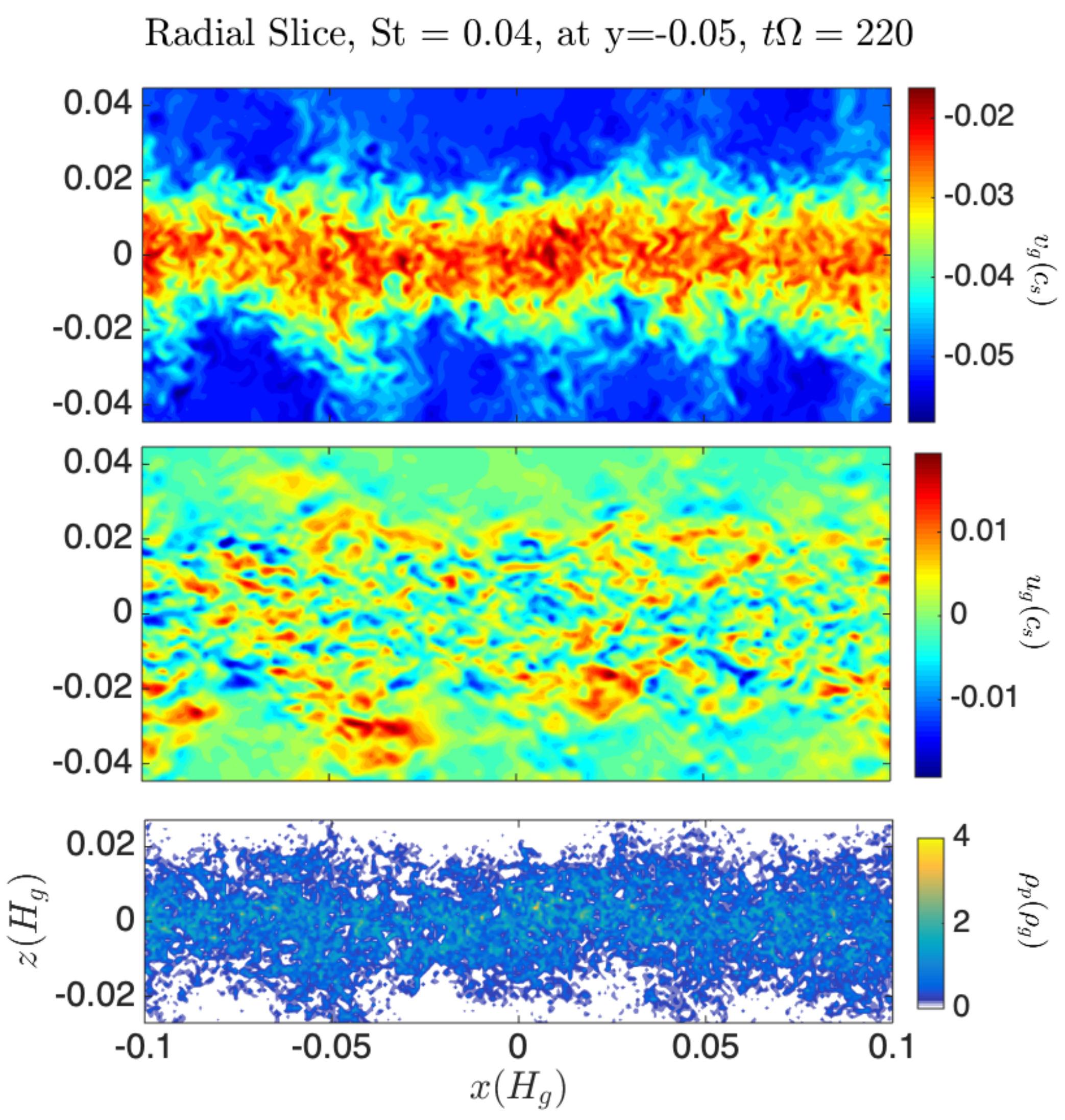}
\end{center}
\caption{Radial Flow slices for St = 0.04, $t\Omega = 220$.  These depict the B3D-04M simulation results.}
\label{Radial_Slice_Late_Stage_St0.4}
\end{figure}

As the final column of Fig. \ref{fig:velocity-profile-st-0.04} shows,
when the flow has fully transitioned into its secondary state ($t\Omega \gtrapprox 300$) the flow fields have transformed as well.  A period 3 midplane symmetric pattern emerges in $\left<v_g\right>_y$ up to the vertical extent of the particle layer whose $H_p \sim 0.015 H_g$.    $\left<u_g\right>_y$ develops into a vertical domain filled with organized structure of zig-zagging contours that extends far away from where the particles are mostly concentrated: $\left<u_g\right>_y$ exhibits an outward pointing chevron pattern within the particle layer but then switches its orientation when moving away about 2-3 $H_p$ from the midplane.  It is also remarkable that the vertical gas velocity $\left<w_g\right>_y$ field is nearly zero within the particle layer but then takes on a period 3 nearly vertically oriented alternating band structure away from the particle layer, as similarly observed by \cite{Li_etal_2018}.  The reasons for this curious feature are not clear.  The bottom panel of Fig. \ref{fig:late_stage_vorticities} shows $\omega_{y,g}$ overlain with  $\left<\rho_p\right>_y$ contours demonstrating the emergence of a period 3 structure here as well. Finally, Fig. \ref{Radial_Slice_Late_Stage_St0.4} shows radial slices during the beginning of the late stage and it is notable that the general turbulent character seen in the earlier stage, especially in the $v_g$, persists as the sinusoidal structure begins to set in.


\subsection{A comparison of 3D axisymmetric dynamics with those uncovered in full 3D flows}

\begin{figure}
\begin{center}
\leavevmode
\includegraphics[width=9cm]{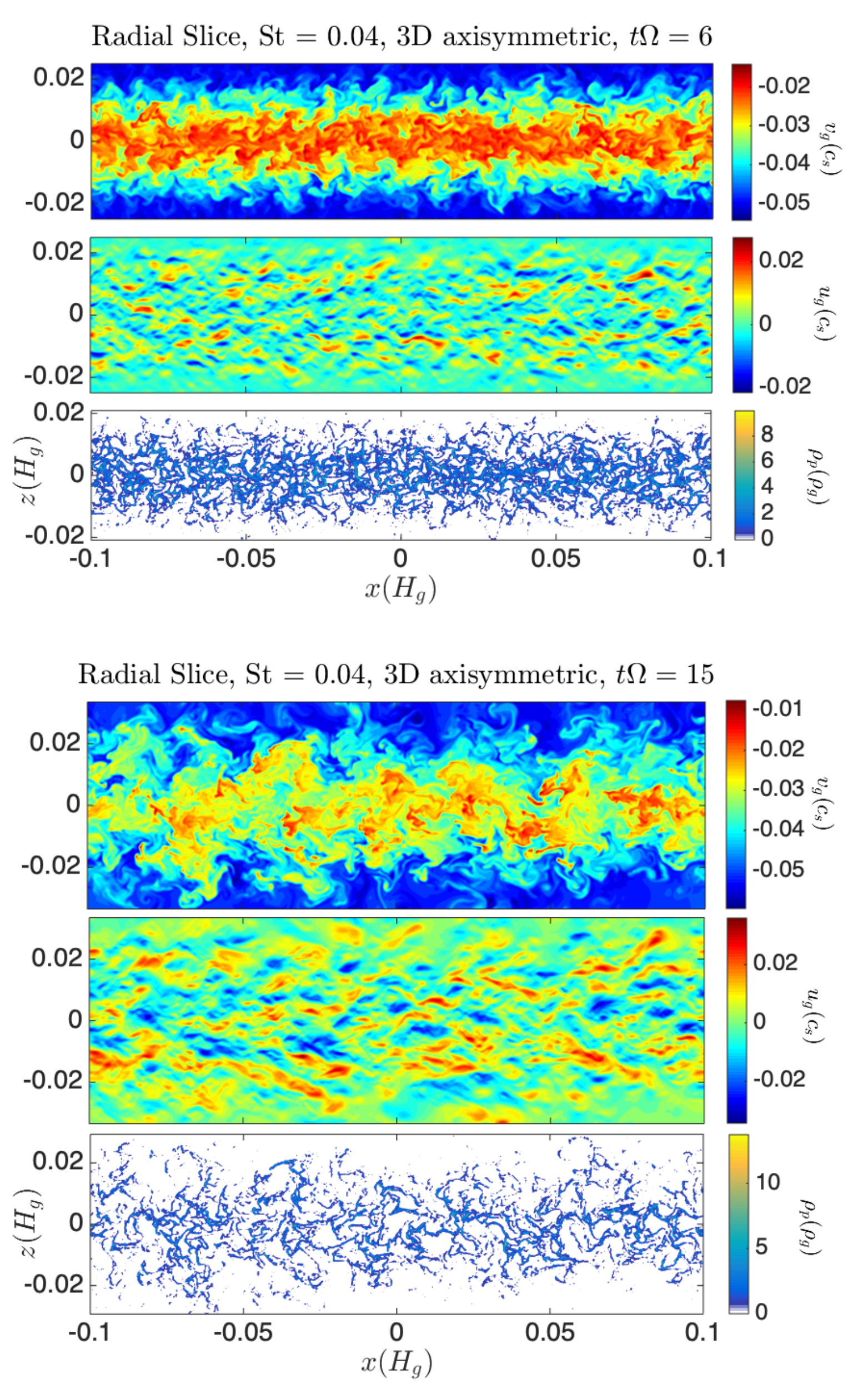}
\end{center}
\caption{Flow slices like in Fig. \ref{fig:slices-st-0.04}, except for 3D axisymmetric simulations with St = 0.04 (simulation A2D-04H).  Top set of 3 panels $\leftrightarrow t\Omega = 6$, while bottom set of 3 panels $\leftrightarrow t\Omega = 15$. }
\label{Comparative_Plottage_2D_St0.4}
\end{figure}

We consider a limited set of 3D axisymmetric simulations in an effort to gain some insight about the emergent turbulent dynamics reported in the previous subsection.  We run these simulations specifically to examine how the transition from the bounce phase into the turbulent state takes shape.  We are wary of running these axisymmetric simulations much farther than these early phases simply because secondary and tertiary transitions involving non-axisymmetric mechanisms likely characterize the true descent into turbulence in the 3D simulations discussed so far.  Thus, any of the interesting features that manifest in the axisymmetric case likely get washed away under the more realistic scenario.  Despite this, some useful insights can be inferred. To be concrete with terminology, 3D axisymmetric means to refer to runs in which all three components of position and velocity are present but are only dependent on the radial and vertical coordinates.  In contrast to this, hereafter we sometimes refer to the full 3D calculations as ``unrestricted 3D".
\par
Fig. \ref{Comparative_Plottage_2D_St0.4} shows the axisymmetric development of $\rho_p,v_g,u_g$ for St = 0.04 in straight analogy with Figs. \ref{fig:slices-st-0.04}a,c and Fig. \ref{Radial_Slice_Late_Stage_St0.4}.  Remarkably, we find that the instability development bounce phase  velocity fields $v_g$ and $u_g$ (i.e., $t\Omega = 6$, top three panels of Fig. \ref{Comparative_Plottage_2D_St0.4}, simulation A2D-04H) look qualitatively identical to the radial flow slices of $v_g$ and $u_g$ {\it at every stage} of the corresponding full 3D simulation (simulation B3D-04M). Even during the emergent phase of the secondary state (e.g., see top panel of Fig. \ref{Radial_Slice_Late_Stage_St0.4}), whilst the layer exhibits a period 2-3 radial sinusoidal variation, the small scale clearly turbulent dynamics exhibited by $v_g$ are essentially the same as in the axisymmetric case.  These trends suggest that the dynamics of the unrestricted 3D case are {\it not} primarily driven by KH-roll-up in the azimuthal direction, as is commonly assumed to be the case; {\it that instability in the St = 0.04 case is primarily an axisymmetric phenomenon.} Moreover, the full 3D simulations also seem to evolve in a way that the flow fields look more like what they look like during its early turbulent phase; perhaps suggesting some type of self-regulation mechanism at work, in which the system is always sufficiently above -- but not too far from -- an instability threshold.  Indeed, by comparison with the later time stamp illustration of the axisymmetric simulation ($t\Omega = 16$, bottom three panels of Fig. \ref{Radial_Slice_Late_Stage_St0.4}), the dynamical zone has puffed out to higher levels in $z$ with attendant appearances of wispy structures and ever finer scale vortex structure.
\par
This direct comparison also shows that the particle densities tend to be higher in the axisymmetric simulations despite the fact that the flow field dynamics are similar to one another. The filaments developing in $\rho_p$ are of finer scale and far more spindly compared to the filaments observed in the corresponding radial slices of $\rho_p$ in Fig. \ref{fig:slices-st-0.04}a and Fig. \ref{fig:slices-st-0.04}c.  Values of $\rho_p/\rho_g$ within particle filaments can get as high as 8-10 in the axisymmetric case while they rarely exceed values of 3-4 in corresponding full 3D simulations.  Also, filament sharpening and attendent void space growth in the axisymmetric case appears to intensify as the simulation evolves.  Overall, this may be a consequence of a downscale forward enstrophy cascade occurring in the axisymmetric case, which should induce sharpening of filamentary structures.  On the other hand, non-axisymmetric motions will readily disrupt such coherent filament development.  However, at this stage this remains a conjecture that should be investigated further.  Nonetheless, these trends behoove exercising caution before interpreting the results of axisymmetric simulations as being applicable to full 3D scenarios.
\par

\begin{figure}
\begin{center}
\leavevmode
\includegraphics[width=9.cm]{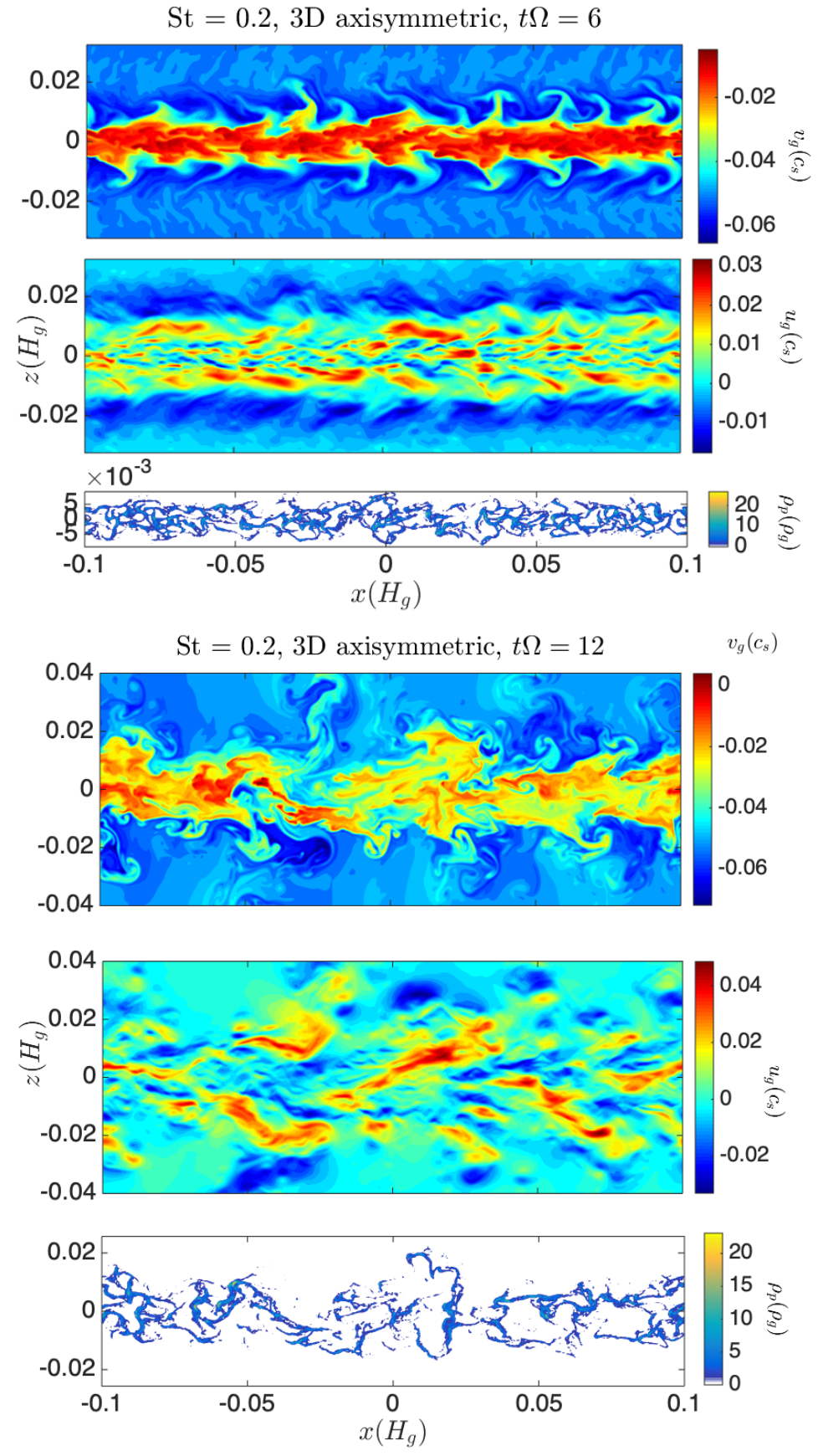}
\end{center}
\caption{Flow slices like in Fig. \ref{fig:slices-st-0.2}, except for 3D axisymmetric simulations with St = 0.2 (simulation A2D-2H).  Top set of 3 panels  $\leftrightarrow t\Omega = 6$, while bottom set of 3 panels $\leftrightarrow t\Omega = 12$. Note the emergence of strong diploar vorticity plumes in $v_g$ for $t\Omega = 6$.}
\label{Comparative_Plottage_2D_St0.2}
\end{figure}

Fig. \ref{Comparative_Plottage_2D_St0.2} shows the analogous axisymmetric development of St = 0.2  that ought to be compared against the results of the corresponding unrestricted 3D flow fields shown in Figs.\ref{fig:slices-st-0.2}a,c.  During the early instability development phase ($t\Omega = 6$, top set of three panels of Fig. \ref{fig:slices-st-0.2}, simulation A2D-2H), $v_g$ shows the emergence of dramatic plumes directed away from the midplane and originating near where the particle layer ends.  Signs of this can be seen in $v_g$ for the full 3D simulation at about the same time (Fig.\ref{fig:slices-st-0.2}a), although the plumes there appear to be somewhat muted in comparison, appearing more wispy.  Unlike the unrestricted 3D case, $u_g$ does not exhibit the same clear signs of emergent cat's eye structure within the off-midplane counterflowing jet layer for $0.005 H_g \lessapprox |z| \lessapprox 0.015 H_g$, although there are clear signs there of large amplitude sinusoidal variation in $u_g$ contours.  Nevertheless, the midplane layers  containing particles exhibit complex textural structure that is qualitatively similar to the emergent unstable dynamics seen in the St = 0.04 case, but to a far more muted extent.  

Also, similar to our concerns above, $\rho_p$ is focused into filaments of stronger relief in the axisymmetric case than compared to what emerges in the full 3D simulations.  The typical density count in the filaments emerging from the axisymmetric simulation are also nearly a factor of two larger than what they are in the corresponding full 3D simulation. 

We observe that by the time the axisymmetric simulation is sufficiently passed the bounce phase, the flow field structure that develops in both $u_g$ and $v_g$ ($t\Omega = 12$, Fig. \ref{Comparative_Plottage_2D_St0.2}) diverges in quality from what normally develops in the full 3D case at similar times.  In particular, several plume-like phenomena in $v_g$ extend significantly away from the midplane with no accompanying particle filaments.  For example, at $t\Omega = 12$ there is a pronounced plume-filament structure in $v_g$ lying between $x = -0.05H_g$ and 0, and above the midplane between $z=0.02H_g$ and $z=0.04H_g$.  Cross-referencing this structure against the map of $\rho_p$ show there are no particles there. There are several other instances of this feature throughout the simulations studied.  Conversely, there are also features in $v_g$ that do correlate with enhanced particle locations as in the case of the dramatic, near-midplane, symmetrically oriented particle filament found in $0\le |z/H_g| \le 0.02$ and $0<x/H_g < 0.02$, which corresponds to a similarly shaped texture in $v_g$ at the same location.  
\par
The situation becomes even more muddled when one attempts to find connections between particles and gas flow fields in the full 3D calculations as no clear correspondences lend themselves to easy visual detection. This observation raises the question of how exactly do the particles influence the turbulent dynamics once the turbulence sets in?

\begin{figure*}
\begin{center}
\leavevmode
\includegraphics[width=14cm]{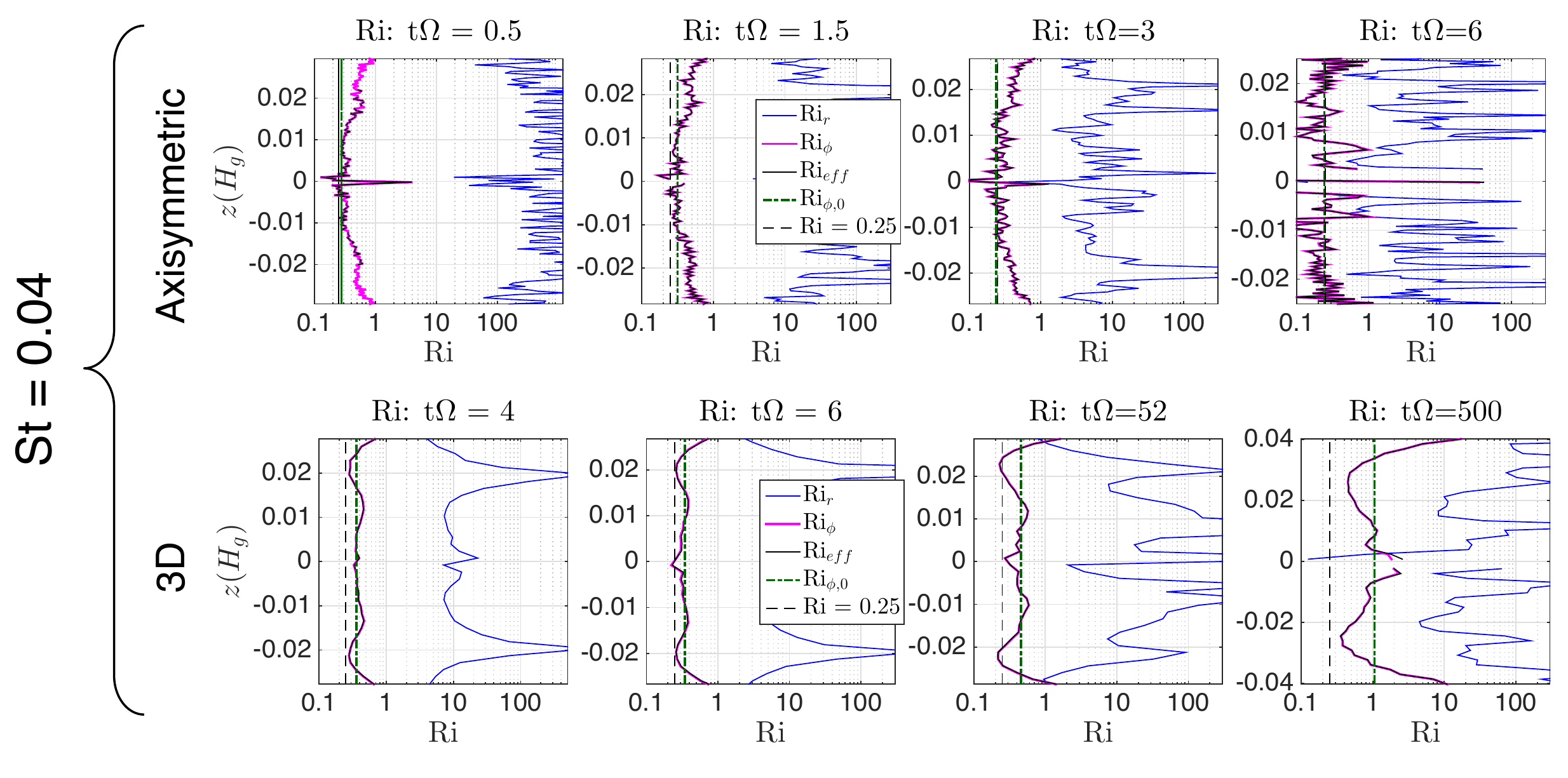}\par
\includegraphics[width=14cm]{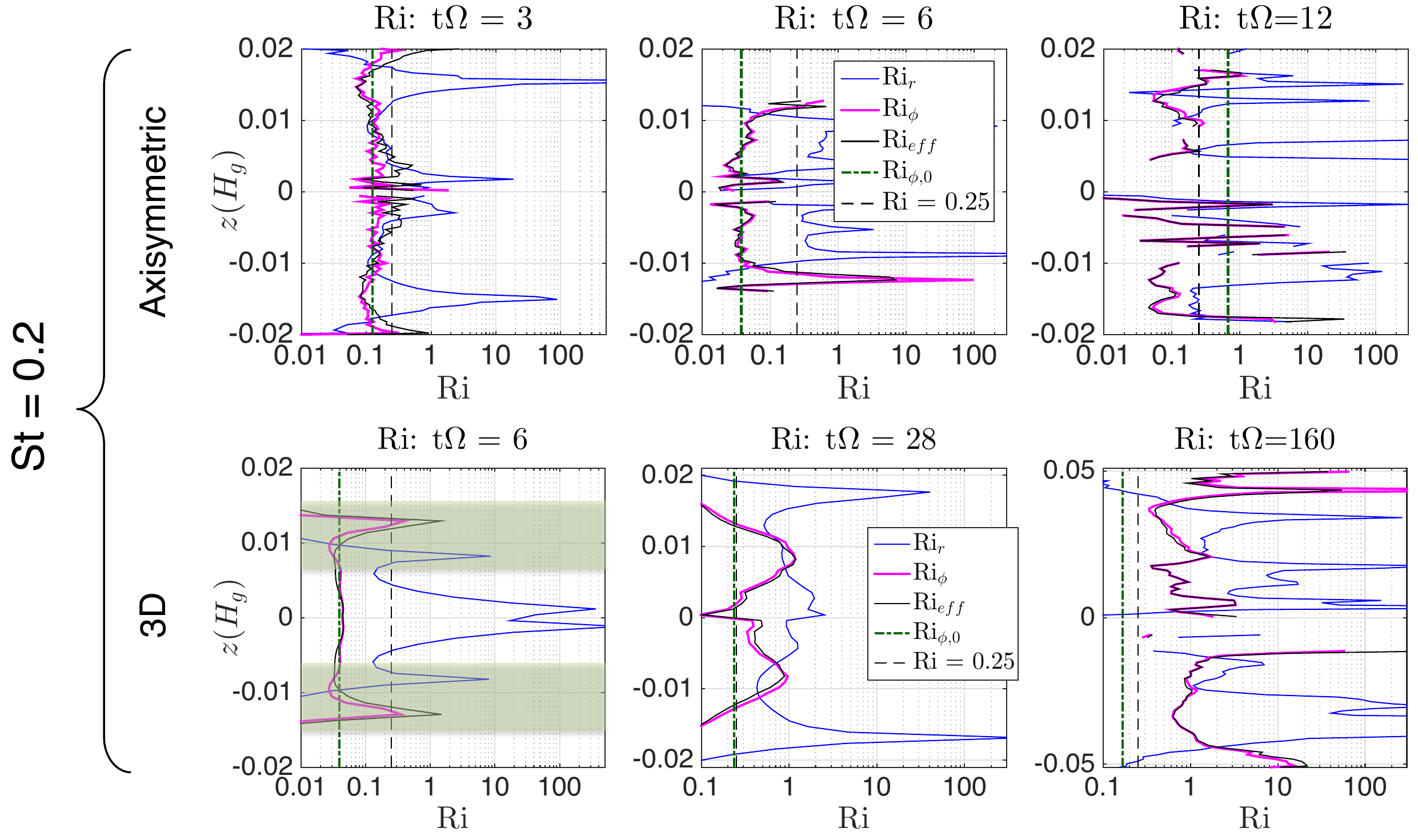}\par
\end{center}
\caption{ Richardson number plots.  See text for details. The shaded region in the bottom left figure coincides with the location of the jets appeared in figure \ref{fig:velocity-profile-st-0.2} for $t\Omega=5$}.
\label{fig:Richardson_Number_Plots}
\end{figure*}

\subsection{Richardson Numbers}\label{sec:Richardson_Numbers}

The Richardson number (in general denoted as ``Ri") is the non-dimensional quantity measuring the destabilizing role of shear against the stabilizing influence of buoyancy oscillations.  In the protoplanetary disk settings considered here, it is assessed on the basis of a radially and azimuthally uniform but vertically varying mean velocity profile generically denoted here by ${\cal U}(z)$. While a formal effective Ri characterizing non-steady particle laden flows in accretion disks is not currently formulated, we adopt the following {\emph{effective}} definition,

\beq
\Ri = \Ri_{{\rm eff}} \equiv
-\frac{\Omega^2 z\rho_p}{{\rho}_g + \rho_p}
\left(\frac{1}{\rho_p}\frac{\partial\rho_p}{\partial z}\right)\Bigg/
\left(\frac{\partial {\cal U}}{\partial z}\right)^2.
\label{Rieff_def}
\eeq

\noindent as promoted by \citet{Sekiya_1998} and \citet{Chiang_2008}.
Implicit in this definition is the assumption that the particle layer is thin enough that the background gas density is unvarying over the vertical scales of interest, which is certainly the case here.  
\par
 The Miles-Howard theorem states that a sufficient condition for the stability of a parallel stratified flow against infinitesimal perturbations is if ${\rm Ri} > 1/4$ everywhere within \citep{Miles_1961,Howard_1961}. If on the other hand there are locations/regions where ${\rm Ri} < 1/4$, then the flow is a candidate for {\it{classic}} stratified shear flow instability \citep[e.g.,][]{Chandrasekhar_1961, Drazin_Reid_2004}, which we generically refer to it as leading to KH roll-up. We note that we consider here the classical criterion for KH-roll-up and discuss further in Sec. \ref{On_Ri} the effect strong rotation has on this criterion especially in light of other previous studies \citep[e.g.,][]{Gomez_Ostriker_2005,Barranco_2009}.
\par
\par
 There are several possible choices for ${\cal U}$ to use in the definition found in Eq. (\ref{Rieff_def}) using the radial-azimuthal mean quantities introduced in Sec. \ref{simulation_results_description}.  However, given recent single fluid descriptions of particle coupled disk gas dynamics \citep[e.g.,][]{Lin_Youdin_2017}, we also think it justified to consider calculating Ri in terms of {\emph{ center-of-mass}} velocities defined (respectively) for the radial and azimuthal component.  As such we motivate
 
\beqa
& & \ U_{{\rm cm}} \equiv
\frac{\rho_{g,0} \left<u_g\right>_{xy} + \left<\rho_p\right>_{xy} \left<u_p\right>_{xy}}
{\rho_{g,0} + \meanrhop}, \nonumber \\
& & \ V_{{\rm cm}} \equiv
\frac{\rho_{g,0} \left<v_g\right>_{xy} + \left<\rho_p\right>_{xy} \left<v_p\right>_{xy}}
{\rho_{g,0} + \meanrhop},
\label{cm_velocities_def}
\eeqa
in which $\left<u_p\right>_{xy}$ and $\left<v_p\right>_{xy}$ are the radial-azimuthal averages of the particle-fluid velocity fields based on their reconstruction described at the end of Sec. \ref{setup}. Note also that in the above we use a constant value $\rho_{g,0}$ instead of an analogously defined radial-azimuthal gas mean $\left<\rho_g\right>_{y}$ simply because the vertical box and particle extents are so close to the midplane that there is hardly vertical variation of the gas density, i.e., it can be easily shown that $\left<\rho_g\right>_{y} \approx \rho_{g,0}$. We consider three instances of Ri all evaluated based on the above center of mass velocities.  For the first, denoted by $\Riphi$, we follow the traditional approach in considering only the vertical variation of the azimuthal velocity component $V_{{\rm cm}}$, i.e.,

\beq
\Riphi \equiv 
-\frac{\Omega^2 z\meanrhop}{{\rho}_{g,0} + \meanrhop}
\left(\frac{1}{\meanrhop}\frac{\partial\meanrhop}{\partial z}\right)\Bigg/
\left(\frac{\partial V_{{\rm cm}}}{\partial z}\right)^2.
\label{Riphi}
\eeq
\noindent In the same vein, we consider a Ri defined on the vertical variation of the radial velocity component $U_{{\rm cm}}$,

\beq
\Rir \equiv 
-\frac{\Omega^2 z\meanrhop}{{\rho}_{g,0} + \meanrhop}
\left(\frac{1}{\meanrhop}\frac{\partial\meanrhop}{\partial z}\right)\Bigg/
\left(\frac{\partial U_{{\rm cm}}}{\partial z}\right)^2.
\eeq
For the final version, we adopt $\Rieff$ as given in Eq. (\ref{Rieff_def}), but with ${\cal U}$ replaced according to

\beq
{\cal U}^2 \rightarrow U_{{\rm cm}}^2 + V_{{\rm cm}}^2,
\eeq
together with $\rho_p \rightarrow \meanrhop$ and $\rho_g \rightarrow \rho_{g,0}$.
\par
$\Riphi$, with $U_{{\rm cm}}$ instead replaced by $U_g$, is the same definition used recently by \citet{Gerbig_etal_2020} as well as in several previous disk studies \citep[][to name a few]{Johansen_etal_2006,Barranco_2009,Lee_etal_2010a, Lee_etal_2010b,Hasegawa_Tsuribe_2014}.  In this formulation $\Riphi$ has been used by previous studies to diagnose whether or not a vertically varying azimuthal profile is stable against non-axisymmetric KH roll-up. Adopting $\Rir$ is analogously appropriate for axisymmetric KH-roll up scenarios like considered in \citet{Ishitsu_etal_2009} and \citet{Lin_2021}, and is appropriate for the solutions discussed here.  Finally, the generalized form $\Rieff$ is useful in assessing the shear stability of Ekman flows \citep[e.g.,][]{Mkhinini_etal_2013}.
\par

\begin{deluxetable*}{c c c c c c c c c} 
\label{tbl:Riphinotfits}
\tabletypesize{\scriptsize}
\tablewidth{0pt} 
\tablecaption{Derived parameters for plots shown in Fig. \ref{fig:Richardson_Number_Plots}. \label{tbl:derived_parameters}} 
\tablehead{ 
\colhead{Simulation ID} & $t\Omega$  & Phase & $H_p$\tablenotemark{a} & \colhead{$H_s$\tablenotemark{a}} & \colhead{$\delta v_{00}$\tablenotemark{b}} & \colhead{$\epsilon_0$} & $\Riphinot$ & Ro\tablenotemark{c} } 
\startdata 
\vspace{-0.2cm}\\
A2D-04H 
  & $0.5$ & bounce & 0.0098  & 0.0136  & 0.025  & 1.02 & 0.279 & 0.919\\
  & $1.5$ & bounce & 0.0094  & 0.0137  & 0.025  & 1.06 & 0.315 &0.912\\
  & $3.0$ & bounce & 0.0089 & 0.0126  & 0.027  & 1.13 & 0.238 & 1.071\\
  & $6.0$ & bounce & 0.0084 & 0.0129 & 0.030 & 1.19 & 0.244 & 1.163\\
  & $15.0$ & shear & 0.0117 & 0.0129 & 0.028 & 0.94 & 0.150 & 1.085\\
\vspace{-0.2cm}\\
\hline
B3D-04L & $4.0$ & bounce & 0.0092 & 0.0141  & 0.0257  & 1.08  & 0.361  & 0.911 \\
  & $6.0$ & bounce & 0.0089 & 0.0138  & 0.0265  & 1.12 & 0.350 &  0.960\\
  & $52.0$ & shear & 0.0096 & 0.0153  & 0.0257 & 1.011  & 0.455 &  0.840\\
  & $500.0$ & pattern & 0.0134 & 0.0218 & 0.0225  & 0.745  & 1.062 &  0.516\\
\vspace{-0.3cm}\\
\hline
\hline
A2D-2H & $3.0$ & bounce & 0.0056 & 0.0091  & 0.0335 & 1.790  & 0.125 & 1.841\\
  & $6.0$ & bounce & 0.0037 & 0.0059 & 0.0419  & 2.521  & 0.038  & 3.551\\
  & $12.0$ & shear & 0.0073 & 0.0165  & 0.0350 & 1.509  & 0.670  & 1.061\\
 \vspace{-0.3cm} \\
 \hline
B3D-2M & $6$ & bounce & 0.0031 & 0.0056  & 0.0445  & 3.211  & 0.039 & 3.973\\
  & $28.0$ & shear & 0.0058 & 0.0101  & 0.0289  & 1.730  & 0.237  & 1.431\\
  & $160.0$ & pattern & 0.0170 & 0.0162  &0.0223  & 0.543  & 0.168 & 1.377\\
  \vspace{-0.3cm}\\
 \enddata 
  \tablenotetext{$\rm{a}$}{In units of $H_g$.}
 \tablenotetext{$\rm{b}$}{In in units of $c_s$.}
 \tablenotetext{$\rm{c}$}{Rossby number defined in Eq. (\ref{Ro:def}).}
\end{deluxetable*}

In the spirit of \citet{Johansen_etal_2006}, Fig. \ref{fig:Richardson_Number_Plots} shows Ri plotted as a function of disk height at various turbulent development epochs for both the full 3D and axisymmetric simulations conducted here.  
As a reference we overlay the  Ri=0.25 line in all the figures, keeping in mind that the actual critical Ri value for a disk setting that includes rotation may be different from
the classical criterion \citep{Gomez_Ostriker_2005,Barranco_2009}.
The top two rows of Fig. \ref{fig:Richardson_Number_Plots} show the results for St = 0.04.  The axisymmetric runs are shown up to the main bounce phase and two main things are evident: first, the radial velocity fields do not satisfy the condition for KH-roll-up as $\Rir$ never gets near the critical values 0.25 and, secondly $\Riphi$ appears to hover about 0.25 and rise to nearly 1 at distances from the midplane both containing the particle layer and exhibiting turbulent dynamics (i.e., for $0<|z/H_g| < 0.025$).  By the time the axisymmetric simulation reaches its strongest turbulent transition point ($t\Omega \approx 6$) $\Rir$ remains mostly greater than 0.25 -- despite its large amplitude fluctuations -- over the bulk of the vertical extent except for a few grid points in the midplane region, while $\Riphi$ have smaller amplitude fluctuations dropping occasionally below 0.25 across significant vertical stretches of domain.  In any event, the axisymmetric simulations demonstrate that something other than KHI is operative here.
\par
The situation is more stark in the full 3D case.  In the lead up to turbulent transition and continuing well beyond it $\Rir$ remains far above 0.25.  Similarly, except for a very localized excursion below 0.25, $\Riphi$ essentially remains greater than the condition for radial KHI across the vertical extent of interest.  Moreover, not only is $\Riphi > 0.25$, but its value is closer to 0.32 at transition over the vertical extent, only dropping close to 0.25 in specific locations of narrow vertical extent -- e.g., near $z/H_g \approx  \pm 0.02$ for $t\Omega = 6$, and a bit higher up for $t\Omega = 52$. When the simulation is well within its turbulent phase $\Riphi$ near the midplane gets even larger increasing beyond 0.4 over the bulk of the layer. This includes the midplane although, once again,  $\Riphi$ hovers near but always above 0.25 even with the most extreme cases (e.g., near $|z| \sim 0.25 H_g$ at $t\Omega = 52$).  Once the simulation has transitioned into its secondary pattern state, $\Riphi$ is everywhere far removed from 0.25 but lies primarily under 1 for the bulk of the turbulent layer with the exception of regions near the midplane ($|z/H_g|< 0.005$ at $t\Omega = 500$) where $\Riphi > 1$ in fact.  These features strongly indicate that the classical non-rotating KHI -- either as radial or azimuthal roll-up -- does not play the primary role in the development nor maintenence of turbulent motions in these simulations where St = 0.04.  It his possible, however, that a rotationally modified form of KHI is operating based on a previous linear study \citep[][also, see discussion in Sec. \ref{On_Ri}]{Barranco_2009}.
\par
The bottom two rows of Fig. \ref{fig:Richardson_Number_Plots} show Ri for St$ = 0.2$.  In both full 3D and axisymmetric cases we see that by the time the simulations reach their deep bounce phase ($t\Omega = 6$) $\Riphi$ dips below 0.25 across the full vertical extent containing particles.  At this time marker $H_p \approx 0.003 H_g$ so that the particle layer is mostly confined to $|z/H_g| < 0.01$ (e.g., see top left panel of Fig. \ref{fig:Flow_Sandwich_Plotter_t6}).  By reference we see that the value of $\Riphi$ lurks around 0.034 up to $z\approx \pm 0.01H$, beyond which it precipitously drops in magnitude. Except for short-ranged dips below 0.25 (e.g., near $z \approx \pm 0.06 H_g$), $\Rir$ mainly remains above 0.25 up to about $z \approx \pm 0.01 H_g$ before similarly dropping precipitously in magnitude like $\Riphi$. The dynamically developing off-midplane jet flow layers, with their incipient cat's eye formations, coincide to where both Ri numbers drop in magnitude for $|z/H_g| > 0.01$.  We therefore conclude that these layers really are undergoing KH-roll-up dynamics.  However, within the particle-containing midplane layer the situation is different as the $\Rir$ values there remains significantly above the criterion for radial KHI.  On the other hand, $\Riphi$ does remain below 0.25 suggesting that this part of the layer is susceptible to azimuthally directed KH-roll-up -- although evidence for such formation is hard to discern from the snapshots shown for this case (e.g., see Fig. \ref{fig:slices-st-0.2}c).  However, we also cannot rule-out the possibility that this part of the midplane is not also subject to the same non-KHI unstable dynamics characterizing the turbulent dynamics in the St = 0.04 case discussed above.
\par
By the time the system has moved well into the midplane turbulent phase, the situation for Ri has changed.  We focus here only on the full 3D calculation by referring to the middle panel of the bottom row of  Fig. \ref{fig:Richardson_Number_Plots}, corresponding to the time stamp $t\Omega = 28$.  It is evident that $\Rir > 1/4$ across that part of the midplane containing most of the particles, and only when $|z/H_g| > 0.02$ does $\Rir$ cross below $1/4$ indicating that the layer gets even more stable against radial KH-roll-up as the system evolves. $\Riphi$ also remains above 1/4 and largely below 1 across the particle containing part of the midplane, but drops well below 1/4 once $|z/H_g|$ exceeds 0.0125, suggesting that these upper layers may themselves be undergoing azimuthal KH-roll-up.  By the time the system has transitioned into its secondary pattern state (e.g., see right panel of bottom row of Fig. \ref{fig:Richardson_Number_Plots}, for $t\Omega = 160$), except possibly for a narrow range near the midplane, the simulation appears stable against both radial and azimuthal KH-roll-up across fully half of its vertical domain.  Of course, this situation corresponds to the emergence of the heretofore discussed radial sinusoidal period-2 feature in all fields.
\par
In all panels shown on Fig. \ref{fig:Richardson_Number_Plots} we plot an estimated ``effective" midplane value for $\Riphi$, denoted hereafter by $\Riphinot$.  The aim here is to develop a relatively smooth estimate derived from the simulation output in the region primarily containing the bulk of the particles. Toward this end we assume a Gaussian-like model for $\left<\rho_p\right>_{xy}$,

\beq
\left<\rho_p\right>_{xy} \approx \rho_{p,{\rm mod}} \equiv \delta\rho_{p,0}\exp\left(-\frac{1}{2} \frac{z^2}{H_p^2}\right),
\label{model_rhop}
\eeq
and determine the values of the parameters $\delta\rho_{p,0}$ and $H_p$ using standard error minimization techniques \citep[e.g.,][]{Nimmo_etal_2017}.  We note that the values determined for $H_p$ via this approach basically agree with the values calculated for $H_p$ according to the prescription described by Eqs. \ref{eqn:particlescaleheight}-\ref{Hp_def} found at the end of Sec. \ref{initial_conditions}. 
\par
\begin{figure}
\begin{center}
\leavevmode
\includegraphics[width=8cm]{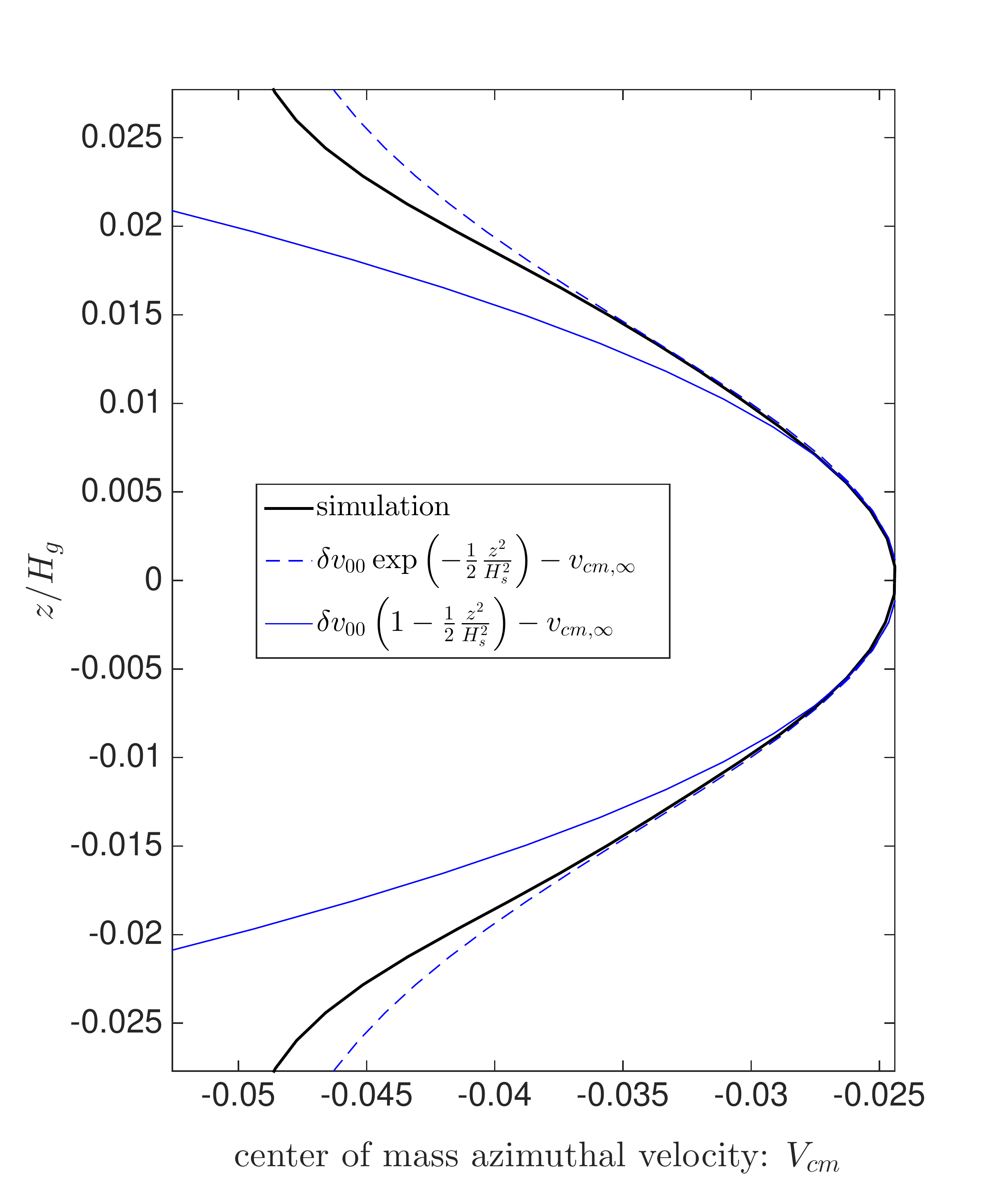}
\par
\end{center}
\caption{An azimuthally-radially averaged plot of the center of mass azimuthal velocity $V_{cm}$ for simulation B3D-04M taken at $t\Omega = 4$.  The corresponding Gaussian fit and its parabolic approximation are shown.  The fit parameters are found in Table \ref{tbl:derived_parameters}.}
\label{facsimile_plot}
\end{figure}

We similarly adopt a Gaussian-like form for $V_{cm}$, in which

\beq
V_{{{\rm cm}}} \approx V_{{{\rm cm,mod}}} \equiv v_{{\rm cm},\infty}
+  \delta v_{00} \exp\left(-\frac{1}{2} \frac{z^2}{H_s^2}\right),
\label{model_Vcm}
\eeq
where $v_{{\rm cm},\infty} = -0.05 c_s$ is the expected asymptotic value far away from the particle layer.  The fit parameters  $\delta v_{00}$ and $H_s$ are also determined via error minimization over a vertical domain of up to 2.5 $H_p$; the aim being to best represent the vertical variation of  $V_{{{\rm cm}}}$ over the bulk of the particle layer.  Fig. \ref{facsimile_plot} shows an example of this approximate fitted form for simulation B3D-04M during its early bounce phase.  We find that this approximate form is satisfactory for our purposes hereafter.

\par

\par 
 With these parameters determined we insert the model forms Eq. (\ref{model_Vcm}-\ref{model_rhop}) into the definition of $\Riphi$ found in Eq.(\ref{Riphi}), followed by evaluating the resulting expression at $z=0$, i.e.,

\beq
\Riphinot \equiv \Riphi\big(z=0\big) =
\frac{\epsilon_0}{1+\epsilon_0}\frac{\Omega^2 H_s^4}{\delta v_{00}^2 H_p^2},\ \ 
\epsilon_0 \equiv \frac{\delta \rho_{p,0}}{\rho_{g,0}}.
\label{Riphinot_def_1}
\eeq
In Table \ref{tbl:Riphinotfits} we summarize the determined fit parameters together with the estimated  value of $\Riphinot$ for each of the simulations and their timestamps shown in Fig. \ref{fig:Richardson_Number_Plots}. $\Riphinot$ is shown on each plot as well.
\par
We note two features. First, we find that the $H_s$ is generally always larger than $H_p$ by up to a factor of 2 or more, which is an unexpected trend. Second, the value of $\Riphinot$ appears to well characterize the behavior of $\Riphi$ in the St = 0.04 simulations through the bounce and early turbulent phase.  This approximation to $\Riphi$ fails to capture its character for full 3D simulations that are in the secondary transition phase.  Similar performance is seen in the St = 0.2 simulations, although it captures the essence of an averaged value across the particle layer in the primary turbulent phase (e.g., for the $t\Omega = 28$ time stamp shown).  This leads us to conclude that during these late stages the simulations for St = 0.2 have undergone a significant transition in character. Despite its limitation, this type of model representation should prove useful in ascertaining the transition to turbulence, especially for cases where St = 0.04, as elucidated further in Sec. \ref{symmetric_instability_problem}.

\section{Turbulence and statistics}\label{sec:statistice}
\subsection{Energy formulation}
It is informative to consider energy balances within the simulated dynamics.
Since the gas component dynamics are largely incompressible, we adopt Eq. (\ref{eqn:pencilgasmomentum}) together with the
incompressibility statement

\beq
\partial_i u_{gi} = 0,
\eeq
in place of mass continuity, Eq. (\ref{eqn:pencilgascontinuity}). 
We designate $u_{gi}$ to be the components of the gas velocity and $u_{pi}$ to be the same for the particle velocities.\footnote{In this section we adopt Einstein notation with the usual convention
of summing over repeated dummy indices $i,j$, where $i,j = 1,2,3$  reference the $x,y,z$ components (respectively).  Thus, $i,j$ here ought not be confused with particle label i or grid label j used in Sec. \ref{setup}.}
This means that the pressure term in the gas momentum equation is replaced with a diagnostic field $P$, thus the equation, with the assumption of Einstein's summation convention, appears rewritten as

\beqa
& &  \partial_t \rho_g u\sub{gi}+ \partial_j \rho_g u_{gi}u_{gj} +  V\sub K \partial_y \rho_g u_{gi} \nonumber \\
& &  \ \ \ \ -2\Omega \rho_g  u_{g2}\delta_{i1} + (\Omega/2) \rho_g u_{g1} \delta_{i2} 
 = -\partial_i p -\overline P_r \delta_{i1} 
  \nonumber \\
& & \ \ \ \ 
 \ \ \  - \Omega_0^2 \rho_g z \delta_{i3}
 -\Omega_0\rho_p(u_{gi} - u_{pi})/\St + D_i(u_{gi}),
 \label{einstein:gas}
\eeqa
in which $\delta_{ij}$ is the Kronecker delta symbol, $D_i(u_{gi})$ is a viscous dissipation function and $\overline P_r$ is the mean radial pressure gradient.  We are reminded
that in these simulations $u_{gi}$ and $u_{pi}$ are deviations atop the base Keplerian flow $V\sub K$.  In this vein, we identify the total
velocities in each fluid component with $U_{gi} = u_{gi} + \delta\sub{i2} V\sub K$ and $U_{pi} = u_{pi} + \delta\sub{i2} V\sub K$,
respectively for the gas and particle components.  
\par
With respect to the energy measures considered in this section, we use the shorthand,
$\displaystyle \big<\bullet\big> \leftrightarrow \int_{{\rm V}} [\bullet] \ d{\rm V}$, to denote volume integrals.
In our domain the volume V will be over the computational domain $L_x,L_y,L_z$.
We define the volume integrated perturbation gas kinetic energy by

\beq
\varepsilon\sub g \equiv \Big<\sfrac{1}{2} \rho_g u_{gi} u_{gi}\Big>,
\label{varepsilon_gas_def}
\eeq
and, similarly, the volume integrated perturbation particle kinetic energy

\beq
\varepsilon\sub p \equiv \Big<\sfrac{1}{2} \rho_p u_{pi} u_{pi}\Big>,
\label{varepsilon_particle_def}
\eeq

\par

%
%
%
%

\subsection{Energy Spectra}\label{sec:spectra}
The energy integral formulation is often times rewritten in Fourier
space.  With $\bf k$ being the three dimensional wavenumber and $k \equiv |{\bf k}|$ its absolute magnitude, it is customary to define a kinetic energy density per unit wavenumber as $\varepsilon\sub{k,g}$, which here is taken to be the total perturbation kinetic energy contained in all wavevectors ${\bf k}$ whose (absolute) wavenumbers  lie in between 
$k$ and $k+dk$. Defining $\tilde u_{gi}$ to be the Fourier transform of $u_{gi}$, this
sum is formally expressed as 

\beq
\varepsilon\sub{k,g} = \frac{\rho_g}{2\delta k}\sum_{k' = k}^{k+\delta k}
\tilde u_{gi}^*\tilde u_{gi},
\label{varepsilon_kg_def}
\eeq
where the star superscript denotes complex conjugation.  The expression is
divided by $\delta k$ to preserve the defined units.  The sum of all of these
contributions must equal the total volume integrated energy of the domain, thus
the discrete infinite sum (i.e., $\forall k$ where $k=0,\delta k, 2 \delta k, \cdots$)

\beq
\sum_{k=0}^\infty \varepsilon\sub{k,g} \delta k = 
 \Big<\sfrac{1}{2} \rho_g u_{gi} u_{gi}\Big> = \varepsilon\sub g,
\eeq
as defined in Eq. (\ref{varepsilon_gas_def}).
Based on this we motivate a similar parsing of the total perturbation 
kinetic energy contained in the
particle fluid.  Unlike the gas component, whose density is treated as constant, the particle component has strongly fluctuating densities and to properly account its partial energies in Fourier space we define a new quantity $\mu\sub i \equiv \sqrt{\rho_p}u_{gi}$, which
is amenable to sensible interpretation and analysis (see Appendix \ref{energy_calculations}).  Similarly denoting
$\tilde \mu\sub i$ to be the Fourier transform of $\mu\sub i$, we define 

\beq
\varepsilon\sub{k,p} = \frac{1}{2\delta k}\sum_{k' = k}^{k+\delta k}
\tilde \mu\sub i^*\tilde \mu\sub i,
\label{varepsilon_kp_def}
\eeq
whose infinite discrete sum over k yield $\varepsilon\sub p$, i.e., 

\beq
\sum_{k=0}^\infty \varepsilon\sub{k,p} \delta k = 
 \Big<\sfrac{1}{2} \rho_p v\sub i v\sub i\Big> = \varepsilon\sub p.
\eeq

\par
An overarching long-term programmatic goal into the future is to assess the dependencies of $\varepsilon\sub{k,g}$
and $\varepsilon\sub{k,p}$ upon $|{\bf k}|$ and to gain some understanding of how energy flows between scales (i.e., what direction does it move, are there multiple cascades involved, etc.?) and what mechanisms are mainly responsible for this transfer. While the latter set of aims is outside the scope of this study, in this preliminary examination we empirically show what the spectrum may possibly look like based on our highest resolution simulations and what various trends occur as simulation parameters change. Under simplifying assumptions (isotropy, single fluid, etc.) the Kolmogorov dependence $\varepsilon\sub{k,g}\sim k^{-5/3}$ falls out of the above equation on the assumption that there exists a range in wave numbers (the inertial range) in which the rate of energy transfer across the sphere of radius $k$, i.e., $\varepsilon\sub{k,g}$, is steady in time.  Typically once a simulation has reached a statistically steady state, in which the energy injected is compensated by losses (see above), an energy spectrum is assessed.  In the simulations we have conducted, any mismatch in this results in a momentary change in the total energy of the system, which average out over long stretches of time.  It is for this reason that spectra produced from simulations are made from composite averages at several timesteps.

\subsubsection{Calibration Spectra}

As we alluded in Sec. \ref{sec:model}, the numerical diffusion in the simulations (namely, hyperdiffusion) restricts the usable domain in k-space to examine turbulent dynamics and, as such, sets a length scale below which the validity of results -- vis-\`a-vis turbulent dynamics and associated structures -- ought be viewed with great caution. So, to identify the reliable simulation sub-domain and identify the location of the dissipation scale set by the numerical methods, we conducted a gas-only $512^3$ simulation (F3D-512)  in a $(2\pi)^3$ periodic domain where turbulence is forced at some larger length scale by a simple forcing function. 

In order to obtain a calibration spectra, we have used the forcing module already existing in the PENCIL code without any modifications.
The temporally random forcing function $\bm{\mc{F}}=\bm{f}(\bm{x},t)$ can be written as \citep{Brandenburg_2001}

\begin{equation}
    \bm{f}(\bm{x},t) \equiv Re\left\{\mc{N}\bm{f}_{\bm{k}(t)}exp\left[i\bm{k}(t)\cdot \bm{x}+i\phi(t)\right]\right\}.
\end{equation}
Here $\bm{k}(t)$ and $\phi(t)$ respectively denote the time dependent wavevector and random phase with $\left|\phi(t)\right|<\pi$. $\mc{N}$ is the normalization factor which varies as $\delta t^{-1/2}$ with $\delta t$ being the timestep. We choose to force the system at $k=2$, in which case, at each step a randomly chosen possible wavenumber with $1.5 < |\bm{k}| <2.5$ is forced. The forcing is executed with the eigenfunctions of the curl operator 

\begin{equation}
    \bm{f}_{\bm{k}}=\frac{i\bm{k}\times (\bm{k}\times \hat{e})-\sigma |\bm{k}|(\bm{k}\times \hat{e})}{\sqrt{1+\sigma^2}\bm{k}^2\sqrt{1-(\bm{k}\cdot\hat{e})^2/\bm{k}^2}}.
\end{equation}
Here $\hat{e}$ is the arbitrary unit vector used to generate $\bm{k}\times e$ which is perpendicular to $\bm{k}$. $\sigma$ denotes the helicity factor which is set to zero in order to make the forcing purely non-helical. Note that this forcing is essentially divergenceless. However, as the fluid equations solved by the code are not strictly incompressible, which is perhaps more applicable for astrophysical systems, a small non-zero divergence is introduced over the course of the simulation.  Nonetheless, the spatio-temporal dynamics of all of our simulations are effectively incompressible, where density variations are extremely weak (e.g., see the quantity $\bar\Pi'$ in Figs. \ref{spacetime-st-0.04}-\ref{spacetime-st-0.2}).

\begin{figure}
\begin{center}
\leavevmode
\includegraphics[width=8.5cm]{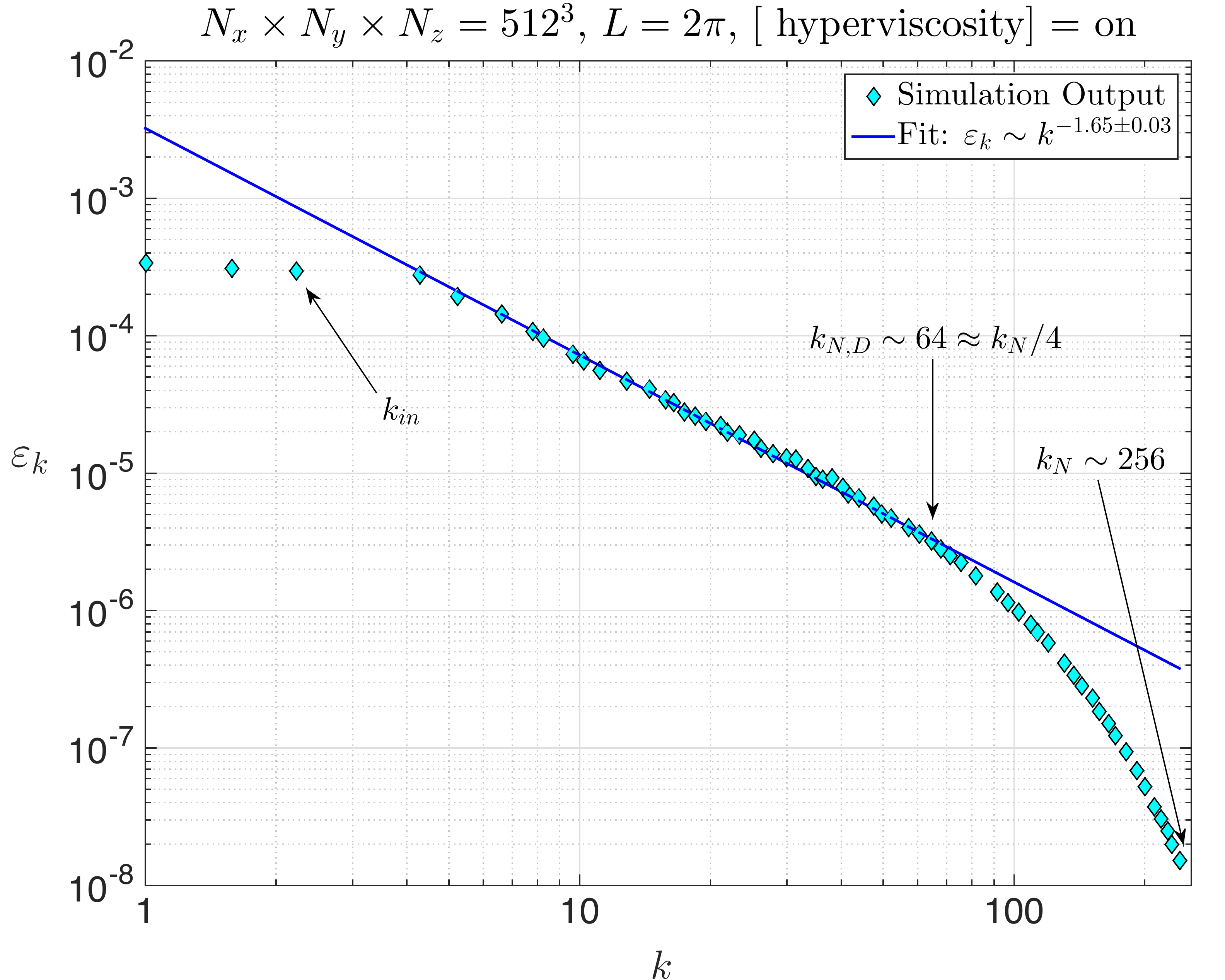}\\
\par
\end{center}
\caption{Kinetic energy spectra $\varepsilon_k$ versus $k$ produced from the $512^3$ element forcing run F3D-512.  Simulation of cube with side $L = 2\pi$ and forcing wavenumbers $ 2.5<k_{in}L < 3.5$. A fit to the simulation output reveals an inertial range Kolmogorov dependence $\varepsilon_k \sim k^{-5/3}$ 
in the range $k_{in} < k \lessapprox k_N/4$, where $L k_N = 256$ is the Nyquist scale ($2dx$).
}
\label{Calibration_Spectra_3D}
\end{figure}

The power spectra obtained from the simulation F3D-512 using the method outlined in Sec. \ref{sec:spectra} and Eq. (\ref{varepsilon_kg_def}) is shown in Fig. \ref{Calibration_Spectra_3D}. It is evident from the figure that in the simulation the turbulence is resolved and a cascade of energy towards smaller length scales (higher $k$) is taking place with an inertial range spanning more than a decade. The energy density behaves like a power-law, i.e., $\varepsilon_k \sim k^{-n}$, with $n = 1.65 \pm 0.03$ best fitting the inertial range, confirming that to within reasonable error this solution is consistent with Kolmogorov's spectra (with an $n=5/3$) expected for homogeneous isotropic turbulence. However, important to note that the actual dissipation scale set by the simulation is placed somewhere around $k_{N,D} \approx k_N/4$, where $k_N$ is the Nyquist wavenumber corresponding to a wave spanning 2 grid points ($2\Delta x$). This trend in kinetic energy power is ubiquitous across all of our science simulations listed in Table \ref{tbl:simulation} (See below for more details).
\par
It is also important to mention that the use of hyperviscosity can lead to a bottleneck effect where energy gets piled up at the smaller scale \citep[e.g.,][]{Haugen_Brandenburg_2004}. This happens particularly when the hyperviscosity is not strong enough to dissipate the energy at those small scales (high wavenumber). This effect is particularly problematic as the accumulated energy tend to scatter back to the larger scale seeking an equi-partition among all wavenumbers, ultimately altering the power-spectrum and the overall gas dynamics.  Note that this numerical effect is not the same as inverse-cascade where an upscale enstrophy cascade takes place.
\par
During the early stages of this investigation we found that choosing the hyperviscosity parameter to too low a value led to the bottleneck effect, which resulted in code blow-up characterized by widespread generation of $2\Delta x$ waves.  Following selection guidelines documented in \citep{Haugen_Brandenburg_2004} as well as in the PENCIL manual, we have carefully chosen the values of hyperviscosity for all our simulation to ensure that the bottleneck effect does not kick in and the gas energy does not show any upward trend in the dissipation range, a feature characteristic of the bottle-neck effect.

\par

\begin{figure*}
\begin{center}
\leavevmode
\includegraphics[width=0.9\textwidth]{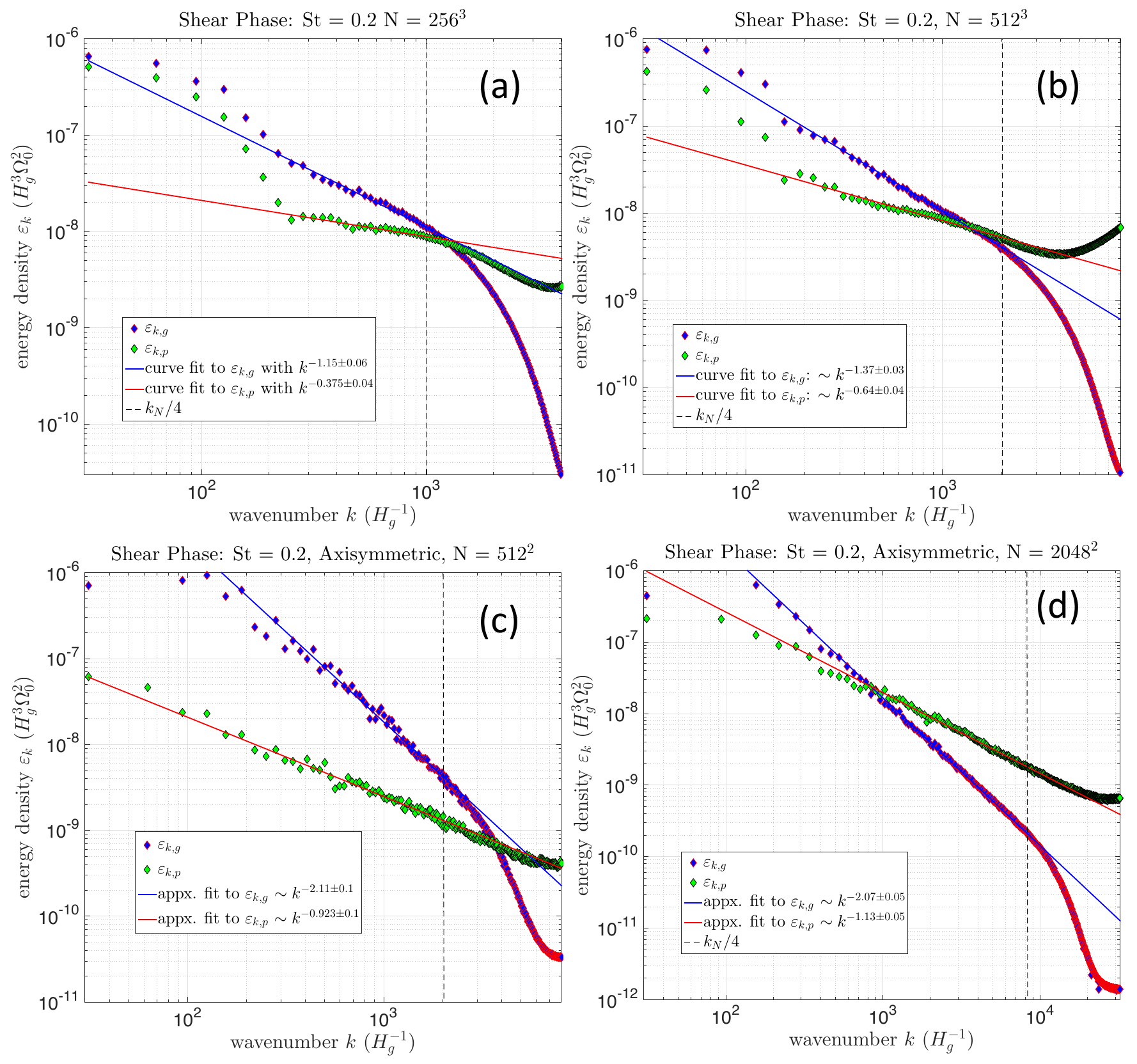}\\
\par
\end{center}
\caption{The kinetic energy spectrum from the science simulations with \St$=0.2$ at the shear phase. The top row shows the spectrum from the $3$D simulations with medium ({\bf{a:}} B3D-2M) and high ({\bf{b:}} B3D-2H) resolution. The purple and green diamonds respectively show the power for gas and solids. The bottom row shows the spectrum for the $3$D axisymmetric  simulations with low ({\bf{c:}} A2D-2H) and high ({\bf{d:}} A2D-2SH) resolution. The pink and green diamonds represent the spectral energy density for gas and solids. The vertical dashed lines correspond to the wavenumber $k_N/4$, the dissipative end of the nominal inertial range. In all four cases, the blue and red solid lines represent power-law fits to the nominal inertial range of each spectra (between $kH_g \approx 350$ and $k_N/4$) of the gas and dust respectively.    The power law index fits to the inertial range of the 3D spectra significantly steepen upon increasing resolution from $256^3$ to $512^3$. However, no such difference is observed in the axisymmetric simulations when going from $512^2$ to $2048^2$ resolution, suggesting that these are effectively converged for $512^2$ resolution. (See text for more discussion) }
\label{Spectra-0p2}
\end{figure*}
\par

\subsubsection{Energy spectra from particle-gas simulations: $\St=0.2$.}

With the calibration established in the previous section, we now move on to the science simulations and discuss the energy spectrum produced by them, along with any possible interpretations that may follow. In Fig \ref{Spectra-0p2} we show the energy spectrum for \St$=0.2$ for both full 3D (\ref{Spectra-0p2}a -- b) and 3D-axisymmetric (\ref{Spectra-0p2}c -- d) simulations for both gas and solid components.\par 
Fig \ref{Spectra-0p2}a shows energy spectra for both the gas and particle fluids for the B3D-2M simulation during its midplane shear turbulent phase.  It is the average of the timesteps $t\Omega = 28,35,55$.  We note several features: starting from $k \approx 220 H_g^{-1}$ and going up to about $k_{N,D} = k_N/4 \approx 1000 H_g^{-1}$ the energy density of the gas component exhibits powerlaw behavior, i.e., $\varepsilon_{k,g} \sim k^{-n_g}$ with $n_g = 1.15\pm0.06$. For this set we established the inferred powerlaw fit using a least squares procedure utilizing energy data starting from $k\approx 350 H_g^{-1}$ up to $k = k_N/4.5$, just shy of the expected cutoff $k_{N,D}$. As expected based on our calibration spectra, $\varepsilon_{k,g}$ steeply plummets beyond $k_{N,D}$.  Up to the beginning of the observed powerlaw behavior $\varepsilon_{k,g}$ carries power that largely lies above the power that might be predicted had the power law been extended to larger scales: that is to say, larger scale modes in the gas component all lie above the blue line.  Similarly, the particle component also exhibits power law behavior in the same $k$ range as the gas component, but its power law index is flatter: i.e., $\varepsilon_{k,p} \sim k^{-n_p}$ with $n_p = 0.375\pm0.04$.  Just like in the gas component, for scales larger than $k\approx 220 H_g^{-1}$, the particle field also contains power larger than that predicted by extending the observed powerlaw behavior into that regime.  With some caution, we therefore {\it nominally} identify $k_{{\rm integral}} = 220 H_g^{-1}$ as the start of an inertial range for both fluids. While we observe that the energy contained in the particle component is generally dominated by the gas component up to the beginning of the numerical dissipation scale $k_{N,D} \approx 1000H_g^{-1}$, the two values appear to be equal to one another at $k \approx 1300 H_g^{-1}$, which lies at slightly shorter scale.  Nonetheless, this equality is confirmed at higher resolution.
\par
Fig \ref{Spectra-0p2}b shows the corresponding energy spectra for the B3D-2H simulation.  It is constructed as the average spectra of only two time stamps, $t\Omega = 35,55$.  We immediately note that energy in the gas and particles are indeed equal at the length scale $k \approx 1300 H_g^{-1}$, which is noteworthy.  As expected, this higher resolution simulation shows power-law behavior for a full decade in scales ranging from $k\approx k_{{\rm integral}}$ up to $k_{N,D} \approx 2000 H_g^{-1}$, extending the resolvable turbulent range by a factor of two in scale. However, we find that the power law index in this higher resolution run has steepened in both quantities where $n_g \approx 1.37 \pm 0.03$ and $n_p \approx 0.64 \pm 0.04$.  Based on this we are led to the tentative conclusion that the $256^3$ element simulation is not statistically converged.  Whether or not the higher resolution $512^3$ element solution is converged cannot be judged at this juncture, requiring a future even higher resolution simulation for confirmation.  However, from our findings for the 3D axisymmetric runs discussed further below, we conjecture that this $512^3$ simulation may have this medium scale inertial range converged. 
\par
\begin{figure}
\begin{center}
\leavevmode
\includegraphics[width=0.48\textwidth]{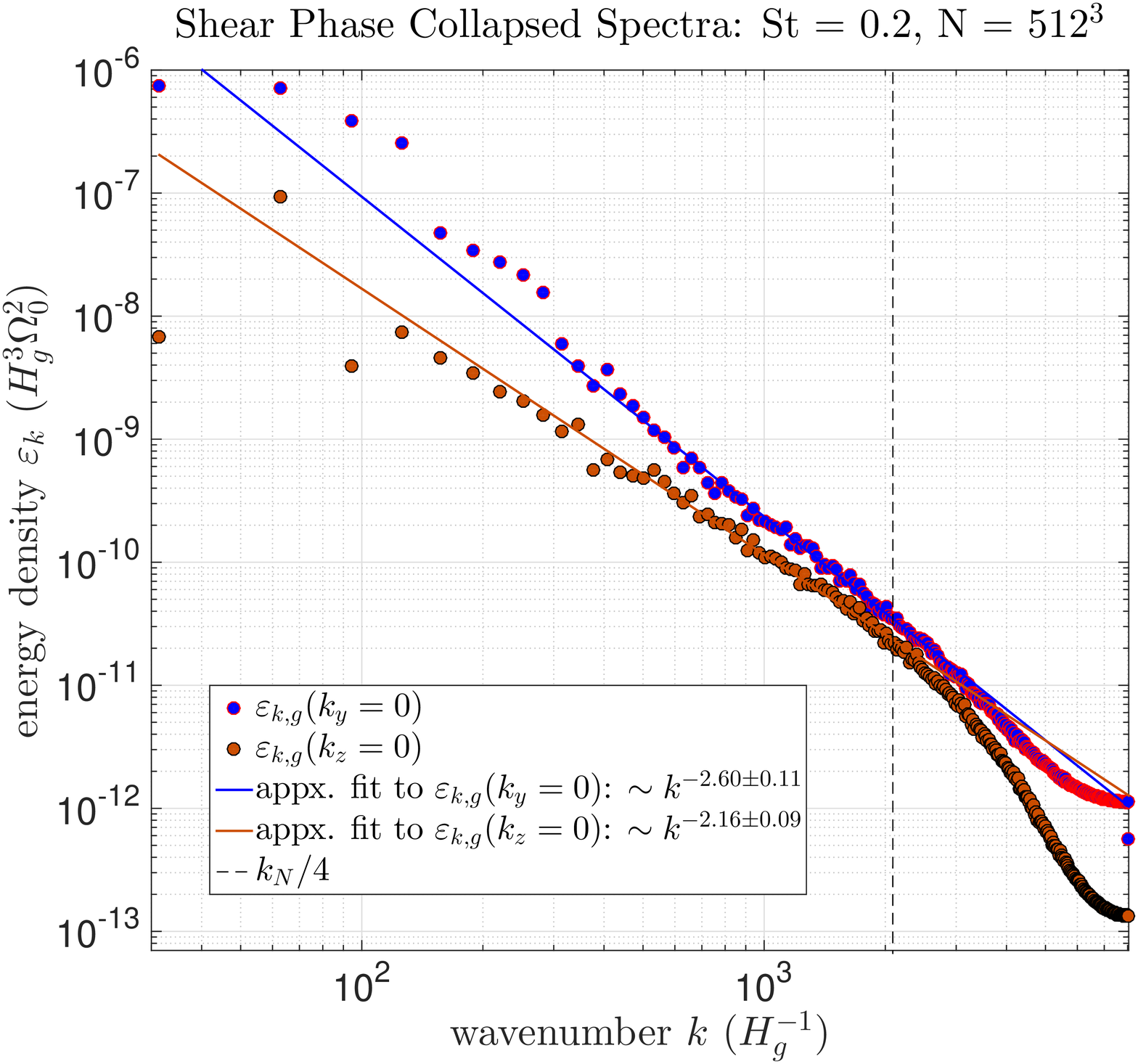}\\
\par
\end{center}
\caption{Energy spectrum $\varepsilon_{k,g}$ for the gas field from simulation B3D-2H collapsed in the azimuthal ($k_y=0$) and vertical ($k_z=0$) direction. A inertial range in the azimuthally collapsed spectra is steeper that the 3D axisymmetric simulation A2D-2H (see Fig. \ref{Spectra-0p2}c) with same resolution, implying the existence of additional energy modes in the axisymmetric direction. }
\label{collapsed_spectra}
\end{figure}
\par
This lack of convergence does not appear to be an issue for the 3D axisymmetric simulations we investigated, where we have conducted two runs from, one being ``high" resolution with $512^2$
elements (simulation A2D-2H) up to ``super-high" resolution with $2048^2$ elements (simulation A2D-2SH).  In Fig \ref{Spectra-0p2}c  and \ref{Spectra-0p2}d , the energy spectrum for the axisymmetric simulations for both the gas ($\varepsilon_{k,g}$; purple diamond) and the solids ($\varepsilon_{k,p}$; green diamonds) are shown. The spectrum for the gas from the high resolution run ($512^2$) follows a power-law $\varepsilon_{k,p}\sim k^{-n_g}$ in the inertial range, where $n_g=2.11\pm 0.1$. The same for the super-high resolution simulation ($2048^2$) comes out as $n_g\approx 2.07\pm 0.05$, lying in the same range of its $512^2$ counterpart withing reasonable errors, indicating a convergence in the simulations. The beginning of the inertial range in both the cases starts at $k_{integral}\approx 200 H_g^{-1}$, extending all the way to $k_{N,D}\approx k_N/4$ in the respective cases, producing an inertial range slightly less than a couple of decades in the super-high resolution run. 

When compared to the full 3D simulations, the 3D-axisymmetric cases produce a much steeper slope for $\varepsilon_{k,g}$, which falls well within our expectation. Throughout this discussion we keep in mind that the energetics and transport characteristics in 2D isotropic turbulence (no rotation, no stratification)  is inherently different from its 3D isotropic counterpart, with the former exhibiting prominent enstrophy cascade towards smaller scales.  Questions like what might the transport characteristics be for flows like these representing a section of disk, where rotation and stratification are dynamically important, and is there a dual cascade of energy and enstrophy in the axisymmetric case, currently remain open. With this in mind, we note that the gas energy $\varepsilon_{k,g}$ at the wavenumber $k_{N,D}\approx k_N/4$ is approximately the same around $3\times 10^{-9}H_g^3\Omega_0^2$, whereas the energy at the integral scale ($k_{integral}\approx 200H_g^{-1}$) is approximately  an order of magnitude more in the 3D axisymmetric run (A2D-2H) compared to the full 3D one (B3D-2H). Whether this extra energy in the axisymmetric simulation is a result of a more efficient extraction of free energy from the background shear at $k_{\rm integral}$ or an outcome of some upscale and -- as yet -- unquantified energy cascade mechanism is not known requiring further investigation. 
\par

\begin{figure*}
\begin{center}
\leavevmode
\includegraphics[width=0.98\textwidth]{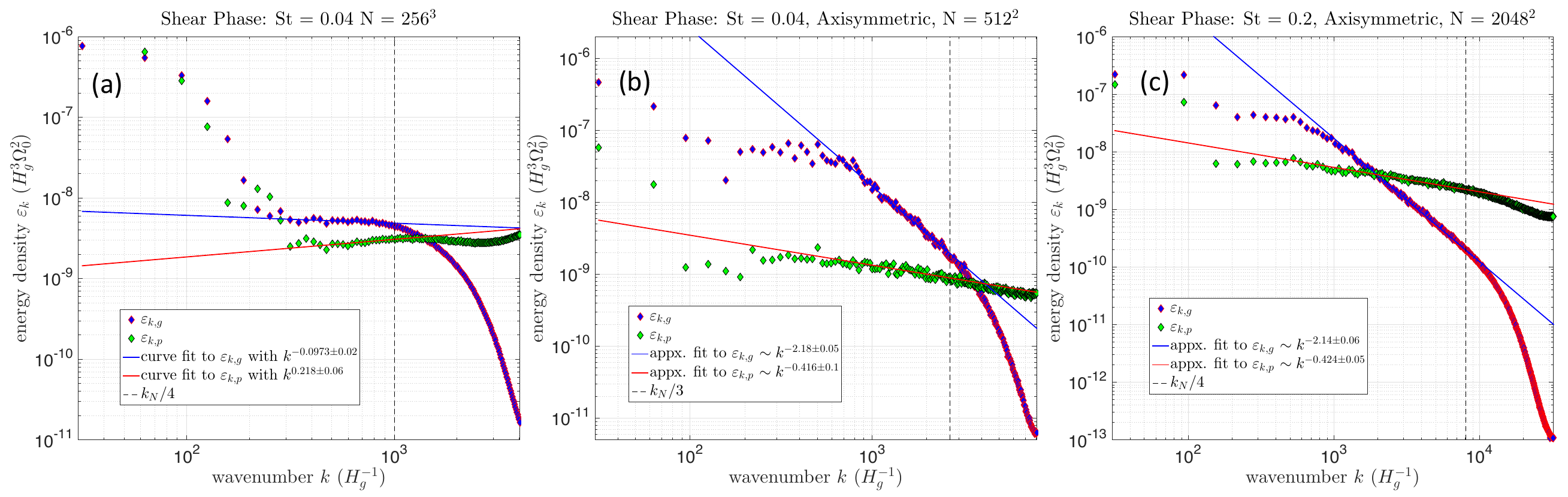}\\
\par
\end{center}
\caption{Kinetic energy spectrum from simulations with \St$=0.04$ at the shear phase for both 3D and 3D-axisymmetric runs. Figure {\bf{a}} shows the spectrum for simulation B3D-04M with the purple and green diamonds representing the gas and particle energy respectively. Figures {\bf{b}} and {\bf{c}} show the energy spectrum from the axisymmetric runs A2D-04H and A2D-04SH respectively. Similar to figure \ref{Spectra-0p2}, the blue and the red solid lines are the power-law fit to the gas and particle spectral energy densities. The vertical black dashed lines denote the positions for $k_N/4$. For both axisymmetric runs, the inertial range for the gas and solids follow similar power-law index. The monotonic decrease of the gas energy beyond the inertial range ensures that the bottle-neck effect is not at play.}
\label{Spectra-0p04}
\end{figure*}
\par

The energy spectrum in the particles $\varepsilon_{k,g}$ in the two simulations, though, show a little difference in the power-law behavior. For simulation A2D-2H with $512^2$ resolution, the inertial range follows a power-law $k^{-n_p}$ where $n_p\approx 0.923\pm0.1$. For simulation A2D-2SH however, $n_p\approx1.13\pm 0.05$. The power-law behavior of the two particle spectrum with a shallower slope extends beyond the wavenumber $k_N/4$, with significantly more energy compared to the gas fields at the small scales, a feature which is still unclear to us. 

In Fig. \ref{collapsed_spectra}, a collapsed gas energy spectrum for the 3D simulation B3D-2H is presented based on the azimuthally averaged velocity fields ($k_y = 0$, purple circles). The power-law index in $\varepsilon_{k,g}$ for the inertial range here comes out as $n_g \approx 2.6\pm 0.11$ which is significantly steeper than the corresponding 3D axisymmetric run A2D-2H. From this result we infer that there are additional modes of energy transfer into and out of axisymmetric structures that are otherwise suppressed in the 3D axisymmetric simulations. We also show the gas energy spectrum for the vertically averaged velocity fields ($k_z = 0$, orange circles), which exhibit power-law behavior with an index $n_g = 2.16\pm 0.09$.  How these may or may not relate to overall composite spectrum remain to be elucidated.  We note that the power-law behavior in both cases here extends somewhat beyond the cutoff wavenumber $k_N/4$, however, we caution inferring anything about the meaning of this until further analysis is done.

\par

\subsubsection{Energy spectra from particle-gas simulations: $\St=0.04$.}

Fig. \ref{Spectra-0p04} shows the energy spectrum $\varepsilon_{k,g}$ and $\varepsilon_{k,p}$ for gas and particle fields respectively for \St$=0.04$. The sub-figure on the left is derived from the $256^3$ 3D simulation B3D-04M. The two sub-figures in the middle and on the right are from the axisymmetric simulations A2D-04H and A2D-04SH respectively with the averages taken with the snapshots at $t\Omega=80,120$ and $220$. 

The energy spectrum $\varepsilon_{k,g}$ for the gas field from the 3D run shows an unexpected nearly flat behavior: $\varepsilon_{k,g} \sim k^{-n_g}$ with $n_g\approx 0.0972\pm 0.02$. The flat region of the spectra starts at $k_{\rm integral}\approx 300 H_g^{-1}$ and extends up to $k_{N,D} \approx k_N/4 \approx 1000 H_g^{-1}$. The lower resolution of the simulation constrains the bandwidth of the inertial range well short of a decade. Beyond $k_{N,D}$, the energy in the gas field drops by nearly a couple of orders of magnitude and no bottle-neck effect is observed either. An unusual pattern is also observed in $\varepsilon_{k,p}$ where the inertial range follows a power-law with a positive index:  $\varepsilon_{k,p}\sim k^{-n_p}$ with $n_p \approx -0.218\pm 0.06$. Similar to the simulations with \St$=0.2$, an increase in the particle energy is also evident here beyond $k_{N,D}$. We are yet to identify the exact reason for the observed behavior in the gas and particle field energy at these small scales. However, it is important to remark that a proper understanding of such systems in the fluid turbulence literature is still in its infancy, and a community wide effort to better understand the particle-gas interplay in a protoplanetary disk setup is warranted. In terms of the convergence of the 3D simulation, it is still too early to assert anything without a high resolution simulation with \St$=0.04$ similar to B3D-02H. Unfortunately, we do not have that at this point due to limited available computational resources.

In case of the 3D axisymmetric simulations (the right two plots of Fig \ref{Spectra-0p04}), we see that  $\varepsilon_{k,g}\sim k^{-n_g}$ where $n_g\approx 2.18\pm 0.05$ for A2D-04H with $2560$ grids per $H_g$ and $2.14\pm 0.06$ for A2D-04SH with $10240$ grids/$H_g$. In both cases, the inertial range starts at $k_{\rm integral}\approx 500 H_g^{-1}$ and extends through $k_{N,D}\sim k_N/4$. However, as is expected from the low resolution run, the bandwidth of the inertial range is significantly shorter compared to its high resolution counterpart. It can be safely asserted that the 3D axisymmetric simulations with \St$=0.04$ are converged with same power-law indices of the inertial range within acceptable uncertainties. When studied in parallel to the ones with \St$=0.2$, it only looks more certain that $512^2$ resolution ($2560$ grids/$H_g$) is probably where a statistical convergence is attained.

The energy spectrum for the particles $\varepsilon_{k,p}$ also behaves in a similar fashion for the 3D axisymmetric simulations where $\varepsilon_{k,p}\sim k^{-n_p}$ with $n_p\sim 0.0416\pm 0.1$ for A2D-04H and $0.424\pm 0.05$ for A2D-04SH. Interestingly, $\varepsilon_{k,p}$ in both the cases do not fall off beyond $k_{N,D}$, but follows the inertial range trend extending through the smallest scales. As we mentioned before, more attention to this matter is needed in order to understand the particle-gas interplay in the smallest scales where the majority of the interesting physics take place in the context of planetesimal formation.

\section{Selected linear analyses}\label{sec:linear_analysis}

\begin{figure*}
\begin{center}
\leavevmode
\includegraphics[width=16.5cm]{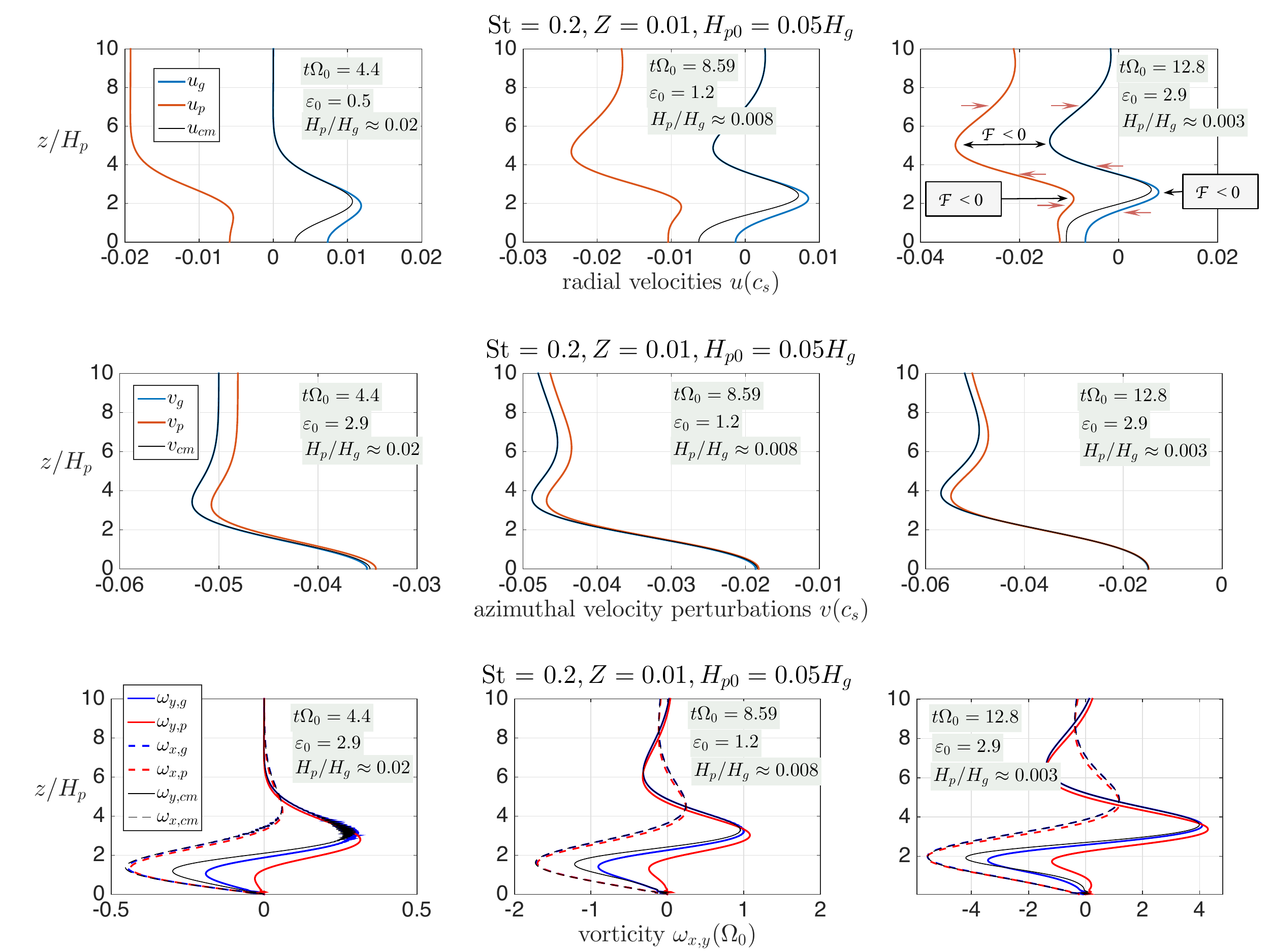}
\par
\end{center}
\caption{Settling solutions for St = 0.2, $Z = 0.01$, and initial particle scale height $H_{p0} = 0.05 H_g$, at three selected times.  Top rows show the radial velocities of the gas (blue), particles (red), and center of mass (grey), (middle row) corresponding perturbation azimuthal velocities, and (bottom row) associated azimuthal (solid lines) and radial vorticities (hatched lines). Panel insets also indicate midplane values of $\varepsilon_0$ and $H_p$.}.
\label{St_0p2_sols}
\end{figure*}

\subsection{Ekman spirals, off-midplane Kelvin-Helmholtz roll-up, and Ri of settling dust}\label{1d_dust_settling_solutions} 

Most of our simulations begin from a laminar state in which dust particles are distributed symmetrically about the midplane using a Gaussian profile with some initial particle scale height $H\sub{p0}$. It is of interest -- in particular, with respect to turbulent development -- to understand how the horizontal velocity components develop as the particles begin their settling toward the midplane. We expect that the system will respond by setting up a velocity field reminiscent of an Ekman layer, as predicted for particle-gas layers in disks \citep{Dobrovolskis_etal_1999}, but modified by epicyclic motions. We therefore consider the horizontally uniform spatio-temporal evolution of the gas and particle velocities as an initial value problem of particles released from rest initially distributed as just described.
\par
In Appendix \ref{particle_settling_solutions} we detail the methods we use to derive the solutions discussed here, which have similarities to the solutions developed in \citet{Lin_2021}. We condense here the basic assumptions of our problem: 
\begin{enumerate}
\item We assume horizontally uniform solutions.  
\item Since the particles are distributed over a layer which is significantly smaller than $H_g$, together with the earlier observation that the evolution of the gaseous component is essentially incompressible in our main simulations, we assume the gas density to be constant and denote it as $\bar{\rho\sub{g}}$. 
\item In steady state the gas has no appreciable vertical motion.
\item The particles are treated as a second fluid with a steady vertical velocity that varies linearly, i.e., $W_p = - \Omega_0 \beta z$, where
$\beta = \beta_-(\St)$ is a constant found in Eq. (\ref{beta_solutions}) in Appendix A \citep[see also][]{Lin_2021}. 
\item The particle density exhibits homologous evolution preserving its Gaussian form defined on the time-dependent particle scale height $H_p(t)$:

\beqa
\varepsilon &\equiv&\frac{\rho_p}{\overline{\rho}\sub{g}} = \varepsilon_0(t)\exp\left(-\frac{z^2}{2H_p^2}\right), \nonumber \\
H_p &=& H\sub{p0} \exp\big(-\beta \Omega_0 t), \nonumber \\
\varepsilon_0(t) &=& \varepsilon\sub{00} \exp\big(\beta \Omega_0 t).
\label{1d_epsilon_solutions}
\eeqa

\item And since the total dust mass is vertically conserved, in which it follows $\displaystyle \varepsilon\sub{00} = Z \cdot \left (H_g/H\sub{p0}\right)$,
solutions are defined by an input value of $\St, \,H\sub{p0}$ and local metallicity $Z$.
\end{enumerate}
\par

\begin{figure*}
\begin{center}
\leavevmode
\includegraphics[width=6.5cm]{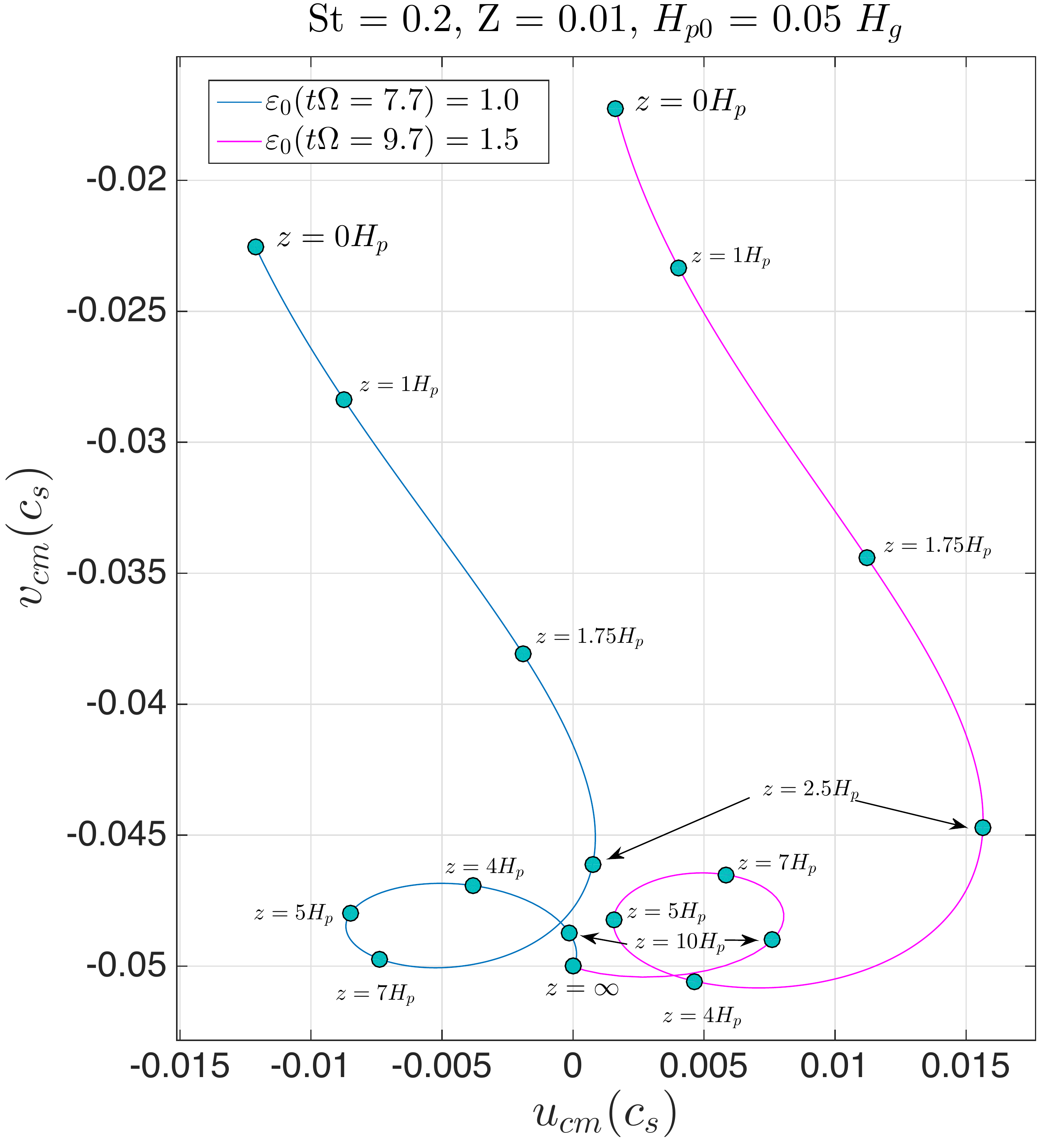}
\includegraphics[width=5.25cm]{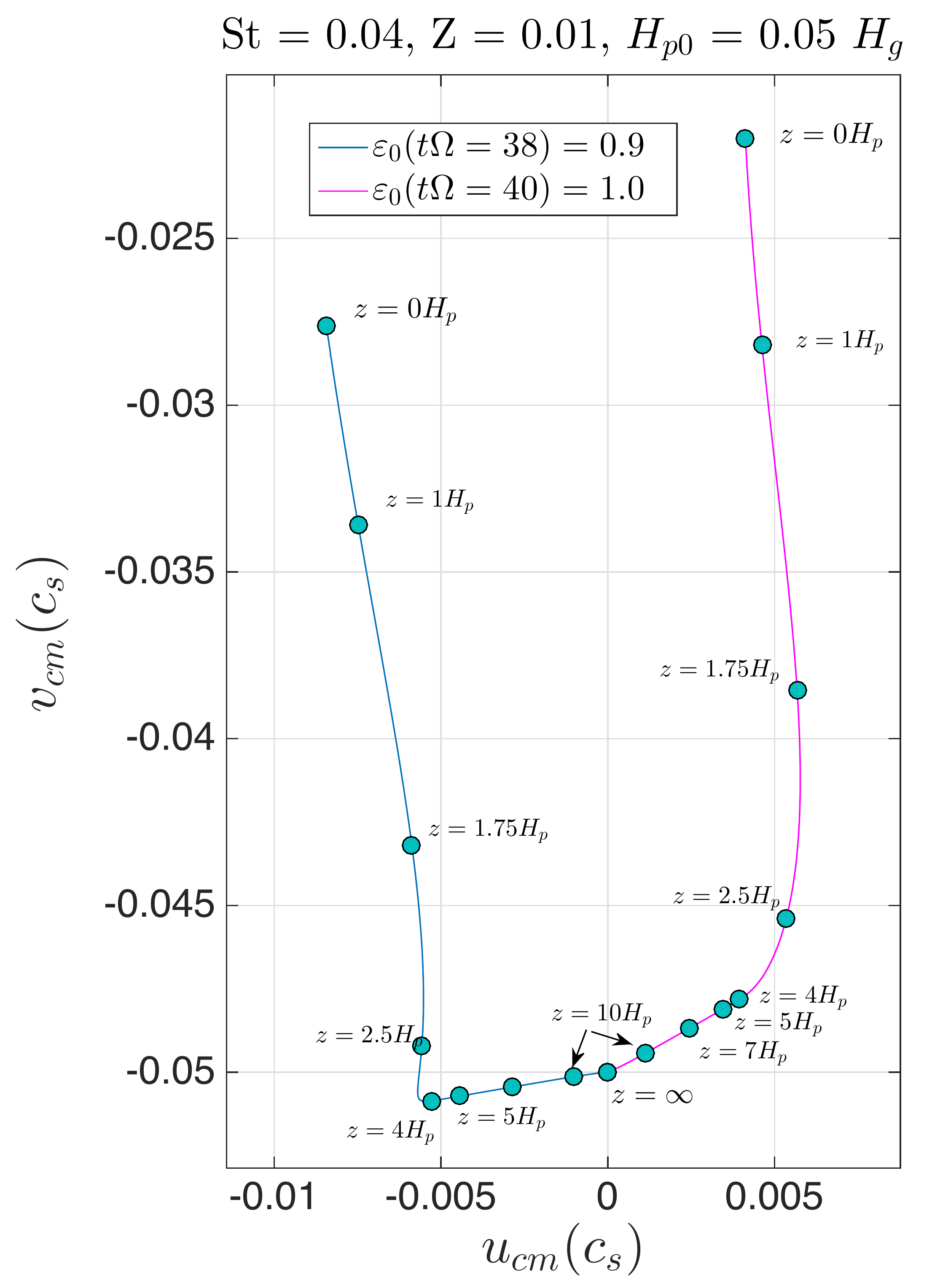}
\par
\end{center}
\caption{Ekman spirals, constructed on the center of mass horizontal velocities, based on the one dimensional settling solutions at selected times: (left panel) St = 0.2, (right panel) St = 0.04. Several fixed values of $z/H_p(t)$ are labeled.}
\label{Figures_Ekman_Spiral}
\end{figure*}
\par

\begin{figure*}
\begin{center}
\leavevmode
\includegraphics[width=17cm]{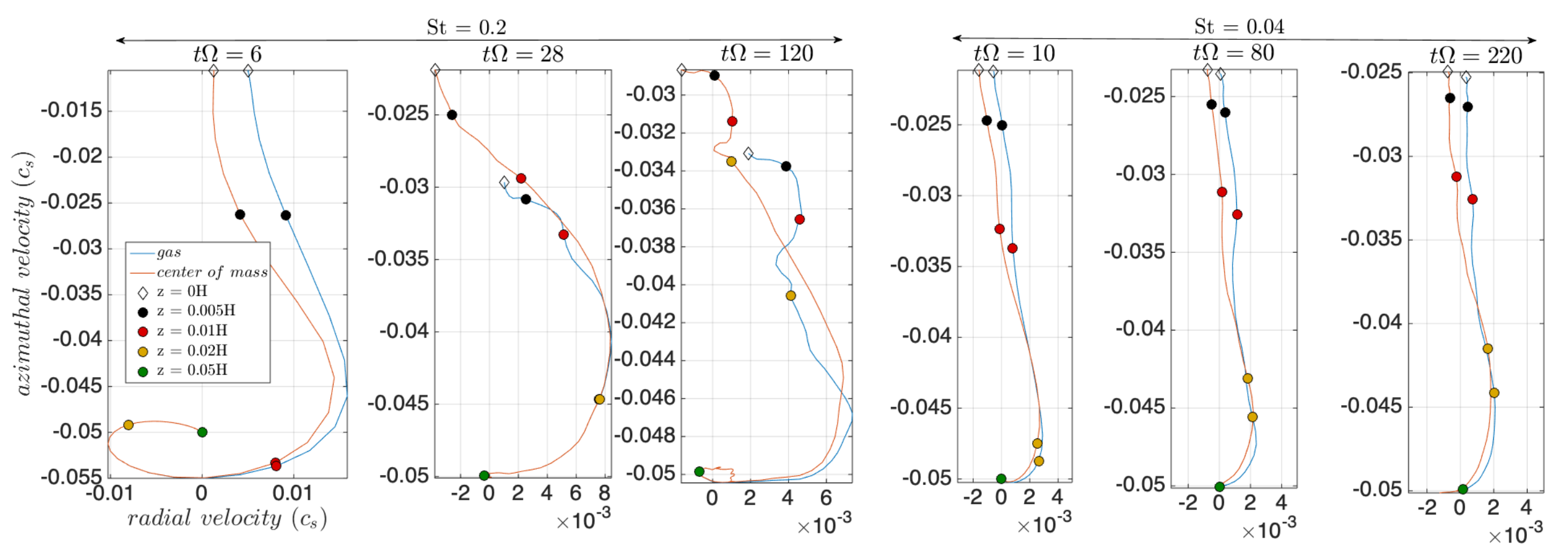}\\
\par
\end{center}
\caption{In comparison to Fig. \ref{Figures_Ekman_Spiral}, we show the emergent
Ekman spirals from the simulations of the previous section at several time stamps.  Here we show radially-azimuthally averaged gas ($\left<u_g\right>_{xy}$ and $\left<v_g\right>_{xy}$) and center of mass velocities -- as defined in Eq. (\ref{cm_velocities_def}) with several layer heights labeled for reference. Simulations shown: B3D-2M for St = 0.2 (left three panels) and B3D-04M for St = 0.04 (right three panels). Note that the Ekman flow structure is strongest for St = 0.2 and that it persists well into the turbulent state.}
\label{Ekman_Plots_From_Sims}
\end{figure*}

 The numerical method employed to solve this problem is different from what is used in PENCIL.  Since our aim is to follow the emergence of horizontal jets as the layer continues its inexorable collapse toward the midplane, it is necessary to resolve the ever-finer developing scales on which particle-gas momentum exchange occurs --  something that cannot be reliably done in simulations with a static vertical grid, especially when the particle scale height approaches the grid scales itself. The method described therefore follows the evolution of the particle-gas layer in a vertically co-moving coordinate frame formulated on a Gaussian grid characterized by the time-dependent particle scale height $H_{p}(t)$ found in Eq. (\ref{1d_epsilon_solutions}).  For further details of the method see Appendix \ref{particle_settling_solutions}.
\par
We discuss the broad properties of the settling solutions by showcasing results for the two values of $\St$ we consider in our large production runs, i.e., $\St = 0.04, 0.2$, together with $Z=0.01$. Our main focus here will be on the latter St value, displaying the former's in Appendix \ref{particle_settling_solutions}. Of the several features we find for St = 0.2, perhaps the main one is the emergence of anywhere from 3-7 counter-flowing radial jets in both the particle and gas fluids.  There is always one main midplane jet whose tip lies at $z=0$, and this is further sandwiched by jet pairs at symmetric locations away from the midplane.  In the simulations involving $\St = 0.2$ particles, 2 symmetric jet pairs form by the time the midplane layer achieves $\varepsilon = 1$.  With increasing distance from the midplane the jets are counter-flowing with respect to one another: the location of the tips of these counter-flowing jets for $\St = 0.2$ are generally found at 2.5 and 5 particle scale heights away from the midplane (e.g., see black arrows in top row of Fig. \ref{St_0p2_sols}), while for $\St = 0.04$ the jets tips are found at 2 $H_p$ and 3.5$H_p$ away from the midplane.  We note also that the amplitude of the jet flow -- especially the midplane jet -- varies epicyclically exhibiting periods of flow reversal (see further below).
\par
The perturbation azimuthal velocity fields (i.e., $v_{g,p}^\prime$) also displays multiple jet flow characteristics like $u$, but by contrast the profile is dominated by the prominent midplane jet. Only by later times (e.g., see top rows of Fig. \ref{St_0p2_sols} and Fig. \ref{St_0p04_sols}, second and third panels) do weaker counter-flowing jets also appear above and below the midplane.
\par
As the particles continue to settle, the radial and azimuthal vorticity components of the jet vorticity -- $\omega_x = -\partial_z v$, $\omega_y = \partial_z u$ (respectively for each fluid) -- steadily increases since $H_p$ continuously shrinks.  When the particles sufficiently settle so that $\varepsilon_0 = \order{1}$, the components of vorticity in each fluid are similarly $\order{\Omega_0}$.
\par
\begin{figure*}
\begin{center}
\leavevmode
\includegraphics[width=7cm]{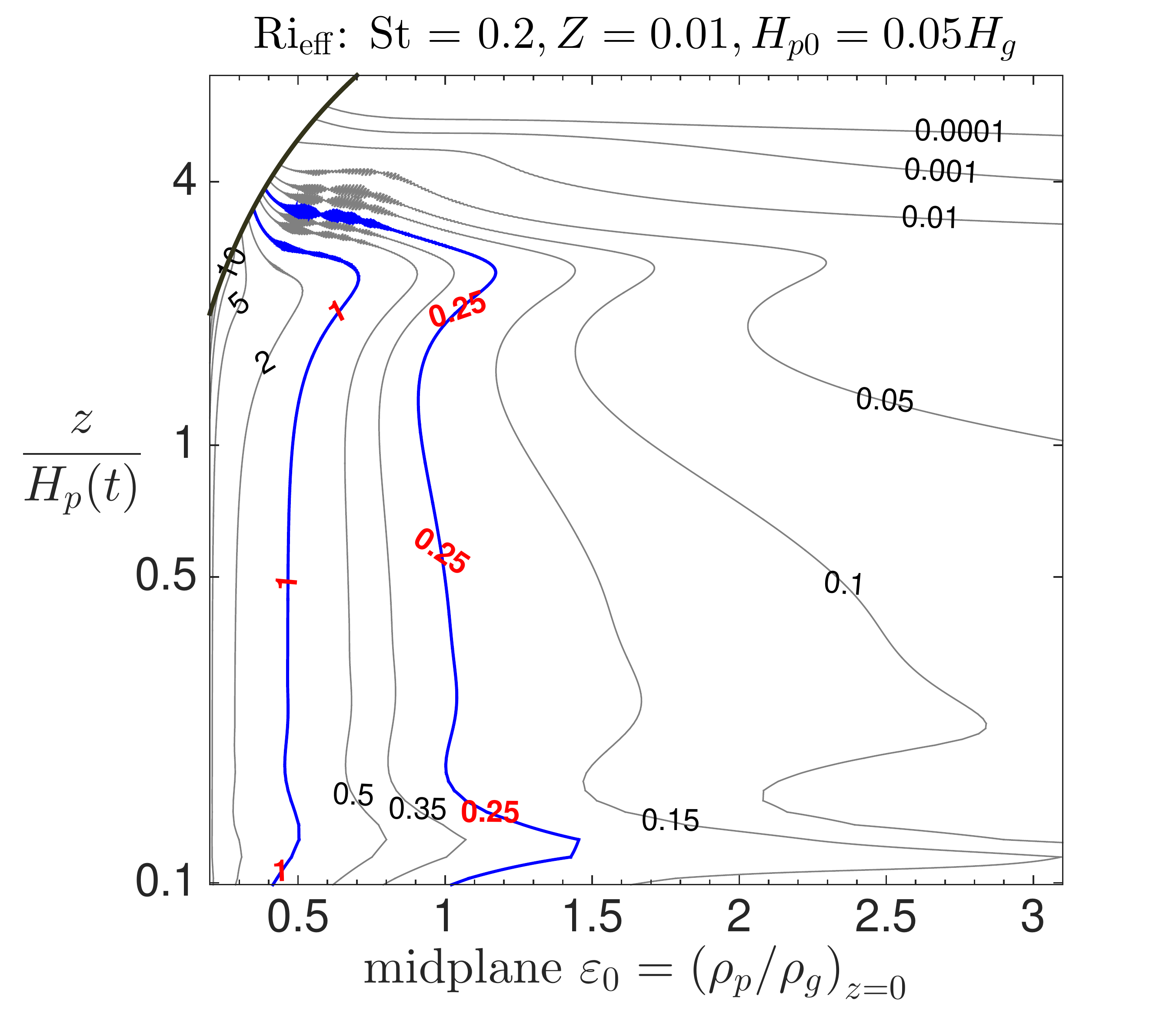}
\includegraphics[width=7cm]{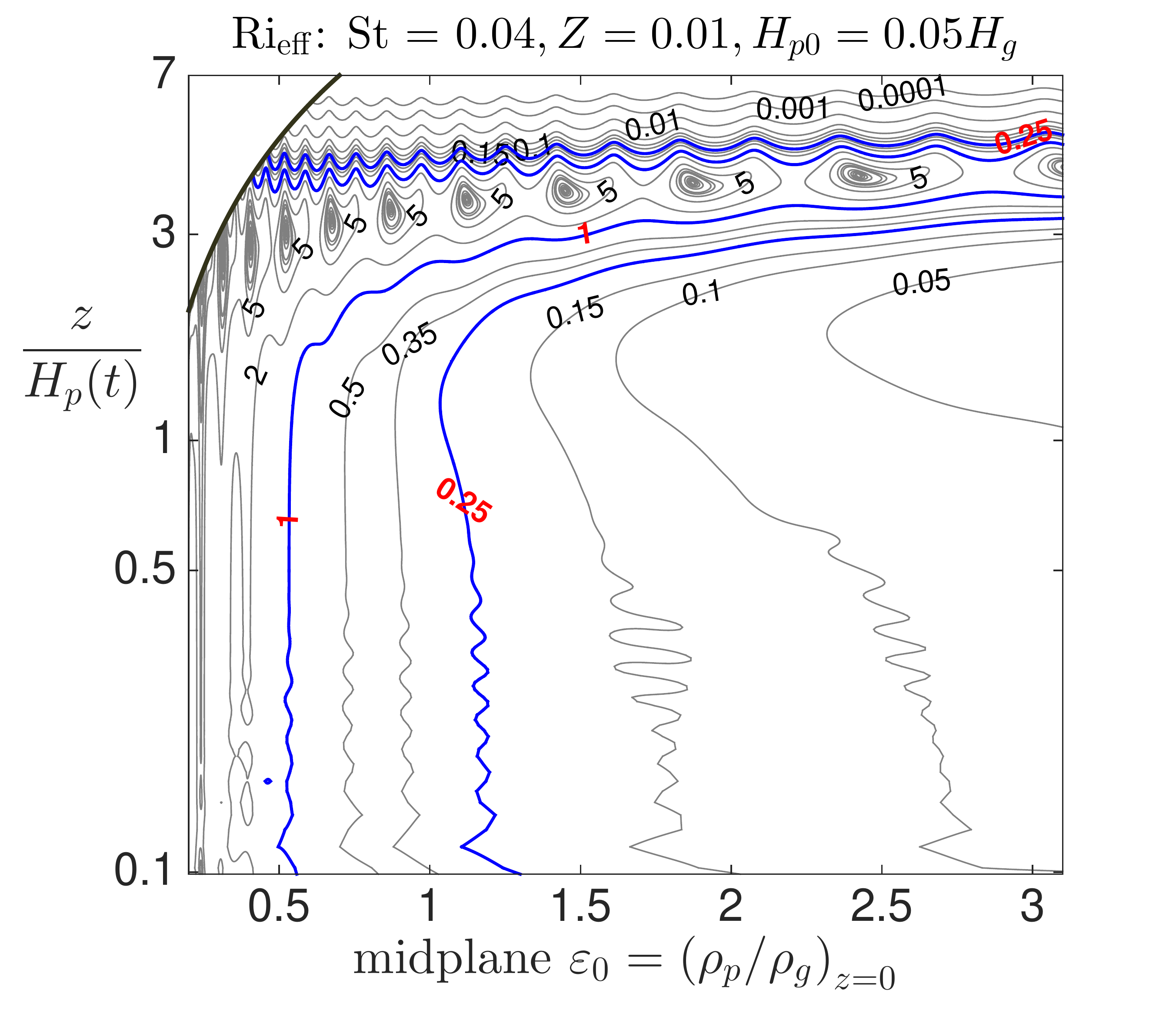}
\par
\end{center}
\caption{Effective values of Ri $\leftrightarrow \Rieff$ as a function of $z/H_p(t)$ and time, which is  proximately represented by the ever increasing values of $\varepsilon_0(t)$ according to Eq. (\ref{1d_epsilon_solutions}).  Left panel shows St = 0.2 while the right panel displays St = 0.04. Various values of Ri are labeled. The special values of Ri = 0.25 and $1$ are shown with blue contours.  The dark-arc bounding the upper left corner of each panel represents the fixed vertical physical scale of our PENCIL code derived solutions, i.e., $z_b/H_p(t)$.}
\label{Figures_Rieff}
\end{figure*}

We find that the particle and gas velocity fields generally follow one another especially at regions well above one particle scale height.  We have defined center of mass velocity quantities,

\beq
u\sub{cm} \equiv \frac{u_g + \varepsilon u_p}{1+\varepsilon};\qquad
v\sub{cm} \equiv \frac{v_g + \varepsilon v_p}{1+\varepsilon},
\eeq
and similarly derive corresponding vorticities,
$\omega_{x,cm} = -\partial_z v_{cm}$ and
$\omega_{y,cm} = \partial_z u_{cm}$.  These are shown throughout all plots in Figs. \ref{St_0p2_sols}-\ref{St_0p04_sols}.  In the same spirit, we define an effective averaged density fluctuation around a mean state in each fluid denoted by subscript ``0" as the following sum

\beq
\frac{\rho\sub{cm}'}{\rho\sub{g0} + \rho\sub{p0}} = \frac{\rho_g'/\rho\sub{g0} + \epsilon \rho_p'/\rho_{p0}}{1+\epsilon}.
\eeq
The buoyancy, $b$, is defined as the vertical gravity multiplying the density fluctuation.   Since for all practical concerns here the gas is effectively incompressible ($\rho_g' \approx 0$), the buoyancy simplifies to

\beq
b\sub{cm} \approx g\frac{\epsilon \rho_p'/\rho\sub{p0}}{1+\epsilon}
= \Omega_0^2 z\cdot \frac{\epsilon }{1+\epsilon}\frac{\rho_p'}{\rho\sub{p0}},
\label{effective_buoyancy}
\eeq
which follows on the physically motivated form introduced by \citet{Chiang_2008}.
\par
One immediately notes how Ekman layer structure develops with increasing severity as the particles continue their collapse. Fig. \ref{Figures_Ekman_Spiral} predicts intensifying Ekman spirals based on the 1D solutions.  This Ekman flow -- that exhibits varying horizontal velocities as a function of height away from the midplane -- is shown for both St values at different times corresponding to values of $\varepsilon_0$ around 1. The emergence of such behavior is not a surprise and is expected to be a generic feature of midplane settled protoplanetary disk dust layers \citep{Cuzzi_etal_1993,Dobrovolskis_etal_1999}. The plots, indicating the center of mass horizontal velocites, also have several values of $z/H_p$ labeled.  Interestingly, the spiral structure is more pronounced for St = 0.2, as the velocity fields exhibit more than one complete loop.  We note that the $z/H_p\rightarrow\infty$ limit corresponds to zero radial speeds and an azimuthal velocity corresponding to $-0.05 c_s$, which is the expected limiting form under a uniform pressure gradient in a region absent of particles according to Eq. (\ref{Asymptotic_Nakagawa_Solutions}) with $\delta = 0.05$ \citep[see also][]{Lin_2021}.  For comparison,  Fig. \ref{Ekman_Plots_From_Sims} shows the Ekman flow structure as appearing in the simulations at various time steps.  The basic qualitative agreement between the restricted 1+1 dimensional calculation and the simulations are evident during the early settling and bounce phases of the simulations. A further detailed comparison to assess how far along in time do the 1D solutions predict the observed mean horizontal flow structure should be done in a future follow-up study.
\par

The radial jet profiles have several inflection points, i.e., locations $z_i$ where $\displaystyle\partial^2_z u\Big|_{z=z_i} = 0$, indicating that the flows violate the Rayleigh criterion for stability against shear roll-up (for instance, see red arrows shown on top row of Fig. \ref{St_0p2_sols}).  Additionally we also observe the \Fjortoft criterion, which  states that unstable linear perturbations of shear flows must have locations $z=z_f$ in which ${\cal F}(z_f) < 0$, where

\beq
{\cal F}(z) \equiv \Big(
u({z}) - u(z_i)\Big)\cdot \partial^2_z u\Big|_{z}.
\eeq
The black arrows on the top row of Fig. \ref{St_0p2_sols} both designate the level of each jet tip
and indicates that these locations clearly satisfy the \Fjortoft criterion.
We see that the jets and the locations where these varied stability criteria are met are significantly removed from the particle layer itself.  In other words, if these jets go unstable they do so free from the direct influence of the particles themselves.  Of course, these jets are a direct result of the passage of particles through that level during settling.  

\par

To this end, we calculate Ri=$\Rieff$ at every vertical level as a function of time in the 1D simulations; these are shown in Fig. \ref{Figures_Rieff}. However, instead of plotting the predicted Ri on the horizontal axis in terms of time, we opt for showing them in terms of the midplane value of  $\varepsilon_0$ -- as defined in Eq. (\ref{1d_epsilon_solutions}) -- which is a more useful proxy for the particle layer's ever shrinking scale height.  We highlight both the contour lines where $\Rieff = 1/4$ and $\Rieff = 1$ based on our discussion above. In all cases considered, $\Rieff$ becomes less than $1/4$ -- especially for regions containing the dust rich layers -- once the midplane value of $\epsilon_0$ begins surpassing $1$.  This is especially evident in the simulation $\St = 0.2$ (Fig. \ref{Figures_Rieff}, left panel) where $\Rieff < 0.25$ over the entire vertical extent of the settling layer. For $\St = 0.04$ (Fig. \ref{Figures_Rieff}, right panel) the criterion begins to be violated at $\varepsilon_0 \approx 1$ and gradually expands toward larger values of $z/H_p$ as the midplane gets more dense.  In fact, a layer in which $\Rieff > 0.25$ appears sandwiched above and below with vertical extent generally diminishing as the midplane density increases. We note the appearance of oscillatory structure in the contour lines which is likely due to the well known and clearly observed epicyclic motions of the jet layer
\citep[e.g.,][]{Li_etal_2018,Li_Youdin_2021}.


\par
It might seem that a similar line of reasoning would lead to the prediction of KH-roll-up by non-axisymmetric perturbations to $v$ owing to its strong midplane jet quality.  However it is likely that such non-axisymmetically driven dynamics get washed out by rapid differential azimuthal stretching experienced by all flow lines owing to the strong radial Keplerian shear
\citep[as suggested in][]{Ishitsu_etal_2009}. However, this remains a conjecture at this point. 
\par
\citet{Gerbig_etal_2020} have suggested that Ri = 1 constitutes a meaningful transition condition for these settled layers; a possibility we further consider in the next subsection.  In anticipation, therefore,
we also label in Fig \ref{Figures_Rieff} those places corresponding to Ri = 1.   We observe that this generally corresponds to values of $\varepsilon_0 = 0.5$.
\par
\par

\subsection{Disk analog of Symmetric Instability}\label{symmetric_instability_problem}
There are recent suggestions that the condition for stability of particle-gas shear flows in strongly rotating protoplanetary disk models, like the ones of concern here, does not need the usual Miles-Howard criterion, Ri$>1/4$, to be satisfied \citep{Gomez_Ostriker_2005,Johansen_etal_2006,Barranco_2009,Gerbig_etal_2020} --  a possibility supported by some theoretical considerations of the nonlinear stability of stratified flows \citep[e.g.][]{Abarbanel_etal_1984,Miles_1986}, and certainly by our numerical simulations reported here.  Model atmospheric flows that support non-barotropic motions can be unstable for ${\rm Ri} > 1/4$. One example of this is the so-called Symmetric Instability (SymI hereafter) -- in which axisymmetric azimuthal mean flows whose flow isolines are misaligned with respect to density/entropy isolines -- are strongly unstable to axisymmetric perturbations for values of Ri up to 1 \citep[e.g.][]{Stone_1966, Vanneste_1993, Stamper_Taylor_2017}.  The total azimuthal flow profiles considered here, which are the sum of the Keplerian flow ($-3\Omega_0 x/2$) and $v_{cm}(z,t)$, certainly exhibit isolines that are misaligned with respect to the particle density isolines that vary only with height.  This is suggestive that axisymmetric disturbances of the flow profiles during particle settling may indeed go unstable by the same type of mechanism that drives the SymI of atmospheric and oceanic flows.
\par
We note that it has recently been argued that the VSI is the disk-analog of the SymI \citep[sometimes referred to as ``sloping convection" in the geophysical fluid dynamics literature][]{Yellin_Bergovoy_etal_2021}.  Indeed, a latitudinal temperature gradient gives rise to a radial variation in planetary zonal flow while a similar vertical variation of Keplerian flow emerges from a radial temperature gradient that leads to the VSI.  The conditions leading to the VSI resemble those giving rise to sloping convection when thermodynamic cooling is instantaneous (zero cooling timescale) within the disk.  In both scenarios, therefore, there is a misalignment between the density and azimuthal/zonal flow isolines.  For further exposition see \citet{Yellin_Bergovoy_etal_2021}.
\par
In the models shown in the figures presented at the end of the last section we find that the condition Ri = 1  is met over significant volumes of the disk's vertical extent once the midplane $\epsilon$ begins exceeding 0.5.  At this stage we conclude that under these conditions the particle layer is quite possibly prone to a disk-analog version of the SymI.

\subsubsection{Motivation}
The turbulent transition in the $\St=0.2$ experiments
unambiguously shows the primary role that the KHI
plays in churning up the particle layer.  This shear roll-up is associated with
the jet pairs located at least 2-3 particle scale heights away from the midplane.  
However, close inspection also reveals that non Kelvin-Helmholtz unstable dynamics are also present in the particle layer itself where the \Fjortoft and Rayleigh criteria 
for shear instability are not clearly met.  This feature is even more prominent
in the $\St = 0.04$ simulation, where KH-roll-up in the layers away from the particle layer is far weaker, the evidence for dynamic activity in the particle layer (within 1-2 $H_p$ of the midplane) falls into even further relief.  In the observed dynamics we have also found that the primary roll-up occurs in exactly axisymmetric or nearly axisymmetric perturbations.  \par
We argue by analogy to submesoscale atmospheric and oceanic dynamics that dynamical
activity in these settled particle layers are driven by the twin action of the SymI and KH-roll-up.  We suspect the significance of the SymI in characterizing the VSI
\citep{Yellin_Bergovoy_etal_2021}
 and the VSSI \citep{Lin_2021} also applies to the turbulent development of the midplane dynamics we report here -- especially for simulations showing sustained turbulent activity in which the SI emerges either weakly or not at all.  \par
 Indeed, analysis of the $\St = 0.04$ simulation after it achieves a quasi-steady shear-driven turbulent state shows that the effective Richardson numbers are near or larger than $1/4$, an important feature also observed in the simulations reported in \citet{Gerbig_etal_2020}.  While it remains to be a circumstantial claim without a more comprehensive analysis, having $\Rieff > 0.25$ maintained in this turbulent state inspires us 
 to consider that the dynamics are not only driven by KH-roll-up
 \citep[while keeping in mind that a rotationally modified KH-roll-up dynamic may also be operating, as suggested originally by][]{Gomez_Ostriker_2005}.
Most importantly, it is our hypothesis that the KH-rollup is associated primarily with the vertical gradient in the radial flow velocities (whether they be expressed as that of the gas or as those of the center of mass), while
SymI action is driven mainly by the vertical gradient of the radial-azimuthal mean of $v$, i.e., the azimuthal velocity departures from Keplerian flow.   As elucidated in submesocale atmospheric dynamical studies and translated to our disk-gas-particle framework, the main ingredients for the SymI are that the total mean flow (azimuthal) velocities have isolines that are misaligned with respect to isodensity lines \citep[e.g.,][]{Vanneste_1993}.
\par
\subsubsection{Model formulation}


\begin{figure*}
\begin{center}
\leavevmode
\includegraphics[width=13.5cm]{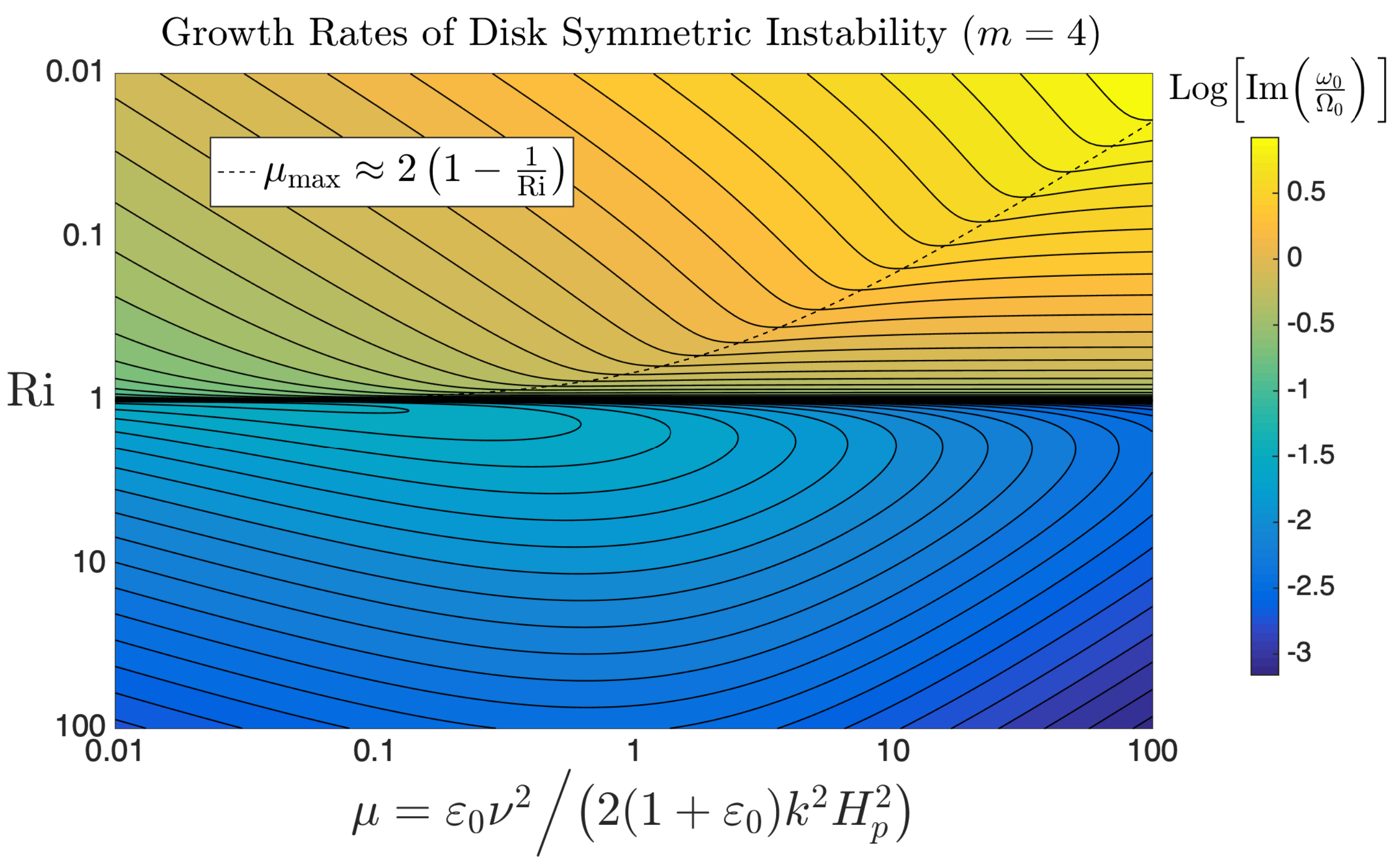}
\par
\end{center}
\caption{Growth rates of the disk analog of the symmetric instability: $m=4$ ($\nu \equiv 2m + 1$).   The general character of these growth rates are the same for all values of $m$, where the critical value of $\Ri$ asymptotically approaches 1. Overplotted is the approximate value $\mu=\mu_{{\rm max}}$ corresponding to the fastest growing mode for $\Ri<1$.  Note that this corresponds to a distinct value $k_{{\rm max}}H_p$ that is implicitly a function of vertical node number $m$ and other properties of the settled particle layer like $\varepsilon_0$ and $\Ri$, see Eq. (\ref{kmax_SymI}). }
\label{SymI_growthrates}
\end{figure*}

We therefore present here a motivated single fluid model (see details in Appendix \ref{disk_symmetric_calculations}) providing an analytically tractable demonstration of the SymI for midplane disk layers. The simplified model can be thought of as the isothermal (i.e., $\gamma \rightarrow 1$) and perfectly coupled terminal velocity limiting form of the single-fluid model for dusty-gas mixtures \citep[e.g.,][]{Laibe_Price_2014,Lin_Youdin_2017}.
This limiting form is the extreme limit of the recast two-fluid equation framework developed  in \citet{Laibe_Price_2014} where the equations of motion are written without approximation instead in terms of center of mass velocities and relative velocities of the two fluid species.  This ``single fluid" limiting form emerges from this recast set of equations after assuming both the terminal velocity approximation together with taking the asymptotic limit of nearly zero particle stopping times.
\par
To isolate the SymI effect from KH-roll-up, we assume there is only a purely azimuthal mean flow state, which is the sum of the Keplerian profile plus a departure $\tilde v\sub 0$ with a parabolic vertical variation given
by

\beq
\tilde v\sub 0 = \delta v\sub{00}\left(1- \frac{1}{2}\frac{z^2}{H_s^2}\right),
\label{model_v}
\eeq
where $\delta v\sub{00}$ characterises the mean azimuthal velocity drop across 2-3 particle scale heights.  
We adopt this as a reasonable facsimile of the center of mass azimuthal flow of the simulations, especially during the bounce and shear-turbulent phase phase of the runs, also see Fig. \ref{facsimile_plot} for more details.
We justify neglecting the vertical variation of the radial flow based on its relatively small amplitude in comparison to the azimuthal flow -- e.g., based on how $\Rir\gg\Riphi$ for all of the shear phase St = 0.04 simulations depicted in Fig. \ref{fig:Richardson_Number_Plots}.

\par
The parabolic form for $v$ adopted in Eq. (\ref{model_v}) is a good fit for the mean profiles that emerge in 3D simulations at various early-to-mid stages of turbulent shear phase development (discussed at the end of Sec. \ref{sec:Richardson_Numbers}) and embodied in the assumed approximate Gaussian form expressed in Eq. (\ref{model_Vcm}). 
We assume a mean density $\rho_0$ whose vertical variation is given by

\beq
\partial_z \ln \rho_0  = - \frac{z}{H_p^2},
\label{density_grad}
\eeq
and, following both \citet{Chiang_2008} and  \citet{Gerbig_etal_2020}, subject to a reduced gravity term acting toward the midplane, and given in magnitude by

\beq
g = g\sub{\rm red} =\frac{\varepsilon_0}{1+\varepsilon_0} \Omega_0^2 z,
\label{geff}
\eeq
where $\varepsilon_0$ is a constant (unlike its intepretation and usage in the previous section), nominally representing the dust-to-gas ratio of the midplane, in the same sense as used in Sec. \ref{1d_dust_settling_solutions}. In this framework given Eqs. \ref{model_v}-\ref{density_grad} together with no radial velocity, we find according to Eq. (\ref{Riphi})'s definition of the azimuthal Richardson number that
 
\beq
\Ri = \Riphi = \frac{\varepsilon_0}{1+\varepsilon_0}\frac{\Omega_0^2 H_s^4}{\delta v\sub{00}^2 H_p^2}.
\label{Ri_simple_model}
\eeq
We note that henceforth we consider only values of $\Ri > 0$, i.e., stably stratified flows.  In order to connect to results discussed in section \ref{sec:Richardson_Numbers}, $\Riphi$ should be considered
as being parallel to and in the same spirit of the definition for $\Riphinot$ expressed in Eq. (\ref{Riphinot_def_1}).
\par
We assume the dynamics are incompressible, which means the radial and vertical velocity perturbations may be written in terms of a single streamfunction $\psi$.  However, vertical density variations are 
advected by the perturbation flow and give rise to buoyancy effects in the dynamics. Except for being set in a shearing box framework, there is no other physics included including drag exchange and fluid viscosity.  \par
All perturbations are axisymmetric, and we assume normal mode solutions that have the functional form for the streamfunction $\psi'$, i.e., $= \hat\psi(z) \exp{\Big(-i\omega_0 t + i k x \Big)} + {\rm c.c.}$, where $\omega_0$ is the normal mode frequency, $k$ is the horizontal wavelength of the disturbance, and $\hat \psi$ is the vertical eigenmode structure function.
\par
In Appendix \ref{disk_symmetric_calculations} we
analyze normal mode solutions of the above described idealized model,
which are expressed in terms of parabolic cylinder functions ${\cal D}_m(z/\tilde\beta)$ \citep{Abramowitz_Stegun_1972}.  Here $\tilde\beta$ is the lengthscale whose real part quantifies the Gaussian decay lengthscale characterizing
${\cal D}_m(z/\tilde\beta)$ -- see  Eq. (\ref{beta_definition}).  The general
solution for the frequency $\omega_0$ is given in (\ref{omega0_sol}) which is a function of three parameters: $\Ri$ as defined in Eq. (\ref{Ri_simple_model}), the non-negative integers $m$ signifying the number of vertical nodes in the disturbances, and finally $\mu$, in which

\beq
\mu \equiv \frac{(2m+1)^2}{2k^2H_p^2}\cdot \frac{\varepsilon_0}{1+\varepsilon_0},
\label{text_mu_definition}
\eeq

\noindent that characterizes a single parameter family of solutions containing $k$ and $\varepsilon_0$.  The general solution for the normal mode response is found in Eq. (\ref{omega0_sol}) reproduced here in slightly rewritten form:

\beq
\frac{\omega_0^2}{\Omega_0^2} = 
\sqrt{
\mu\left(2\frac{{\rm Ri}-1}{{\rm Ri}} + \mu - 
\frac{2i}{\nu} \sqrt{\frac{2\mu}{\Ri}}
\right)
} - \mu + 
\frac{i}{\nu} \sqrt{\frac{2\mu}{\Ri}}
,
\label{omega0_sol_text}
\eeq
where we have introduced $\nu \equiv 2m+1$ for notational convenience. Further analysis found in Appendix \ref{disk_symmetric_calculations} proves that normal modes come in growing/decaying pairs for all finite values of Ri $\neq \Ri_{{\rm c}}(m)$, in which

\beq
\Ri_{{\rm c}}(m) \equiv 1 - \frac{1}{\nu^2}
= \frac{4m(m+1)}{(2m + 1)^2}.
\label{Ri_c_definition}
\eeq
Normal modes are marginal for the countably infinite set of $\Ri = \Ri_c(m)$ values. Inspection of Figure \ref{SymI_growthrates} indeed shows that the growth rates become dramatically large once $\Ri$ passes below $\Ri_c$.  Given $\Ri_c$'s dependence on $m$, nominally speaking we expect strong instability when $\Ri$ passes under $1$, with instability appearing first as $m\rightarrow \infty$.
\par
It is instructive to showcase certain limiting cases for the above solutions which we do in the following subsections.

\subsubsection{$\mu\gg 1$ limiting form, a $\Ri < 1$ criterion}
  We first start by considering the fate of modes in the limit where $\mu \gg 1$ which corresponds to large values of $m$ and/or small values of $kH_p$. According to Eq. (\ref{big_mu_expansion}),
we find

\beq
\frac{\omega_0^2}{\Omega_0^2} \approx 
\left(1-\frac{\Ri_c}{\Ri} \right)
\left(1 + i\frac{1}{\nu}\sqrt{\frac{2}{\mu\Ri}}\right) + \order{\frac{1}{\mu}},
\eeq

\noindent which means that up to $\order{1/\sqrt{\mu}}$ $\omega_0$ is

\beqa
\frac{\omega_0}{\Omega_0} &=& \pm \sqrt{1-\frac{\Ri_c}{\Ri}}
\cdot \left(1  + i\frac{1}{2\nu}\sqrt{\frac{2}{\mu\Ri}}
\right) + \order{\frac{1}{\mu}}.
\eeqa

\noindent There is a dramatic change in the character of the modes when $\Ri - \Ri_{\rm c}$ crosses zero.  When $\Ri > \Ri_c$ the oscillating disturbances, while being unstable, have a growth rate that is weak, and proportional to $1/\sqrt{\mu}$.  However once $\Ri < \Ri_c$, then the modes show order 1 growth.  This says that the character of the disturbances undergo a stark transition at $\Ri \approx \Ri_c$.  We also observe that the modes with the highest values of $m$ are the most unstable when $\Ri < \Ri_c$. For a layer undergoing collapse, where the value of $\Ri$ is steadily going down, the modes with the highest values of $m$ begin to go unstable once $\Ri < 1$ since $\Ri_c(m\rightarrow \infty) = 1$.  Outside of $m=0$, for which the theory predicts stability for all $\Ri$, the critical values for the first few vertical nodes are $\Ri_c(m=1,2,3,4,5,\cdots) = 8/9, 24/25, 48/49, 80/81, 120/121$, and so on. Therefore, for all practical reasons we view SymI to be relevant once the particle layer satisfies $\Ri < 1$ throughout most of its bulk. For the collapsing layer solutions considered in the previous subsection, this would appear to start taking root once the midplane particle layer achieves values of $\varepsilon_0 \approx 0.5$ (for example, see both panels of Fig. \ref{Figures_Rieff}).

\begin{figure}
\begin{center}
\leavevmode
\includegraphics[width=8.5cm]{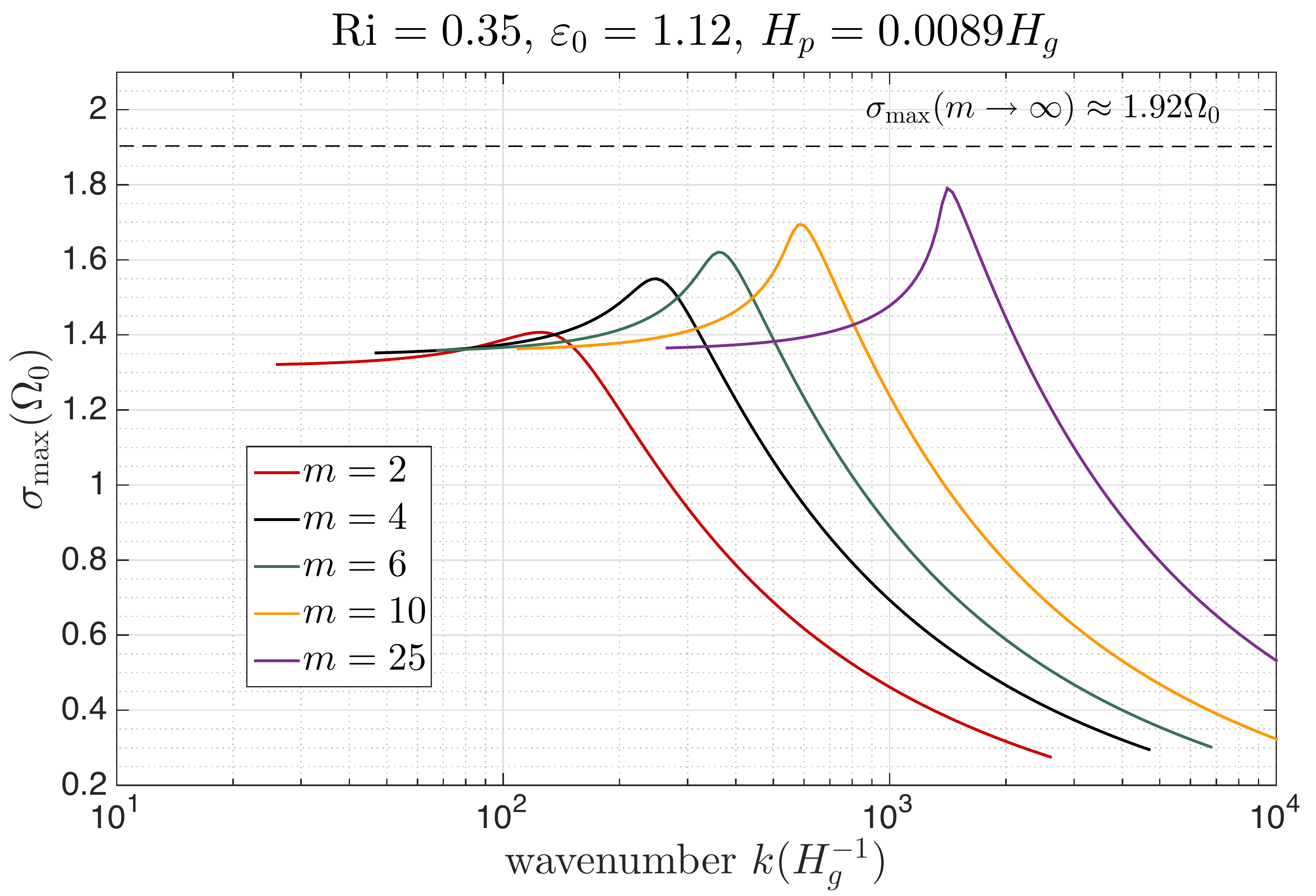}
\par
\end{center}
\caption{Following Eq. (\ref{omega0_sol_text}), this figure depicts predicted growth rates of the SymI shown as a function of radial wavenumber for several vertical parameters $m$ for a given set of values Ri, $\varepsilon_0$ and $H_p$. Following the analysis of section \ref{large_m_asymptotics}, the asymptotic growth rate $\sigma_{{\rm max}}(m\rightarrow \infty)$ is also shown. }
\label{Selected_SymI_growthrates}
\end{figure}

\subsubsection{$0<\mu\ll 1$ limiting form}\label{large_m_asymptotics}
We consider values of $\mu$ that are small, which for all practical purposes corresponds to very large values of $kH_p$ and/or small values of $\varepsilon_0$; however we consider the former case to be of practical use. From Eq. (\ref{small_mu_expansion}) we find to leading order that

\beq
\frac{\omega_0^2}{\Omega_0^2} \approx 
i\sqrt{\frac{2\mu}{\Ri}}\left[ 
 \frac{1}{\nu} - \sqrt{1-\Ri} 
\right].
\eeq
The critical value Ri = Ri$_c$ is easily recovered by equating to zero the expression found within the brackets. The growth rates, while non-zero all throughout $\Ri \neq \Ri_c$, are fairly muted for $\Ri > \Ri_c$ but shoot up with a dramatic change in character once $\Ri < \Ri_c$.  The oscillation frequency also shows a strong shift around $\Ri = \Ri_c$.  
\subsubsection{$m\gg 1$ and $kH_p = \order{m}$: A maximum growth rate}
One can consider the asymptotic limiting form of Eq. (\ref{omega0_sol_text}) in the limit where $m\gg 1$ together with maintaining $kH_p = \order{m}$.  This tandem limit keeps $\mu$ an $\order{1}$ quantity according to its definition in Eq. (\ref{text_mu_definition}).  Thus to leading order it follows that

\beq
\frac{\omega_0^2}{\Omega_0^2} =
\sqrt{\mu^2 - 2\mu\left(\frac{1}{\Ri} - 1\right)} - \mu
+ \order{\frac{1}{m}}.
\eeq
An elementary analysis shows that Im$(\omega_0)$ achieves a maximum value when the expression underneath the squareroot operator equals zero.  Therefore we define a maximally growing value of $\mu$ to be
\beq
\mu_{{\rm max}} = 2\left(\frac{1}{\Ri}-1\right),
\eeq
whereupon it follows that -- provided, of course,  Ri $<1$ -- the fastest growing mode has a growth rate,
$\sigma_{{\rm max}} \equiv $ Im$(\omega_0)$, approximately given by
\beq
\frac{\sigma_{{\rm max}}}{\Omega_0} = \sqrt{\mu_{{\rm max}}} = 
\sqrt{2\left(\frac{1}{\Ri}-1\right)}.
\label{SymI_asymptotic_large_m_growth_rate}
\eeq
We show overlain on Fig. \ref{SymI_growthrates}
the approximate relationship relationship $\mu_{{\rm max}}(\Ri)$.
Finally,
restoring the relationship between $kH_p$ and $\mu$, we find a corresponding
fastest growing horizontal mode, $k_{{\rm max}}$ to be

\beq
k_{{\rm max}} H_p \approx 
\left(m+\frac{1}{2}\right)\sqrt{\frac{\varepsilon_0}{1+\varepsilon_0}
\cdot \frac{\Ri}{1-\Ri}}.
\label{kmax_SymI}
\eeq
Inspection of Figs. \ref{SymI_growthrates}-\ref{Selected_SymI_growthrates}  indicates that Eq. (\ref{kmax_SymI}) does a fairly good job at predicting the fastest growing mode even for $\order 1$ values of $m$.  The asymptotic growth rate expression from Eq. (\ref{SymI_asymptotic_large_m_growth_rate}) works well as an upper bound.  Nevertheless its utility is self-evident as it overpredicts the low order $m$ growth rates by about 25\%, at most.

\subsection{Transition to secondary state via SI?}\label{SI_transition?}
It is worthwhile to ask for the two St numbers simulated, if the transition from the shear driven turbulence phase into the nonlinear pattern state is consistent with the SI.  We consider this by estimating the growth rate of the particle scale height during this transition phase by approximating the domain averaged time dependent scale height $\bar H_p(t)$ as

\beq
\bar H_p(t) = H_{p,{{\alpha}}}
+ \delta h \exp\left(\frac{t}{t_g}\right)
\eeq
being a sum of a mean turbulently set particle scale height, $H_{p,{{\alpha}}}$ that also corresponds to an estimated value of $\alpha \approx \St \left(H_{p,{{\alpha}}}/H\right)^2$ \citep{Dubrulle_etal_1995}, and an exponentially growing perturbation piece characterized by an amplitude $\delta h$ and a growth time scale $t_g$.  The aim here is to compare the simulation derived values of $t_g$ with the prediction made in the theory of turbulent SI developed in \citet{Chen_Lin_2020} and  \citet{Umurhan_etal_2020}.
\par
\begin{figure*}
\begin{center}
\leavevmode
\includegraphics[width=17.0cm]{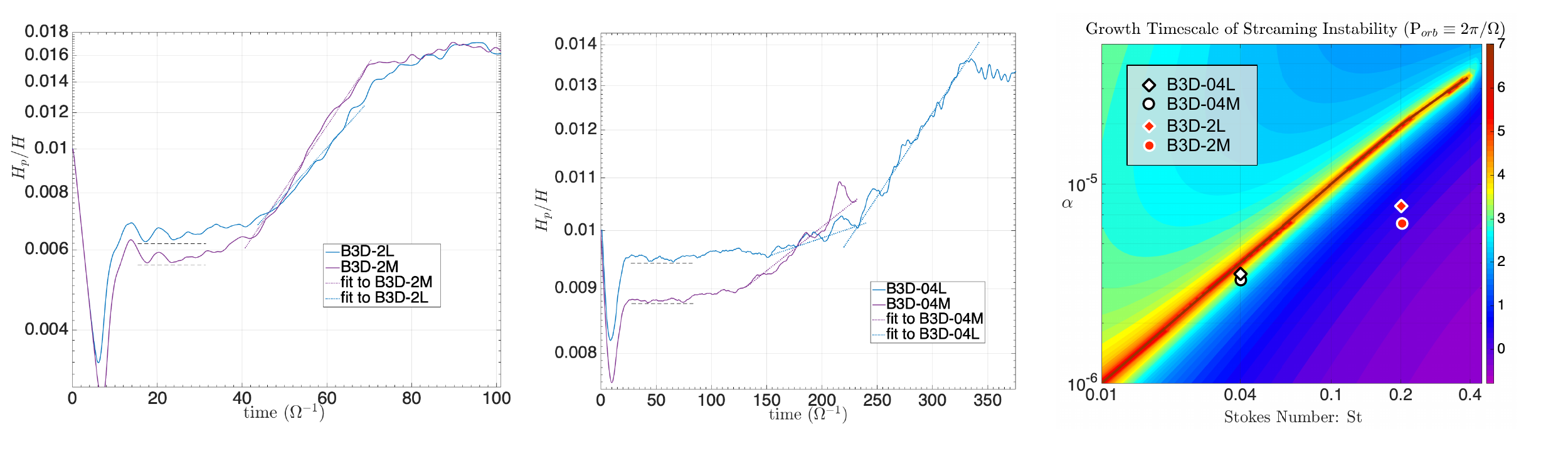}
\par
\end{center}
\caption{Particle scale height versus time for several 3D simulations and a semilog plot
together with approximate baselines (dashed lines, for $H_{p,\alpha}$ and approximate growth lines (dotted lines) described further in text: (left panel) low to medium resolution runs for St = 0.2 , (middle panel) low to medium resolution runs for St = 0.04.  (Right panel) predicted growth rates for the SI maximized over wavenumber \citep{Umurhan_etal_2020} as a function of St and $\alpha$, where the four simulations shown in the left two panels are represented with symbols.
Note all simulations appear on the relatively active side of the SI.}
\label{Hp_Theory_Growth}
\end{figure*}

The left and middle panels of Fig. \ref{Hp_Theory_Growth} shows a log-linear plot of the time series for $\bar H_p$ for four simulations.  Various dashed lines show our estimated fitted values for $H_{p,{{\alpha}}}$ and corresponding estimated slopes (dotted lines) drawn over the secondary growth phase where the approximate values of $\bar H_p(t) \approx \tilde H_p(t)$ may be read off of these lines for any input time $t_i$ falling nominally in the growth rate time range. Based on this understanding of $\tilde H_p(t)$ it follows

\beq
\ln \left[\tilde H_p(t_i)-H_{p,{{\alpha}}}\right]
\approx \ln \delta + \frac{2\pi t_i}{t_g},
\eeq
which, after
selecting two times from the fitted lines, allows one to estimate $t_g$, i.e.,

\beq
t_g \approx (t_2 - t_1)\bigg/\ln\left[
\frac{\tilde H_p(t_2)-H_{p,{{\alpha}}}}{\tilde H_p(t_1)-H_{p,{{\alpha}}}}\right].
\eeq
Taken from the theory developed in \citet{Umurhan_etal_2020}
the right panel of Fig. \ref{Hp_Theory_Growth} shows the predicted growth timescales of the SI for $Z=0.01$ and $\beta = 0.05$ as a function of $\alpha$ and St as maximized over wavenumber.  Note that the growth timescales in that study are quoted in units of local orbit times, i.e., $P_{{\rm orb}} = 2\pi\Omega^{-1}$.  Thus, we read off the quoted values from that graph and multiply it by $2\pi$ and call the result $t_{g,{{\rm th}}}$ in order to compare the predicted growth rates against the values of $t_g$ derived from the simulations presented here, the latter of which are in units of $\Omega^{-1}$.  The results of this exercise are summarized in Table \ref{SI_Fitting}.

\begin{table}[ht!] 

\centering
\caption{Simulations and Predicted SI growth rates maximized over wavenumber}
\vspace{0.1in}
\begin{tabular}{c | c c c c | c}
\hline
\hline
 $\displaystyle 
 \begin{array}{c}
 {\rm Simulations} \\
 {\rm Identifier}
 \end{array}
 $
 & St
 &  $\displaystyle \frac{H_{p,\alpha}}{H}$ 
 & $\alpha\times 10^6$ 
 &   $\displaystyle {t_g\Omega}$ 
 &   $\displaystyle {t_{g,{{\rm th}}}}\Omega $
 \\
 \hline
B3D-04L & 0.04 & 0.0095 &  3.57 & 	2306 & 	15800 \\
B3D-04M & 0.04 & 0.0088 &  3.10 & 	59.7 & 	2513 \\
B3D-2L & 0.2 & 0.00625 &  7.83 & 	10.5 & 	16.5 \\
B3D-2M & 0.2 & 0.0056 & 6.30  & 	9.11 & 	11.2 
\end{tabular}
\label{SI_Fitting}
\par
{\small{
$^\dagger$Based on theoretical predictions of \citet{Umurhan_etal_2020}.
}}
\\
\end{table}
Examining the right panel of Fig. \ref{Hp_Theory_Growth} depicting the results of the low to medium resolution 
runs with St = 0.2, we find that the level set during the onset of the shear turbulent phase has a lower value of $H_{p,{{\alpha}}}$ (by about 15\%) for the higher resolution of the two, a trend observed in other simulations of the SI \citep[e.g.,][and others]{Yang_etal_2017,Li_etal_2018}.  The lower resolution run therefore operates at a higher effective $\alpha$.  Theoretically it would imply a longer growth rate for the lower resolution run and this trend is borne out by the prediction for turbulent SI.  However, a comparison of ${t_{g,{{\rm th}}}}$ against the measured values $t_g$ values found in Table \ref{SI_Fitting} shows that the theory appears to systematically underpredict growth rates.
\par
The middle panel of  Fig. \ref{Hp_Theory_Growth} shows the corresponding runs for St = 0.04.  The same trends, i.e., vis-\`a-vis shear turbulent phase, are manifest here as well: the lower resolution run corresponds to higher turbulence levels compared to the higher resolution, with concomitant faster growth rates of the latter compared to the former.  However, the growth rates are vastly shorter than those predicted for the turbulent SI according to \citet{Umurhan_etal_2020}.
\par
At this stage it is difficult to conclude as to what may be driving this secondary growth, and we can at best only speculate.  It might be possible for these runs that the classical SI is acting in conjunction with a secondary effect that derives from the turbulent vertical shear profile \citep[e.g., the VSSI,][]{Lin_2021}, but that the effect of the classical SI is stronger than this secondary effect in the St = 0.2 case -- resulting in a better match based on just the SI growth rates alone -- while the secondary effect is stronger than the SI in the St = 0.04 case in which the actual observed growth timescales are much shorter than those predicted from the turbulent SI prediction. We think the VSSI is the leading candidate mechanism to explain this effect; however, an investigation of this hypothesis is outside the scope of this paper but should be considered in follow-up work.

\section{Discussion}\label{sec:discussion}

\subsection{On turbulence in particle laden sheared midplane layers}
The simulations conducted here appear to indicate that midplane settled particle layers that are not strongly susceptible to the SI instead erupt via other stratified axisymmetric fluid instabilities involving the vertical shear of both the radial and azimuthal velocities.   These instabilities appear to lead to an early turbulent phase from which some secondary processes takes root, possibly involving the SI, that leads to a sinusoidal pattern state of varying degrees of coherence and characterized by some amount of radial drift.\par
The instability mechanisms leading to the early turbulent phase involve the particle component as a collective agent, and mainly in a stabilizing role.  The flow analysis conducted indicates that the primary mechanism driving instability depends on the St number of the particle component and although we have examined only two values of St, we can confidently say that there are at least two types of destabilizing processes: Based on the St = 0.04 suite of simulations it appears that these settled layers primarily experience the protoplanetary disk analog of the SymI, whose dynamic relies on the vertical shear of the azimuthal velocity profile and requiring Ri $< 1$.  Based on the St = 0.2 suite of numerical experiments the settled layers experience radial KH-roll-up within layers located about 1-2 particle scale heights away from the midplane.  Further, the assessed Ri values within the settled particle layers for St = 0.2 should support, in principle at least, azimuthal KH-roll-up as well, but visualization of the flow fields do not present clear evidence for it either.\par
During the review phase of this manuscript one of the reviewers commented on the possibility that the non-turbulent SI could explain the early bounce phase turbulent development.  We have examined this possibility by doing a spot-check on the predicted growth rates for the St = 0.04 case using the laminar SI theory of YG2005.  For the apparent lengthscales ($\lambda \sim 0.01 H$) emerging during the early bounce phase (e.g., see the $w_g$ field in Fig. 6 at $t\Omega = 6$), the predicted SI growth rate in the zero turbulent limit falls in the range $0.05-0.1 \Omega_0$, which is at least a factor of 5 weaker than the corresponding growth rate of the SymI under those similar conditions.  A detailed examination of this matter, to better illustrate this case, deserves to be done in the future.
\par
In either case, however, the midplane layers exhibit unsteady motions -- apparently turbulent -- that eventually leads into a secondary transition resulting in a pattern forming state.  We have examined whether the drifting patterned state is an instance of the SI and the predictions also appear mixed: for the St = 0.2 case the measured growth rates between the initial putative turbulent phase and the final pattern state appear to be consistent with SI under turbulent conditions  if, however, systematically somewhat faster-acting than predicted
\citep[e.g., following the theory of,][]{Umurhan_etal_2020,Chen_Lin_2020}.  On the other hand, in the St = 0.04 case, the corresponding measured transition phase growth rates are far more rapid than those predicted by one to two order of magnitude (see Table \ref{SI_Fitting}).  What is responsible for this growth, and what is the cause for the dynamical differences between these two St values, remains uncertain.  Is it the SI working in tandem with a secondary process, which senses the underlying turbulent state in the mean?  Might the secondary transition be an instance of the proposed Vertical Shearing Streaming Instability (VSSI), a process recently examined in a theoretical model by \citep{Lin_2021}?  Furthermore, is it possible that this emergent drifting pattern state is a numerical artifact that goes away if the radial and/or azimuthal box scales are made larger?  These are all questions that need immediate resolution in future studies.  In Sec. \ref{On_Ri} we further reflect on the possible action of KH-roll-up for Ri $>1/4$.
\par
Often rooted in simulations of relatively large St simulations (e.g., St $ > 0.2$), it is a commonly held assumption that settled particle layers subject to the SI are likely also susceptible to azimuthal KH-roll-up \citep[e.g.][]{Barranco_2009,Lee_etal_2010a,Lee_etal_2010b}.  Our findings suggest that there is, in fact, a diversity of processes at play; and which actor dominates depends on the layer's St and local metallicity $Z$.  Indeed, the St = 0.04, $Z=0.01$ simulations examined here appear to be driven entirely by the axisymmetric SymI effect.  Increasing St to 0.2 while keeping $Z$ fixed appears to lead to both the SymI and radial KH-roll-up as the primary dynamical mechanisms.

\subsection{On the relationship to the findings of Ishitsu et al. (2009)}

In their unpublished study \citet{Ishitsu_etal_2009} examined the fate of particle-gas setups like considered here, in which an imposed radial pressure gradient induces a relative streaming between gas and particles.  Their simulations were 3D axisymmetric and considered two sets of St = 0.001 and 1.0.  The early development of the layer in the St = 0.001 case is highly reminiscent of the transition observed in both of our St suite of runs.  In particular, the upper right panel of \citet{Ishitsu_etal_2009}'s Figure 6 shows off-midplane filament development where the particle layer's vertical gradient is greatest.  This character is very similar to the dynamical structure observed throughout the early development of both of our 3D axisymmetric {\it and} full 3D simulations.  For the St $=1.0$ simulation shown, their simulation erupts into the kind of pattern characteristic of KH-roll-up as reported in \citet[][e.g., see Figure 7 of that work]{Barranco_2009}.  No such KH-roll-up pattern is discernible in the St = 0.001 run reported in \citet{Ishitsu_etal_2009}.
\par
In this sense we think these authors are justified in writing,
\begin{quote}
``However, the instability of two-fluid shown in this work has the axis-symmetric unstable mode. As a result, the stabilization caused by the increase of the azimuthal wave number due to the radial shear is not effective. We expect that the instability occurs in the radial direction, and then the perturbation with small azimuthal wave number grows," 
\citep[][pg. 14]{Ishitsu_etal_2009}.  
\end{quote}
Not only do our 3D axisymmetric simulations confirm this notion, but our full 3D investigations suggests that non-axisymmetric dynamics -- e.g., like azimuthal KH-roll-up -- are not the primary instability mechanism driving turbulence in low St midplane-settled particle layers, i.e., our findings implicate axisymmetric modes as the primary driver of turbulence in full 3D scenarios.

\subsection{This study in relation to Lin (2021)}

\citep{Lin_2021} undertake a two-fluid stability analysis of a model similar in set-up to what has been considered here.  In that study the author considers the stability of an already settled particle layer whose particle equilibrium is established within the framework of a local $\alpha$-disk shearing box model often used to capture the essence of particle profiles within gas driven turbulence \citep[e.g.,][]{Dubrulle_etal_1995,Youdin_Lithwick_2007,Laibe_etal_2020} and as recently used in the analysis of the SI by \citet{Chen_Lin_2020} and \citet{Umurhan_etal_2020}.  For input values of $Z$, St and $\alpha$ the particle gas equilibrium in the radial-azimuthal mean admits solutions with non-trivial vertical variations in all quantities on which a numerical stability analysis is done.  For conditions examined most similar to ours (c.f., case C with the St = 0.04 runs treated here) \citet{Lin_2021} report that the fastest growing mode has a horizontal wavenumber $kH_g \approx 1100$ with growth rate $\sim 0.6 \Omega$ and, most importantly, with a mode amplitude becoming greatest at about 1-2 scale particle scale heights.  Perhaps most importantly, \citet{Lin_2021} show that the vertical shearing instability revealed in the two-fluid setup is recovered within a single fluid framework as well.
\par
While our findings agree in spirit with those reported in \citet{Lin_2021} there are notable differences likely attributed to our differing steady states, since ours are drawn directly from those exhibited by actively collapsing solutions as opposed to those equilibria resulting from their adopted putative turbulence model. Since our analysis includes neither particle diffusion nor a model for turbulent viscosity, we predict that instability persists as $m\rightarrow \infty$ with corresponding fastest growing radial wavenumber similarly diverging, and this is similar in quality for the VSSI in the inviscid limit as reported in \citet{Lin_2021}.  However, a notable difference in predictions of the inviscid limit is that we predict growth rates that asymptote to $\order 1$ values of $\Omega$ while the VSSI seem to grow without bound as $kH \rightarrow \infty$ \citep[e.g., see Figure 15 of][]{Lin_2021}. Of course, viscosity due to any developing turbulence should temper growth rates at short wavelengths, at least in reality.  
\par
Yet, the simple setup that went into motivating the SymI analysis developed in section \ref{symmetric_instability_problem} captures the essence of what occurs during the settling and transition phase of the numerical experiments conducted here, especially the St = 0.04 simulations.  Both studies demonstrate that an effectively single-fluid process can lead to strong instability and quite likely explains the dynamical source of turbulence in these midplane settled layers.  Moreover, by being an effectively single-fluid model, it shows that the process does not need a relative stream between the two fluids in order to become active like required for the SI \citep[in this regard see recent conceptual advances of][]{Squire_Hopkins_2018a,Squire_Hopkins_2018_MNRAS}. Despite this broad conceptual agreement,  further work is needed to reconcile these two theoretical approaches.  However, it would seem that the VSSI framework might be best applied in analyzing how an already turbulent sublayer further develops as it already has built into it a model of turbulence.  Perhaps it might be used in explaining the secondary transition into the drifting pattern state we have reported on here.  This is grounds for further investigation.

\subsection{The Current work in the context of Garaud \& Lin (2004)}

\cite{Garaud_Lin_2004} conducted similar studies in order to investigate the evolution of the settled dust layer in a two fluid approximation. Their work used a mono-disperse population of small dust grains along with a gas disk which is strictly laminar in the absence of solids. However, in their working perturbation equations (their equations 15-19) the term containing the coriolis force, and hence the effect of rotation in the problem is absent. This is a major and the most important difference between their work and the current one. The emergence of SymI in the work presented here is axisymmetric in nature and emerges solely from the vertical variations of the azimuthal flow velocity that has its root in the cross-velocity components in the Coriolis term, whereas \cite{Garaud_Lin_2004} put their emphasis on the growth of non-axisymmetric perturbations absent Corliolis influences. Hence the evolution of the shear layer reported in their study belongs to a general class of KH instability without any possibilities of SymI. Apart from this point, \cite{Garaud_Lin_2004} put significant effort on the effect of cooling time of the stability of the shear layer, whereas we adhere to a strictly isothermal condition for all our analysis. It is noteworthy that they observe the boundary of instability to extend beyond the classical limit of Ri =1/4 when cooling effects are included in simplified analytical modeling. Whether these effects play a role in realistic disks remain to be explored. 




\subsection{A note on dust settling instability}

Settling dust grains can also give rise to the dust settling instability (DSI), a bi-product mechanism of the general class of resonant drag instabilities
 \citep[RDI,][]{Squire_Hopkins_2018a,Krapp20}. As noted in \citep{Squire_Hopkins_2018a}, the settling instability has two regimes: (i) for modes in which $k_x w_{sx} \neq -k_z w_{sz}$, where ${\bf w}$ is the streaming velocity, the growth rates are generally   $\le \Omega$ but, (ii) as $k_x w_{sx} \rightarrow -k_z w_{sz}$ growth rates show divergence with increasing absolute wavenumber $k$. In the latter case and so long as $w_{sz}\neq 0$, \cite{Squire_Hopkins_2018a} predict that a rapid eruption of activity will appear at the smallest lengthscales, with growth rates $\sim k^{1/3}$ (for $\epsilon \ll 1)$. \cite{Krapp20} observe the development of fine scale structure in their highly resolved simulations centered on disk sections centered at heights $z\sim H_g$ with box sizes $\sim 0.1 H_g$. Most of the activity of the DSI is expected to happen far from the midplane, nominally at locations where the product combination of settling velocities and local mean dust-densities are maximized. Moreover, the unrestricted growth rates for increasing wavenumber suggests that resolving DSI may need more resolution than what we have in our simulations. In our simulation results we conjecture that the DSI may be operating, especially far from the midplane, although, we have not notice any obvious signature in our simulation domain spanning only $0.2~H_g$. However, in our analytical model, the primary equations do not have any vertical velocity under equilibrium (see equations \ref{u_prime_eqn} -- \ref{Delta_prime_eqn}) that can give rise to DSI.  This lack of signs of the DSI in the simulations reported here may be because the simulations take place close to the midplane where the settling velocities are expected to be small since $w_{sz} \sim \Omega H_p \St$, e.g., see Eq. (\ref{settling_Wp_eqn}). However, with sufficiently high numerical resolution the fastest growing mode ought to appear as very short vertical wavenumber midplane parallel banded structures, which given our current resolution capabilities, are not resolvable. Additionally, there is as yet no Richardson Number analysis for the DSI based on which, a more quantitative distinction between the two processes might be made.  Lastly, while the DSI can lead to disorder in the flow as the flow settles, whether or not it can drive sustained midplane turbulence once the main settling phase has receded is not expected.

\subsection{On spectra and simulation convergence}

The St = 0.2 kinetic energy power spectrum from Fig. \ref{Spectra-0p2} makes it evident that for the 3D simulations, the power-law index of the inertial range changes when we increase the resolution. The root cause of this effect, not being properly pinpointed in this work, could be multifaceted. It is possible that the 3D simulations are not yet resolved at $512^3$ resolution ($2560$ grids $/H_g$) and that higher resolution is required. Another possibility could be due to the number of particles used in the simulations with moderate and high resolutions: In order to minimize the computation expense, the number of particles used per grid point in the high resolution run (B3D-02H) is only $0.125$ (i.e., 1 particle per 8 grids) compared to $1.0$ in the moderate resolution run (See table \ref{tbl:simulation}). The gas and the particle fields in the system communicate with each other through a drag term with a finite relaxation time -- Eqs. (\ref{eqn:gasmomentum}-\ref{eqn:dustmomentum}) --  where the momentum exchange through the density fluctuations are nonlinear, implying scale-to-scale energy transfer. So, the implementation of a comparatively low number of super particles may influence the gas dynamics. We disfavor this explanation because when settling is present, particles cluster towards the disk midplane effectively reaching more than one particle per grid. We also do not see any bottleneck effect in the gas kinetic energy beyond the dissipation scale, so the possibility of back-scattering of energy can be ruled out. 


It is also interesting to note that for 3D-axisymmetric simulations, the energy spectrum for both high and super-high resolution simulations are converged with similar power-law index for the inertial range. For both \St$=0.2$ and $0.04$, the power-law index hovers between $2.1$ and $2.2$. The small difference here for the two different St is consistent with the trend reported by \citet{Pandey_etal_2019} where the authors found a dependence of the energy spectrum on the overall mass loading effect, especially at the higher wavenumbers. Note that the highest 3D axisymmetric simulations have $10240$ grids/$H_g$, so it is possible that for the full 3D simulations, going beyond $512^3$ may show convergence in the energy spectrum.  Given the relatively minor change in power-law slope in the 3D-axisymmetric suite of runs going from high ($512^2$) to super-high ($2048^2$) resolution runs, we cautiously conjecture that the $512^3$ resolution full 3D simulation might be close to convergence for the dynamically resolved scales.
Unfortunately, we were limited by available computational resources to carry out simulations with resolution higher than $512^3$ and are unable to verify this conjecture at this time nor characterize the turbulent behavior at high resolution for St = 0.04. 
\par
It is also important to note that the character of the 3D axisymmetric simulations in the inertial range are fundamentally different from the full 3D ones, both in terms of slope of power spectrum and the location of the integral ($k_{\rm integral}$) scale. Classic two-dimensional  turbulence is characterized by simultaneous downscale enstrophy cascade and upscale energy cascade together with a $E_k\sim k^{-3}$ behavior in the inertial range \citep{Kraichnan_Montgomery_1980}, while in full 3D isotropic turbulence energy cascades toward smaller scales with $E_k\sim k^{-5/3}$.  3D axisymmetric scenarios in a disk might exhibit cascade properties similar to purely 2D flows, but this is not quite certain as yet. The statistical behavior under conditions where rotation and stratification are on equal footing is complicated by wave dynamics, which present alternate pathways for scale-to-scale energy exchange beyond just nonlinear velocity advection (i.e., inertial effects). Little is known about how this unfolds under this dual influence in both atmospheric/geophysical flows \citep[e.g., see discussion of this for atmospheric flows in Sec. 4.5 of][]{Alexakis_Biferale_2018} and much less in dust-laden protoplanetary disk models.  It is known that strongly stratified flows, or dimensionally constrained settings can exhibit mixed/split and anisotropic energy cascades while strong rotation can support some amount of inverse cascade behavior under suitable forcing and/or vertical scales.  In these cases new exchange pathways emerge due to wave dynamics and produce inertial range spectral slopes that deviate from that expected from pure 2D turbulence ($k^{-3}$) or from full 3D Kolmogorov ($\sim k^{-5/3}$).  For example, under relatively strong stratification and moderate wavenumbers in atmospheric models there exists the so-called Bolgiano-Obukhov scaling where the kinetic energy behaves like $E_k \sim k^{-11/5}$ before it eventually turns down to $E_k\sim k^{-5/3}$ at sufficiently high wavenumbers.  In this scenario energy exchange carries kinetic energy into potential energy through wave transfer via gravity waves upon which nonlinear potential energy advection forward spreads energy toward smaller scales.  In our particle-laden sheared setting there are nonlinearities in the gas-dust drag exchange terms.  Exactly how this dynamical pathway mitigates and/or directs the cascade of energy -- and how it ultimately shapes the resulting energy spectra --  is yet to be systematically examined in either 3D axisymmetric and/or full 3D settings.  
\par
In short: the energy content in these flows depends upon whether the equations are being treated as full 3D or not, and that the spatial energy distribution may not be Kolmogorov, especially at intermediate scales bridging energy injection and the very short scales that should exhibit $k^{-5/3}$ Kolmogorov character. Therefore, looking back upon our results: while large scale structures look qualitatively similar in both axisymmetric and full 3D, a one-to-one correspondence between the two may not be an accurate representation on the small scales were particles are expected to cluster. As a result, we are cautioned to infer very much about what happens on the small scales based on the results of currently available high resolution axisymmetric simulations.  We expect better light on this will be shed upon the advent of high resolution 3D experiments or new simulation tools with wider dynamical scale resolution.

\subsection{On turbulent activity where Ri $>1/4$}\label{On_Ri}
The persistence of activity in disks where the Richardson number exceeds the classical critical value of $1/4$ remains enigmatic.  Here, we briefly review what is known about this feature of protoplanetary disk models and place our findings with regards to the SymI into that context.
\par
\citet{Gerbig_etal_2020} unequivocally demonstrated the emergence of sustained (and probably turbulent) 3D activity in midplane settled dust layers exhibiting Ri $>1/4$.  The antecedents to this can be found in the three studies of \citet{Gomez_Ostriker_2005}, \citet{Johansen_etal_2006}, and \citet{Barranco_2009} where,  in order to isolate and better understand operative physical effects, the stability and nonlinear response of a midplane-settled particle disk setting was examined within a restricted two-dimensional, azimuthal-vertical slice.  Initiated with vertical shears in the azimuthal velocity field, over time it can be seen that the particle layer develops sinusoidal undulations that eventually grow in amplitude and finally erupting to generate strong vertical mixing
\citep[see especially][]{Barranco_2009}.  
\citet{Johansen_etal_2006}, who examined this dynamic for the two-fluid model, calculate the effective Richardson number as a function of disk height for simulations after reaching their putative equilibrated statistical state, for which they find that Ri $\approx 1$ within the confines of the disk containing most of the dust.   \citet{Barranco_2009} similarly report effective minimum Ri that significantly exceed 1/4 for simulations in their well-developed stage.
\par
\citet{Gomez_Ostriker_2005} perform a corresponding stability analysis in a tractable one-fluid physical model and find that values of Ri as high as 5 could be linearly unstable.  Similarly, \citet{Barranco_2009} report the possibility of linear instability for values of Ri as high as 1.25.  For the nonlinear models presented in these three studies, the period of time from initiation of a model run until the manifestation of full-fledged nonlinear layer development ranges from 5 to 10 orbit times or, in our units, after $t\Omega = 30$-60.
Similar roll-up times are reported for KH-roll-up in \citet{Johansen_etal_2006}.
\par
\citet{Gomez_Ostriker_2005} and \citet{Barranco_2009} attribute to Coriolis effects the tendency for such layers to be unstable even when $\Ri > 1/4$.  Indeed, the classic KH-roll-up analysis involves the analysis of a two-dimenensional vertically sheared stratified fluid.  The Ri = $1/4$ criterion is technically only appropriate to that simple setup sans rotation.  The corresponding incompressible linear stability problem, in the guise of the Taylor-Goldstein equation \citep[e.g.,][]{Garaud_Lin_2004}, is technically second order in time.  Introduction of Coriolis effects -- i.e., in the way they appear in the disk problem considered here and the aforementioned studies -- raises the problem's temporal order by one, which certainly enriches the range of normal mode behavior.  As we have seen, the added complexity introduced by rotation is also reflected in the dynamics responsible for the SymI.  Indeed, the structure of the perturbation equations for the SymI -- i.e., Eqs. (\ref{u_prime_eqn}-\ref{Delta_prime_eqn}) --  are similar to the corresponding ones for KH-roll-up considered in the single-fluid setup of \citet{Barranco_2009}.  \par

In light of these reflections, one may define a characteristic Rossby number in terms of the amplitude and scale of the azimuthal velocity shear via

\beq
\Ro = \delta v_{00}\big/2\Omega H_s,
\label{Ro:def}
\eeq
e.g., like those based on averaged fits to $\left<v_g\right>_{xy}$ or $V_{{\rm cm}}$ discussed in Sec. \ref{sec:Richardson_Numbers} and Eq. (\ref{model_Vcm}) in particular.  Indeed, activity persists for Ri $>1/4$ for Ro $\gtrapprox 1$ -- especially throughout all phases of development in the St = 0.04 suite of simulations (see last column of Table \ref{tbl:Riphinotfits}).  
This should be considered in comparison to the Rossby number of Keplerian flow, $\Ro_{K} = 3/4$. Thus it seems that for problems of this sort the critical Richardson number should in general be a function of the Rossby number, i.e., $\Ri_{{\rm c}} = \Ri_{{\rm c}}(\Ro)$, in which 
the classical stratified limit is recovered when rotation goes away

\beq
 \Ri_{{\rm c}}\left(\Ro\rightarrow \infty \right) = 1/4.
\eeq
We think this is a worthwhile program 
 -- one that was instigated in \citet{Barranco_2009} -- 
for future clarification in the context of midplane-settled protoplanetary disk modeling.
\par
Nevertheless, we think the SymI dynamics unfolding in the St = 0.04 simulation dominates the rotationally modified KH-roll-up likely simultaneously present.  While we have not done a detailed KH-roll-up analysis to complement that done in \citet{Barranco_2009}, we can see from that study that the predicted growth rates for a configuration that most resembles our St = 0.04 simulations, with $H_p \approx 0.01 H_g$, predicts growth rates $\sim 0.1 \Omega$
\citep[see 3$^{\rm rd}$ column, 3$^{\rm rd}$ row of][]{Barranco_2009} which is a factor of 5 times slower than what is seen in our simulations.
\par 
We think the SymI is a part of the explanation for why such settled particle layers go turbulent in the simulations conducted both here and in \citet{Gerbig_etal_2020}. All of our simulations manifest clear layer transition by 1 orbital period and fully developed nonlinear activity by 2-2.5 orbital periods (i.e., $t\Omega = 12$-15), which is far shorter than the time it takes non-axisymmetric KH-roll-up to develop according to the above mentioned studies. 
Therefore, we conjecture that 3D axisymmetric dynamics like the SymI and radial KH-roll-up are the main drivers of activity, at least for the range of St numbers considered here.
A comprehensive study in this respect should start with a theoretical single-fluid model that parses the relative importance of KH-roll-up and the SymI in 3D-axisymmetry.  This then should be followed with understanding how these dynamics play out in the presence of two-stream dynamics including the SI.  In principle, the results of the single-fluid theory ought to be contained in the generalized VSSI framework, particularly within its single-fluid limit \citep{Lin_2021}.

\section{A Hard-boiled Summary with some final remarks}\label{sec:summary}
We provide summary bullet points of the major findings and conclusions of this concerning small midplane sections of protoplanetary disks not subject to an external source of turbulence:
\begin{enumerate}
\item  Midplane settled particle layers impart both upon itself (as a second fluid) and upon the gas-fluid a complex Ekman pattern of radial and azimuthal jets in the azimuthal-radial mean.  The emergent epicyclically oscillating jet profiles are a result of the momentum exchange between the gas and particles, and come about while the particles are settling toward the midplane. 
\item  The jet flows are subject both to classic KH roll-up as well as being unstable to the so-called Symmetric Instability. The SymI relies on a mismatch in density and azimuthal velocity isolines in a rotating atmosphere and is well known to be relevant to mixing-layer dynamics in the ocean and frontogenesis in the atmosphere, among other geophysical phenomena \citep[e.g.,][]{Hoskins_1974,Bennetts_Hoskins_1979,Thomas_etal_2013,Zeitlin_2018,Stamper_Taylor_2017,Zhou_etal_2022}.  
\item In the St = 0.2 simulations radial KH-roll-up appears to act in distinct layers that are at least 2-3 $H_p$ away from the midplane and with relatively \REV{short} growth rates $\sim \Omega$.  This axisymmetric dynamic appears to be present in the previously unpublished study of \citet{Ishitsu_etal_2009}, which has recently been examined in a new theoretical framework in \citet{Lin_2021}. These KH-roll-up dynamics play a primary role in driving instability of the midplane region for St = 0.2, but appears to operate together with SymI.  Nonetheless, we note its significance in that this classical shear instability acts on layers substantially removed from where the particles reside, contrary to previously held assumptions or expectations.
\item The axisymmetric SymI acts primarily within 1-2 $H_p$ of the midplane.  Its growth rate is relatively short at $\approx$1-2$\Omega$, it appears to be the other main driver of midplane turbulence in the St = 0.2 simulations, and it might be the primary --if not sole -- driver of instability in the St = 0.04 runs as the Ri numbers are near 1 in this latter case.  In the \St$=0.04$ case, we find that the SymI erupts into unsteady dynamics by $t\Omega = 6$, which is comparable to the KH-roll-up timescale in the St = 0.2 simulation, suggesting that both processes are active and act on similar timescales. Thus, for the numerical experiments conducted here  we posit that while the shape of the mean flows that develop in settling layers are St dependent, the following unstable dynamics taking root in them are that of a single perfectly-coupled fluid responding to a St-dependent flow profile.
\item We note that the SymI is distinct from the SI as the latter emerges as a result of momentum exchange between components drifting through one another, while the former only requires strong vertical gradients in both the perturbation azimuthal gas velocity ($v_g$) and mean density together with isoline misalignment between particle density and total azimuthal gas velocity ($V_K + v_g$).  In essence, the SymI is effectively a single-fluid baroclinic dynamic.
\item Furthermore, we have reason to suspect that the SymI we observe in these layers is of the fundamentally same kind of mechanical process that drives the VSI based on the connection of the latter with the ``sloping convection" effect well-known in geophysical fluid flows \citet{Yellin_Bergovoy_etal_2021}.

\item The analytical theory we developed indicates that a layer will experience strong widespread
instability in the particle layer due to the SymI once $\Ri < 1$.  The fastest growing mode's growth rate is well approximated by Eq.  (\ref{SymI_asymptotic_large_m_growth_rate}), 
$\sqrt{2/\Ri -2}\Omega$, where Ri is based on the azimuthal velocity's vertical shear.  The fastest growing wavelength depends on integer vertical mode number $m>0$ and is given in Eq. (\ref{kmax_SymI}).

\item We believe this finding goes toward explaining the
results reported in \citet{Gerbig_etal_2020} where they find that settled turbulent particle
layers have effective values of $\Ri$ to be well above $1/4$ but less than $1$, especially for 
simulations where $Z>0.01$ \citep[e.g., see Figure 6 of][]{Gerbig_etal_2020}.  We therefore suspect
that the SymI plays a role --and a prominent one depending on St -- in driving midplane turbulence in all numerical computational studies of the SI in otherwise laminar disk models.
\item Given the above findings, we therefore conclude that these particle settled midplane layers are subject to three simultaneously acting instabilities: the above mentioned two, and the SI.   Based on the simulations with the specific input parameters we have conducted here, we conjecture that the SI is the weakest of the three with the slowest growth rates \citep{Chen_Lin_2020,Umurhan_etal_2020}.  We conjecture that whenever the parameters for $\St$ and $Z$ \citep[or $Z/\Pi$ as proposed in][]{Sekiya_Onishi_2018} permit the SI with relatively fast growth rates, it does so out of a turbulent state driven by the other two shear instabilities.  

\item We have produced space-time diagrams of the azimuthally averaged particle surface density field and have examined the azimuthally averaged particle surface densities at late times.  As has been done before, in those simulations that admit the SI and allow for its nonlinear development and saturation (i.e., the $Z=0.01, \St = 0.2$ simulation), its signature is clearly visible by the emergence and maintenence of radially drifting coherent particle overdensities.  In the high resolution simulation of
 $Z=0.01, \St = 0.04$, the spacetime diagram indicates intermittent manifestation of overdensities -- that appear to form, drift and later dissipating only to reform and repeat this dynamic once again.  At this stage it is unclear whether this observed intermittency is a result of the SI struggling to emerge, or if it is some other collective turbulent effect that momentarily drives particle overdensities only to be later destroyed \citep[e.g.,][]{Yang_etal_2018}.  This requires further analysis.
 
  \item We have calibrated the PENCIL code used here by simulating 3D Kolmogorov turbulence in a non-rotating frame free of particles by forcing the simulation at approximately 1/3 the box size.
  We recover the expected $n_g = 5/3$ inertial range power law behavior (i.e., $\varepsilon_{k,g}\sim k^{-n_g}$) down to a wavenumber $k_{N,D}=k_N/4$, where $k_N$ is the simulation's Nyquist wavenumber.  At lengthscales smaller than $2\pi/k_{N,D}$ the gas kinetic energy plummets super exponentially.  It therefore means that down-scale propagating turbulent kinetic energy does not appreciably reach scales less than 6-8 grid points in these simulations.  All subsequent analyses, including statistical measures of underlying turbulence, are therefore here restricted to scales larger than $2\pi/k_{N,D}$.
 
 \item For the particle-gas simulations done here with box sizes $=0.2 H$ we conjecture that all medium sized $256^3$ simulations are not statistically converged.  We have confirmed this to be the case in for the St = 0.2 simulation as the nominal inertial range power-law slopes for both the particle and gas kinetic energies ($\varepsilon_{k,p}\sim k^{-n_p}$ and $\varepsilon_{k,g}\sim k^{-n_g}$, respectively) steepen for the corresponding high resolution $512^3$ run.  Unfortunately due to limitations of available resources we were not able to confirm this trend for a concomitant high resolution run in the St = 0.04 case.
 \item We have run a corresponding series of high ($512^2$ element) and super-high ($2048^2$ element) resolution 3D-axisymmetric simulations for which we find that the power-law slope of the gas kinetic energy appears to be converged at $512^2$:  For St = 0.2 we find $n_g \approx 2.1 \pm 0.05$; while for  St = 0.04, $n_g \approx 2.15 \pm 0.05$.  The inertial range particle kinetic energies appears converged for the St = 0.04 case with $n_p \approx 0.42 \pm 0.05$.  However, the corresponding inertial range power-law slope shows steepening for St = 0.2 with $n_p = 0.92 \pm 0.1$ for high resolution and $n_p = 1.1 \pm 0.05$ at super-high resolution.
 \item Based on this 3D-axisymmetric finding we cautiously conjecture that the inertial range power slope reported for the St = 0.2 high resolution full-3D simulation might be converged or close to convergence with $n_g = 1.37 \pm 0.03$ and $n_p = 0.64 \pm 0.04$.  In that case the inertial range appears to emerge at $k\approx 200 H_g^{-1}$ corresponding to about 1/6 the box size, which is roughly 3 particle scale heights viewed symmetrically from the midplane.  What these inertial range behaviors say about the nature of turbulent kinetic energy cascade remains to be understood.

\end{enumerate}

Much of the findings reported in this study owe their illumination to the earlier axisymmetric study of \citet{Ishitsu_etal_2009}.  It therefore strongly suggests that axisymmetric instabilities primarily operate in midplane settled particle layers in gaseous protoplanetary disks not subject to external sources of turbulence {\it and} where the SI is at best only weakly operative.  The recent results of \citet{Gerbig_etal_2020} appears to confirm that such layers are sufficiently active even though the SI is weakly operating.
If such weakly turbulent disk conditions are realizable in realistic protplanetary disk models, then it remains a challenge to understand how low St number conditions can lead to planetesimal formation. Indeed, global evolution modeling of particle growth in turbulent disks show that disks maintain low St particles for weak-to-moderate levels of external turbulence during the first million years after disk formation
\citep[e.g.,][]{Estrada_etal_2016, Sengupta_etal_2019}.  If such externally driven turbulence scenarios shut-off for some reason, then such low St number particles would settle to the midplane, then to be subjected to the particle-shear driven turbulent state (discussed thus far) that could conceivably further act to thwart planetesimal accumulation.
We are therefore swayed by
the concerns stated in the conclusions of
\citet{Ishitsu_etal_2009}, where they write 
``Thus, even though the global turbulence is weak in the dead zone, the turbulence due to the instability described in this paper may play the role of avoiding planetesimal formation and floating dust in the disk."\\
\section{ACKNOWLEDGEMENT}

We are grateful to P.E. Estrada, J.N. Cuzzi and K. Shariff of NASA-ARC for many illuminating discussions and institutional support.    D.S. was supported by the NASA Postdoctoral Program (NPP) fellowship, and partly by the NASA Astrobiology Institute. O.M.U. acknowledges the NASA Planetary Science Division ISFM for Planet Formation and Exoplanets Theory at NASA/Ames for financial and computational support, and also, the NASA ROSES TCAN grant  ``Dynamical instabilities in the aid of planet formation in circumstellar disks" (20-TCAN20-0011). We thank the anonymous referee for many insightful comments and suggestions which significantly improved the quality of this manuscript. We are also indebted to insights shared with us by our TCAN collaboration colleagues W. Lyra, J. Simon, C.C. Yang, and A. Youdin. All the simulations presented in this paper are performed on the NASA Advanced Supercomputing  (NAS) facility with generous computational resources provided through NPP and ISFM allocations.  

\appendix

\section{Reformulating particle fluid evolution equations in terms
of $\mu$}\label{energy_calculations}

Our aim is to rewrite the particle-fluid evolution equations
Eq.(\ref{eqn:dustcontinuity}-\ref{eqn:dustmomentum}) in terms
of the variable $\hat\mu\sub i \equiv \sqrt{\rho_p}U_{pi}$,
i.e., in terms of the full velocity field including the background Keplerian flow.  We note that within this section we use the usual Einstein index convention and, as such, the dummy indices $i$ and $j$ used here are not to be confused with the particle labeling indices $i$ or with the grid labeling indices $j$ used extensively in Section. \ref{setup}.
Leaving the RHS of Eq.(\ref{eqn:dustmomentum}) in terms of a forcing function $F\sub i$ we input
this definition into the total momentum conservation and manipulate the result accordingly.
The purpose of this procedure is to have the equations appear in terms of conjugate symmetric variables in Fourier space that
make assessing the energy contained in the particle component straightforward.  
We will take the result we develop in terms of $\sqrt{\rho_p}U_{pi}$  and then restore the definition $U_{pi} = V\sub {{\rm K}} \delta\sub{i2} + u_{pi}$ to recover the form we seek in terms of the perturbation velocities and $\mu\sub i \equiv \sqrt{\rho_p} u_{pi}$. Starting with

\beqa
\partial\sub t \rho_p U_{pi} + \partial_j \rho_p U_{pj} U_{pi} + \epsilon_{ijm}\rho_p 2\Omega_j 
U_{pm} =
F_{pi}
\eeqa

\noindent in which $\Omega_i = \delta\sub{i3}\Omega_0$ and where $\epsilon_{ijk}$ is the Levi-Civita symbol. We have after rewriting the above in terms of $\hat\mu\sub i$ that

\beqa
F_{pi} &=&
\rho_p^{1/2}\partial\sub t \hat\mu\sub i + \sfrac{1}{2}\rho_p^{-1/2}\hat\mu\sub i \partial_t \rho_p
+\partial_j \hat\mu\sub j \hat\mu\sub i + 
\epsilon_{ikm}\rho_p^{1/2}2\Omega_j \hat\mu\sub m.
\label{substep:appendix:C}
\eeqa

\noindent We similarly re-express the dust continuity equation as

\beq
\partial_t \rho_p + \hat\mu\sub j \partial_j \rho_p^{1/2} + \rho_p^{1/2}\partial_j \hat\mu\sub j = 0.
\label{rewritten:dust:continuity}
\eeq

\noindent Replacing the time derivative of $\rho_p$ in Eq. (\ref{substep:appendix:C}) by the above expression, followed by factoring out $\rho_p^{1/2}$ reveals

\beq
F_{pi}= \rho_p^{1/2}
\left[
\partial_t \hat\mu\sub i + \rho_p^{-1/2} \partial_j \hat\mu_j \hat\mu_i - \sfrac{1}{2} U_{pi} U_{pj}\partial_j \rho^{1/2} - \sfrac{1}{2} U_{pi} \partial_j\hat\mu\sub j
+
\epsilon_{ijm}2\Omega_j \hat\mu\sub m
\right].
\eeq

\noindent We may rewrite the third term on the RHS of the above equation through a number of derivative by parts maneuvers to find

\beqa
 - \sfrac{1}{2} U_{pi} U_{pj} \partial_j \rho_p^{1/2} &=& -\sfrac{1}{2}\partial\sub j \Big(U_{pi} \hat\mu\sub j\Big) + \sfrac{1}{2}\rho_p^{1/2}\left(U_{pi}\partial_j U_{pj}  + U_{pj}\partial_j U_{pi}\right)
 \nonumber \\
 &=& -\sfrac{1}{2}\Big(\hat\mu\sub j \partial_j U_{pi} + U_{pi} \partial_j \hat\mu\sub j\Big)
 +\sfrac{1}{2}\Big(\hat\mu\sub j \partial_j U_{pi} + \hat\mu\sub j \partial_j U_{pi}\Big)
 \nonumber \\
&=& -\sfrac{1}{2}U_{pi} \partial_j \hat\mu\sub j + \sfrac{1}{2} \hat\mu\sub i \partial_j U_{pj},
\eeqa

\noindent while

\beq
\rho_p^{-1/2} \partial_j \hat\mu_j \mu_i =  U_{pj}\partial_j\hat\mu\sub i + U_{pi}\partial_j \hat\mu\sub j.
\eeq

\noindent Putting these all together yields a reworked momentum evolution equation

\beq
\partial_t \hat\mu\sub i + U_{pj}\partial_j\hat\mu\sub i + \sfrac{1}{2}\hat\mu\sub i \partial_j  U_{pj}
+ \epsilon_{ijm}2\Omega_j \hat\mu\sub m = F\sub i\Big/\sqrt{\rho_p}.
\label{hatmu:momentum}
\eeq

\noindent Finally, making the replacements $\hat\mu\sub i \rightarrow \mu\sub i + \delta\sub{i2} \sqrt{\rho_p} V\sub{{\rm K}}$ and
$ U_{pi} \rightarrow  u_{pj} + \delta\sub{i2} V\sub{{\rm K}}$,
we have

\beq
\partial_t \mu\sub i + u_{pj}\partial_j\mu\sub i   + V\sub{{\rm K}}\partial_y\mu\sub i  
 +
\sfrac{1}{2}\mu\sub i \partial_j v\sub j
+\biggl(V\sub{{\rm K}}\partial\sub t \sqrt{\rho_p}
+\sfrac{1}{2}\sqrt{\rho_p} V\sub{{\rm K}} \partial_j u_{pj}
+u_{px}\partial_x \sqrt{\rho_p}V\sub{{\rm K}} 
\biggr) \delta\sub{i2}
+ \epsilon\sub{ijm}2\Omega_j \mu\sub m  - 2\Omega_0\sqrt{\rho_p} V\sub{{\rm K}} \delta\sub{i1}
= F\sub i\Big/\sqrt{\rho_p},
\nonumber
\eeq

\noindent which after making use of Eq. (\ref{rewritten:dust:continuity}) becomes finally

\beqa
& & \partial_t \mu\sub i + u_{pj}\partial_j\mu\sub i   + V\sub{{\rm K}}\partial_y\mu\sub i  
 +
\sfrac{1}{2}\mu\sub i \partial_j u_{pj} + \epsilon\sub{ijm}2\Omega_j \mu\sub m
 \nonumber \\
& & \hskip 2.0cm   
+\biggl(u_{px}\partial_x \sqrt{\rho_p}V\sub{{\rm K}} -V\sub{{\rm K}}u_{pj}\partial_j \sqrt{\rho_p} -V\sub{{\rm K}}^2\partial_y \sqrt{\rho_p} 
\biggr) \delta\sub{i2}
- 2\Omega_0\sqrt{\rho_p} V\sub{{\rm K}} \delta\sub{i1}
= F\sub i\Big/\sqrt{\rho_p}.
\label{mu:momentum}
\eeqa

 Energy statistics on the particle fluid may be constructed based on either of the two formulations found in Eq. (\ref{hatmu:momentum}) or Eq. (\ref{mu:momentum}) depending upon the context of interest; whether it be the total energies (the former) or the perturbation energies (the latter). In particular, these expressions will be Fourier transformed (e.g., the transform of $\mu_i({\bf x})$ = $\tilde\mu\sub{k,i}({\bf k})$) and partial energies and their scale-to-scale transfer can be assessed within the usual prescribed narrow wavenumber bins.

\section{Time dependent settling solutions}\label{particle_settling_solutions}
We consider solutions to the horizontally uniform steady settling solutions of the system. We assume that
the gas density is constant and write it here as $\overline\rho_g$, and that the gas field has no vertical velocity component.  We use
capital letters to designate these solutions, e.g., for the horizontal gas velocity we have
$u_g = U_g(z,t)$, and so forth for the other variables.  
Since by assumption $W_g = 0$, the evolution equations for the gas are

\beqa
\partial_t U_g + 2\Omega_0 V_g &=& -\Pi_r - \frac{\Omega_0 \varepsilon}{\St}\Big(U_g - U_p\Big), 
\label{Ug_eqn}
\\
\partial_t V_g  - \sfrac{1}{2}\Omega_0 U_g &=& - \frac{\Omega_0 \varepsilon}{\St}\Big(V_g - V_p\Big),
\label{Vg_eqn}
\eeqa

where the particle to gas ratio, written here as $\varepsilon = \rho_p(z,t)/\overline{\rho_g}$, differs from the its general definition in the text (i.e., $\epsilon$), since $\rho_g$ is also a variable.  
In this way $\varepsilon$ is a proxy for the particle density $\rho_p$.  For the particle component, the particle density continuity equation for the particle density is

\beq
\partial_t{\varepsilon} + \partial_z\left(W_p \varepsilon\right) = 0,
\label{settling_varepsilon_eqn}
\eeq

while for the particle momentum equations we have

\beqa
\partial_t U_p + W_p\partial_z U_p + 2\Omega_0 V_p &=& -\Pi_r - \frac{\Omega_0}{\St}\Big(U_p - U_g\Big), 
\label{Up_eqn}
\\
\partial_t V_p  + W_p\partial_z V_p - \sfrac{1}{2}\Omega_0 U_p &=& - \frac{\Omega_0}{\St}\Big(V_p - V_g\Big), 
\label{Vp_eqn}
\\
\partial_t W_p + W_p\partial_z W_p  &=&  - \frac{\Omega_0}{\St}W_p - \Omega_0^2 z.
\label{settling_Wp_eqn}
\eeqa

We begin by analyzing Eq. (\ref{settling_Wp_eqn}) by noting that $W_p = -\beta \Omega_0 z$ is a solution in which
$\beta$ is the solution of the time evolution equation.

\beq
\partial_t\beta = \Omega_0\left(\beta^2 - \frac{\beta}{\St} + 1 \right).
\eeq

Solutions to this equation depend upon whether or not $\St \le 1/2$ \citet{Lin_2021}.  We define

\beq
\displaystyle
\beta_\pm = \frac{1 \pm \delta}{2 \St}; \quad \delta \equiv 
\sqrt{1 - 4 \St^2}.
\label{beta_solutions}
\eeq

When $\St \le 1/2$ the time-asymptotic stable settling solution is
given by $\beta = \beta_-$ \citet{Lin_2021}
\footnote{For $\St \le 1/2$ the solution $\beta = \beta_+$ corresponds
to exponentially growing perturbations.}.
When $\St > 1/2$ the solution is given by

\beq
\beta = \frac{1}{2\St}\left[ 1 + |\delta| \tan \left(\frac{|\delta| \Omega_0 t}{2 \St}\right)
\right], \qquad \left|\frac{|\delta| \Omega_0 t}{2 \St}\right| \le \frac{\pi}{2},
\eeq
while noted here, we will not consider these solutions any further in this study.
With the purely linear dependence on $z$ for $W_p$, the solution to Eq. (\ref{settling_varepsilon_eqn}) has a Gaussian form

\beq
\varepsilon = \varepsilon_0(t)\exp\left(-\frac{z^2}{2H_p^2}\right),
\label{varepsilon_solution}
\eeq
in which

\beq
\frac{d \ln \varepsilon_0}{dt} = - \frac{d \ln H_p}{dt} = \beta.
\label{varepsilon_parameter_time_dependencies}
\eeq
The proof of this solution follows by inserting the
solution form Eq. (\ref{varepsilon_solution}) together with $W_p = -\beta \Omega_0 z$ into Eq. (\ref{settling_varepsilon_eqn}), collecting the resulting expression into like powers of $z$ followed by setting their coefficients to zero, which produces the time dependencies 
on $\varepsilon_0$ and $H_p$ found in Eq. (\ref{varepsilon_parameter_time_dependencies}).  We note that the total vertically integrated mass is conserved as the product 
$\displaystyle \int_{-\infty}^\infty \rho_d(z,t) dz = \sqrt{2\pi} \rho_g \varepsilon_0 \cdot H_p$ is always time-independent constant for $\rho_g$ constant.  Because we will restrict our attention to problems with $\St < 1/2$, in all of our subsequent analyses we will assume that $\beta$ achieves its time-asymptotic value $\beta_-$, in which case

\beq
\varepsilon_0 = \varepsilon\sub{00}e^{\displaystyle \beta \Omega_0 t}, \qquad
H_p=H_{p0} e^{\displaystyle -\beta \Omega_0 t}.
\eeq

The constant coefficients $\varepsilon\sub{00}$ and $H_{p0}$ relate to one another on the assumption that the vertically integrated particle density $\Sigma_p = \sqrt{2\pi} \varepsilon\sub{00}H_{p0} \rho_g(0)$ is constant, in which case

\beq
\varepsilon\sub{00} = Z \left(\frac{H_g}{H_{p0}}\right), \qquad
Z \equiv \frac{\Sigma_p}{\Sigma_g},
\label{varepsilon_00_solution}
\eeq
where $Z$ is the local model disk metallicity (see text).  Thus,
solutions to the remaining equation set are 
defined by the parameters $Z,H_{p0}$ and $\St$.
\par\medskip
The solutions for the remaining flow variables, $U_g,U_p,V_g,V_p$, may
be determined from solving 
equations (\ref{Ug_eqn}-\ref{Vg_eqn})
and (\ref{Up_eqn}-\ref{Vp_eqn})
after inserting for $\epsilon$ the solution for $\varepsilon$ found in Eq. (\ref{varepsilon_solution}), as well as
replacing $W_p$ with $-\beta \Omega z$ where $\beta=\beta_-$,
(Eq. \ref{beta_solutions}).  Since the problem involves the solution of a settling layer, it proves beneficial to go into a spatial coordinate frame that follows this descent as the particle scale height $H_p(t)$ shrinks over time.  For example

\beq
U_g(z,t) \rightarrow U_g(\varphi,t); \qquad 
\displaystyle \varphi = \exp\left(-\frac{z^2}{2H_p^2}\right),
\eeq
and similarly for $U_p,V_g,V_p$.  $\varphi$ is now an independent variable that varies between $0$ and $1$.  In these new coordinates, partial derivatives are replaced according to their coordinate transformed forms

\beqa
& & z\partial_z \longrightarrow z\left(\frac{\partial \varphi}{\partial z}\right) \partial_\varphi= -2 \varphi \ln{\varphi} \partial_\varphi, \nonumber \\
& & \partial_t \longrightarrow 
\partial_t + \left(\frac{\partial \varphi}{\partial t}\right)\partial_\varphi = 
\partial_t - 2 (\beta \varphi \ln{\varphi}) \partial_\varphi.
\eeqa
In these new coordinates Eqs. (\ref{Ug_eqn}-\ref{Vg_eqn},
\ref{Up_eqn}-\ref{Vp_eqn}) are now reexpressed as

\beqa
\partial_t U_g - W_\varphi \partial_\varphi U_g
-2\Omega_0 V_g &=& -\Pi_r - \frac{\Omega_0}{\St}\varepsilon_{0}\varphi\Big(U_g-U_p\Big), 
\label{Ug_mu_eqn} \\
\partial_t V_g - W_\varphi \partial_\varphi V_g
+(1/2)\Omega_0 U_g &=&  \frac{\Omega_0\varepsilon_{0}}{\St}\varphi\Big(V_g-V_p\Big), 
\label{Vg_mu_eqn} \\
\partial_t U_p 
-2\Omega_0 V_p &=&  \frac{\Omega_0}{\St}\Big(U_p-U_g\Big), 
\label{Up_mu_eqn} \\
\partial_t V_p 
+(1/2)\Omega_0 U_p &=&  \frac{\Omega_0}{\St}\Big(V_p-V_g\Big), 
\label{Vp_mu_eqn}
\eeqa
where the speed $W_\varphi \equiv -2\beta \varphi \ln \varphi$ is
always greater than zero.  Note how in this transformed
coordinate system the vertical advection by $W_p$ of the horizontal particle velocities $U_p,V_p$ is now transformed into
an effective upward advection of the horizontal gas velocities, with no more vertical advection in the particle component.  This makes sense because we have moved into a reference frame that follows the evolving particle scale height.  
The solution to the above set of equations are sought subject to the condition that $\partial_z U_g\big|_{z=0} = \partial_z V_g\big|_{z=0} = 0$.  In all cases we have determined, these gas velocity conditions automatically impose the same conditions at the midplane  for $U_p,V_p$ (i.e., at $\varphi = 1$).  
We use the convention described in \cite{Umurhan_etal_2020} and write $\Pi_r = -2\delta c_s \Omega_0$, where $\delta$ is the local disk opening angle.
In the limit as $z\rightarrow \infty$ we make sure the solution behave according to the Nakagawa solutions in the limit $\epsilon \rightarrow 0$ \citep{Nakagawa_etal_1986, Youdin_Goodman_2005, Umurhan_etal_2020,Lin_2021}, i.e.,

\beq
U_g(\varphi\rightarrow 0) = 0, \quad
V_g(\varphi\rightarrow 0) = -\delta c_s , \quad
V_p(\varphi\rightarrow 0) = -\frac{\delta c_s}{1+\St^2}, \quad
U_p(\varphi\rightarrow 0) = -\frac{2 \St \delta c_s}{1+\St^2}, \label{Asymptotic_Nakagawa_Solutions}
\eeq
Evidently, all velocity variables are scaled by $\delta c_s$.  Henceforth we assume $\delta = 0.05$, noting that this choice has little influence on the qualitative solutions we demonstrate forthwith.
\par
We note that a quick eigenvalue analysis of Eq. (\ref{Up_mu_eqn}-\ref{Vp_mu_eqn}) -- setting the forcing terms  $U_g,V_p \rightarrow 0$ shows
that the eigenvalues, $\sigma$, are decaying epicyclic oscillations: $\sigma = \pm i\Omega_0 - \Omega/\St$.  Thus, drag forcing by the gas induces temporally decaying epicyclic motions in the particle fluid.   The memory of the forcing decays away on a time scale $\St/\Omega_0$.
A cursory inspection of  Eq. (\ref{Ug_mu_eqn}-\ref{Vg_mu_eqn}) indicates that  qualitatively similar response occurs in the gas fluid due to the drag forcing by the particles, except that
the decay timescales depend upon the local value of $\varepsilon$, which is a function of height.
\par
{\emph {Numerical Method.}} We use a third order correct upwind differencing scheme to stably calculate the advection terms in 
(\ref{Ug_mu_eqn}-\ref{Vg_mu_eqn}).  
Given the sign of the derivative terms, this amounts to a forward differencing scheme on a grid $0\le \varphi_i\le 1$.  To better capture the decaying behavior of the solutions as
$\varphi \rightarrow 0$, we further move into a stretched coordinate system in which 

\beq
\varphi_i = \exp\left(1 - \frac{1}{\zeta_i} \right), \label{stretched_coordinate_zeta}
\eeq
where $\zeta_i$ are points on a uniform grid between 0 and 1, endpoints included.  We also note that

\beq
\frac{\partial \zeta}{\partial \varphi} = 
\frac{1}{\varphi}\left(\frac{1}{1-\log\varphi}\right)^2,
\label{stretched_coordinate_zeta_chain_rule}
\eeq
which is used in the actual differencing scheme described further below.
\par
We discretize in time with time step $\Delta t$, where the time at the $n^{\rm th}$ time step is given by $t_n = n\Delta t$.  All dependent variables at time step $n$ are denoted with superscripts ``$n$''.  Spatial values of the dependent variables are designated by the subscript ``$i$", where $i$ runs from $1$ to $N$, the total number of grid points in the domain. We implement an exponential time integrator common in varied applied mathematical studies \citep{Cox_Matthews_2002} including some astrophysical applications \citep[e.g.,][]{Umurhan_Regev_2004,Umurhan_etal_2007}.  Thus for the particle component the evolution follows

\beqa
U_{p,i}^{n+1} &=& \exp\left(-\frac{\Delta t \Omega_0}{\St}\right)\cdot\Big[U_{p,i}^n \cdot \cos\left(\Delta t \Omega_0\right) + 2 V_{p,i}^n\sin\left(\Delta t \Omega_0\right) \Big]
+ \nonumber \\
& & \ \ \ \ \ \ \ \ \frac{\Delta t \Omega_0}{\St}\cdot\exp\left(-\frac{\Delta t \Omega_0}{2\St}\right)\cdot\left[U_{g,i}^n \cdot \cos\left(\frac{\Delta t \Omega_0}{2}\right) + 
2 V_{g,i}^n\sin\left(\frac{\Delta t \Omega_0}{2}\right)\right],
\\
V_{p,i}^{n+1} &=& \exp\left(-\frac{\Delta t \Omega_0}{\St}\right)\cdot\Big[-\frac{1}{2}U_{p,i}^n \cdot \sin\left(\Delta t \Omega_0\right) + V_{p,i}^n\cos\left(\Delta t \Omega_0\right) \Big]
+ \nonumber \\
& & \ \ \ \ \ \ \frac{\Delta t \Omega_0}{\St}\cdot\exp\left(-\frac{\Delta t \Omega_0}{2\St}\right)\cdot\left[-\frac{1}{2}U_{g,i}^n \cdot \sin\left(\frac{\Delta t \Omega_0}{2}\right) 
+ V_{g,i}^n\cos\left(\frac{\Delta t \Omega_0}{2}\right)\right].
\eeqa
The evolution of gas velocity quantities follows

\beqa
U_{g,i}^{n+1} &=& \exp\left(-\frac{\Delta t \Omega_0}{\St}\varepsilon(t_n)\varphi_i\right)\cdot U^n_{g,i} + \Delta t \exp\left(-\frac{\Delta t \Omega_0}{2\St}\varepsilon(t_n)\varphi_i\right)\cdot F^n_{g,i}, \\
V_{g,i}^{n+1} &=& \exp\left(-\frac{\Delta t \Omega_0}{\St}\varepsilon(t_n)\varphi_i\right)\cdot V^n_{g,i} + \Delta t \exp\left(-\frac{\Delta t \Omega_0}{2\St}\varepsilon(t_n)\varphi_i\right)\cdot G^n_{g,i},
\eeqa
where

\beqa
F^n_{g,i} &=& W_{\varphi,i}\Big(D_\varphi U^n_g\Big)_i + 2\Omega_0 V^n_{g,i} - \Pi_r + \frac{\Omega_0}{\St} \varepsilon\sub{0}(t_n)\varphi_i U^n_{p,i} \\
G^n_{g,i} &=& W_{\varphi,i}\Big(D_\varphi V^n_g\Big)_i - (1/2)\Omega_0 V^n_{g,i}  + \frac{\Omega_0}{\St} \varepsilon\sub{0}(t_n)\varphi_i V^n_{p,i}.
\eeqa
The derivative stencil is given by a third order forward differencing scheme in $\varphi$.  Because we go into the stretched coordinates defined
in equations (\ref{stretched_coordinate_zeta}-\ref{stretched_coordinate_zeta_chain_rule}), this derivative operation is written out as

\beq
\Big(D_\varphi U^n_g\Big)_i \longrightarrow \left(\frac{\partial \zeta}{\partial \varphi} \right)_i \left(-\frac{11}{6} U^n_{g,i} + 
3 U^n_{g,i+1} - \frac{3}{2} U^n_{g,i+2} + \frac{1}{3} U^n_{g,i+3}\right),
\eeq
and similarly for $\displaystyle \Big(D_\varphi V^n_g\Big)_i$.  
All numerical solutions for $U_g,V_g,U_p,V_p$ are initiated with the spatially uniform
solutions found in Eq. (\ref{Asymptotic_Nakagawa_Solutions}).  Also, all
solutions displayed in the text were derived on a uniform grid $0\le \zeta_i\le 1$ either $N=512$ or $N=1024$ points (the latter usually to check for convergence) and all time steps were in the range by $0.01 \St < \Delta t < 0.025 \St$.  
And finally, $\displaystyle \varepsilon\sub 0(t_n) = \varepsilon\sub{00} \exp\big(\beta t_n\big)$. 
\par
\begin{figure*}
\begin{center}
\leavevmode
\includegraphics[width=18cm]{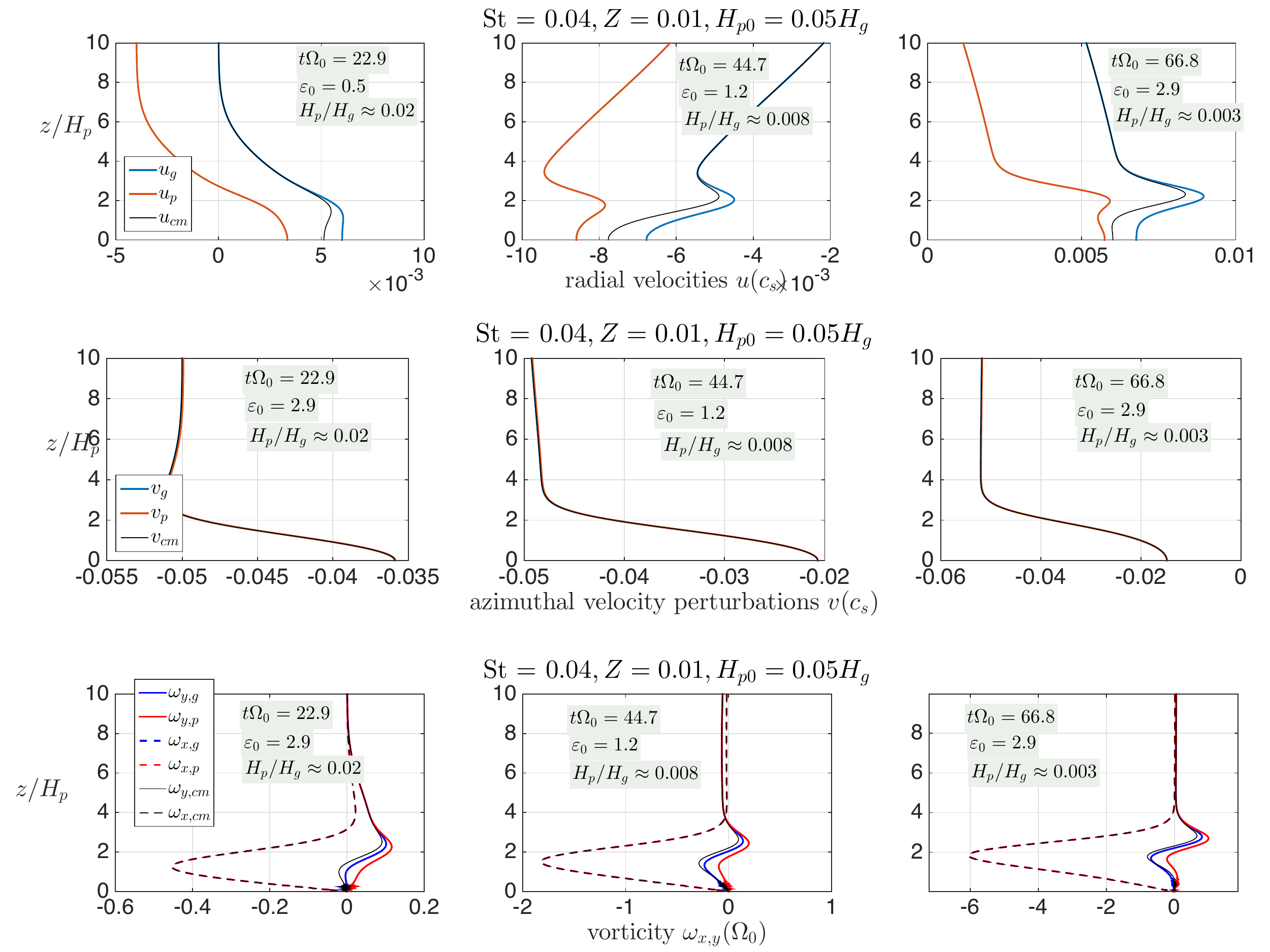}
\par
\end{center}
\caption{Like Fig. \ref{St_0p2_sols}; settling solutions for St = 0.04, $Z = 0.01$, and initial particle scale height $H_{p0} = 0.05 H_g$, at three selected times.}.
\label{St_0p04_sols}
\end{figure*}

Figs. \ref{St_0p04_sols} and \ref{St_0p2_sols} display a sample set of generated solutions of the horizontal velocities and associated vorticities. A detailed discussion of these solutions may be found in the main body of the text.
Note, by the Howard Semicircle Theorem, the magnitude of the azimuthal vorticity denote upper bounds of the growth rates of shear induced instabilities when, of course, the Richardson criterion is met.

\section{A motivated single fluid model problem for the disk-analog of the symmetric instability}\label{disk_symmetric_calculations}

We consider the linear normal mode response of a simplified single fluid ``toy" problem that exemplifies the essential processes we believe to be responsible for the layer overturn during the bounce-phase of the simulations.  As observed in the simulations, settling dust imparts momentum upon the gas with a vertical dependence.  The resulting collective motion of the gas and dust shows a net vertical variation in the azimuthal flow in the center of mass frame, like depicted in Fig. \ref{Ekman_Plots_From_Sims} as well shown in Fig. \ref{St_0p2_sols} and Fig. \ref{St_0p04_sols}.  In the terminal velocity and zero stopping time limit of the recast two-fluid equations, the momentum exchange forcing will manifest as an effective non-zero vertically varying external force acting in both the radial and azimuthal directions.  In the following examination we consider only the radial component of this force, $f_r(z)$, as present and we say that it is not dynamically active.  This immediately leads to a new effective steady state azimuthal velocity $\tilde v_0$ that adjusts to the force, i.e.,
\beq
-2\Omega_0 \tilde v_0(z) = f_r(z),
\eeq
\par\noindent
the analysis considered henceforth assumes single-fluid perturbations atop this basic state.
\par
We view the instability as being primarily axisymmetric, driven by the shear in that part
of the mean azimuthal flow that departs from the basic Keplerian ($=-(3/2)\Omega_0 x$), which is driven by the background pressure
gradient and the particle stream, defined in the text as $\tilde v_0(z)$.  We treat buoyancy effects driven
by the particle layer in the Boussinesq approximation, namely that density fluctuations -- whether they are driven by advective motions or are a result of weak compressibility dependencies on temperature -- are dynamically significant only
when coupled to the external gravitational field \citep{Spiegel_Veronis_1960}.  The background density field
of the model single fluid is represented by $\rho_0(z)$.  It is also assumed that the gas
fluid is incompressible.  Thus the simple model for the perturbation quantities $u',v',w',\Delta'$

\beqa
\partial_t u' -2\Omega_0 v' &=& -(1/\rho_{00})\partial_x p', \label{u_prime_eqn} \\
\partial_t v' +(1/2)\Omega_0 u' +  w'\partial_z \tilde v_0&=& 0, \label{v_prime_eqn} \\
\partial_t w'  &=& -(1/\rho_{00})\partial_z p' - g(z) \Delta', \label{w_prime_eqn} \\
\partial_t \Delta' + w' \rho_0^{-1}\partial_z \rho_0 &=& 0. \label{Delta_prime_eqn}
\eeqa
\noindent where $\rho_{00}$ is taken as a reference constant scale measure of the density.  These equations are supplemented with the statement of incompressiblity,

\beq
\partial_x u' + \partial_z w' = 0. \label{incompressibility}
\eeq
\noindent The definition found in Eq. (\ref{incompressibility}) motivates the definition of a stream function ($\psi$) and vorticity ($\omega'$) where

\beq
u' = \partial_z \psi', \quad w' = -\partial_x \psi', \qquad
\omega' \equiv \partial_z u' - \partial_x w' = \Big(\partial_x^2 + \partial_z^2\Big)\psi'.
\eeq
\noindent The vertical component of gravity is taken to be

\beq
g(z) = -\frac{\varepsilon_0}{1+\varepsilon_0} \Omega_0^2 z
\eeq
\noindent where we have included the factor $\varepsilon_0/(1+\varepsilon_0)$, where $\varepsilon_0$ is the midplane value of $\epsilon$, to represent an effective reduced gravity in a single fluid formulation of the gas-particle fluid
 \citep[e.g.][]{Chiang_2008}.  We note that as per our observations in the simulations, the mean gas density varies little and its dynamics are largely incompressible, especially on the scales on which the particles cluster about the midplane.  We therefore represent the mean density layer profile and its gradient via
 
\beq
\rho_0^{-1}\partial_z \rho_0 = \partial_z \ln \rho_0 = -\frac{z}{H_p^2}.
\label{particle_density_representation}
\eeq
\noindent The set of Eqs. (\ref{u_prime_eqn}-\ref{particle_density_representation}) may be viewed as the linearized perturbations of the dual terminal velocity (i.e., ``strong-drag/small-grain") and isothermal limit
of the one-fluid equivalent representation of a two-fluid system, as motivated by both \citet{Laibe_Price_2014} and \citet{Lin_Youdin_2017}.  We note that this limit contains no two-stream effects.\par

We rewrite all perturbation quantities as normal modes, e.g., $\psi' \rightarrow \psi\sub k(z) \exp\left[{ikx - i\omega_0 t}\right]$, and similarly for all the other quantities.
We derive an equation for the perturbation vorticity by operating on Eq. (\ref{u_prime_eqn}) by $\partial_z$ and subtracting
from it the result of operating on Eq. (\ref{w_prime_eqn}) by $-\partial_x$.  After inserting the assumed normal mode
form into the result, as well as into Eq. (\ref{v_prime_eqn}) and Eq. (\ref{Delta_prime_eqn}) we find the following three equations

\beqa
-i\omega_0 \Big(\partial_z^2 - k^2\Big)\psi\sub k - 2\Omega_0 \partial_z v\sub k &=& ik g(z) \Delta\sub k, \\
-i\omega_0 v\sub k -ik \psi\sub k\partial_z \tilde v_0 + (\Omega_0/2)\partial_z \psi\sub k &=& 0, \\
-i\omega_0 \Delta\sub k - ik \psi\sub k \cdot \partial_z \ln\rho_0 &=& 0.
\eeqa
\noindent The above three equations may be further combined into a single equation for $\psi_k$:

\beq
{\cal L}\psi\sub k = 
\Big(\Omega_0^2-\omega_0^2\Big)\partial_z^2 \psi\sub k  +
2\Omega_0 ik \partial_z\Big(\psi\sub k \partial_z \tilde v\sub 0\Big)
+ \omega_0^2 k^2 \psi\sub k
+ k^2 \psi\sub k\Big(g\partial_z \ln\rho_0\Big) = 0.
\label{main_perturbation_eqn}
\eeq
\noindent Our aim is to assess the growth rate, Im$(\omega_0)$, and frequency response, Re$(\omega_0)$, to perturbuations to Eq.
(\ref{main_perturbation_eqn}) subject to the condition that all perturbation quantities (i.e., $\psi\sub k, \partial_z\psi\sub k$) decay as $z\rightarrow \pm\infty$.
To facilitate analytical treatment, we consider a model for the background functions that permit us to seek solutions
of Eq. (\ref{main_perturbation_eqn}) in terms of parabolic cylinder functions ${\cal D}_m(\zeta)$.  We therefore adopt the following

\beq
g\partial_z \ln\rho_0 = -\frac{\varepsilon_0\Omega_0^2}{1+\varepsilon_0}\frac{z^2}{H_p^2},
\qquad \tilde v\sub 0 =  v_{{\rm cm},\infty} + \delta v\sub{00}\left(1-\frac{1}{2}\frac{z^2}{H_s^2}\right),
\label{model_tractable_flow}
\eeq
\noindent where $v_{{\rm cm},\infty}$ is the asymptotic constant value far from the particle layer and $H_s$ is the scale factor associated with the vertical variation of the azimuthal perturbation flow field as uncovered and discussed in the text.  Typically it is slightly larger than the particle scale height $H_p$, but for our purposes here it is immaterial to the final outcome below.  It is important that the variation in $v$ be parabolic for what follows.
Thus,
Eq. (\ref{main_perturbation_eqn}) may be rewritten in terms of a standard parabolic cylinder equation form
if we introduce

\beq
\psi_k = e^{-az^2/4} \Psi_k; \qquad
a = -\frac{2ik \delta v\sub{00}\Omega_0}{H_s^2(\omega_0^2-\Omega_0^2)}.
\eeq
\noindent and the vertically stretched coordinate $z=\tilde\beta\zeta$, where the scale $\tilde\beta$, satisfies

\beq
\frac{\tilde\beta^4\gamma^2}{\Omega_0^2-\omega_0^2}\left(\Ri
-\frac{\Omega_0^2}{\Omega_0^2-\omega_0^2}
\right) = \frac{1}{4}; \qquad
\gamma^2(k) \equiv \frac{\delta v\sub{00}^2 k^2}{H_s^4}.
\label{beta_definition}
\eeq
\noindent We note that $\tilde\beta$ may be complex and that we have re-expressed constants in terms of  the azimuthal form of the Richardson number, $\Riphi$ defined in Eq. (\ref{Riphi}), i.e.,
 
\beq
\Ri \longleftrightarrow \Riphi = \displaystyle{
-\frac{\displaystyle{g\partial_z \ln \rho_0}}{\displaystyle{\big(\partial_z \tilde v_0}\big)^2} =
\frac{\displaystyle{\frac{\varepsilon_0\Omega_0^2}{1+\varepsilon_0}\frac{z^2}{H_p^2}}}{\displaystyle{\frac{\delta v\sub{00}^2z^2}{H_s^4}}}
}
= \frac{\varepsilon_0}{1+\varepsilon_0}\frac{H_s^4}{H_p^2}\frac{\Omega_0^2}{\delta v\sub{00}^2}.
\eeq
Eq. (\ref{main_perturbation_eqn}) now simplifies into the canonical form in terms
of the non-dimensional independent variable $\zeta$

\beq
\partial_\zeta^2 \Psi_k + \left[
\frac{\omega_0^2 k^2 - i\Omega_0\gamma}{\Omega_0^2-\omega_0^2}\tilde\beta^2
-\frac{1}{4}\zeta^2
\right]\Psi\sub k = 0.
\eeq
The solution to the above are the standard Parabolic Cylinder functions ${\cal D}_m(z/\tilde\beta)$ 
\citep{Abramowitz_Stegun_1972}, provided
certain conditions ensuring that either
$\psi\sub k$ rapidly decays  as $z \rightarrow \pm\infty$ or the quantization condition is satisfied: (i) first,
the quantization condition is

\beq
\displaystyle
\frac{\omega_0^2 k^2 - i\Omega_0\gamma}{ \Omega_0^2-\omega_0^2}\tilde\beta^2 = m+\frac{1}{2},
\label{condition1}
\eeq
which must be satisfied for non-negative integer values of $m$; (ii)
the asymptotic behavior of $\psi_k$ for $|z|$ large is given by

\beq
\psi_k\Big(|z|\rightarrow \infty\Big) \sim \exp\left[-\left(a + \frac{1}{\tilde\beta^2}\right)\frac{z^2}{4}\right],
\label{condition2}
\eeq
which means that we require that

\beq
{\rm Re}\left(a + \frac{1}{\tilde\beta^2}\right) > 0.
\label{condition2actual}
\eeq
The quantization condition Eq. (\ref{condition1}) together with Eq. (\ref{beta_definition}) straightforwardly leads to solutions for $\omega_0^2$ given by

\beqa
\frac{\omega_0^2}{\Omega_0^2} &=& 
\sqrt{
\mu\left(2\frac{{\rm Ri}-1}{{\rm Ri}} + \mu - 2 i \frac{\sqrt{2\mu/\nu^2}}{\sqrt{{\rm Ri}}} \right)
} - \mu +  i \frac{\sqrt{2\mu/\nu^2}}{\sqrt{{\rm Ri}}} ,
\label{omega0_sol}
\eeqa
in which

\beq
\nu \equiv 2m + 1, \qquad
\mu\equiv \frac{1}{2} \frac{\varepsilon_0}{1+\varepsilon_0} \frac{\nu^2}{k^2 H_p^2}.
\label{mu_definition}
\eeq
We have deliberately rewritten the above in order to express all vertical velocity gradient information in terms of Ri.
\par
The above form for $\omega_0^2$ satisfies the condition that solutions exhibit Gaussian
decay as $|z| \rightarrow \infty$. (The algebraic procedure also admits solutions for $\omega_0^2$ where the sign of the first (squareroot) term on the RHS of Eq. (\ref{omega0_sol}) is negative.  However, we find that these solutions violate the large $|z|$ asymptotic decay criterion in both the Ri $\rightarrow \infty$ and  Ri $\rightarrow 0$ limits.  There still remains the possibility a second branch of solutions associated with this possible root might be viable in some intermediate Ri limit, but this is not yet determined at this stage.)
The solutions are characterized by three parameters, the vertical node $m$ (through $\nu$), the parameter $\mu$ which contains information about the stratification as well as the horizontal wavenumber, and Ri containing the relative measure of shear to stratification.

We note several features.  Most importantly we see that $\omega_0 = 0$ solutions exist for specific values of $\Ri$.  Setting $\omega_0$ to zero in both Eqs. (\ref{beta_definition}) and (\ref{condition1}) reveals this is possible if both of the following relationships are simultaneously satisfied

\beq
\tilde\beta^4\gamma^2\big(\Ri -1\big) = \frac{1}{4}, \qquad
i\tilde\beta^2\gamma = -\frac{\nu}{2}.
\eeq
Upon eliminating $\tilde\beta$ in both expressions shows that this is possible only when
$\Ri = \Ri_c$ where,

\beq
\Ri_c \equiv 1 - \frac{1}{\nu^2}.
\eeq
This condition signifies a strong change in the character of these solutions.
Furthermore, a series expansion of Eq. (\ref{omega0_sol}) for $0<\mu \ll 1$ shows

\beq
\frac{\omega_0^2}{\Omega_0^2} \approx 
i\sqrt{\frac{2\mu}{\Ri}}\left[ 
 \frac{1}{\nu} - \sqrt{1-\Ri}
\right] + \order{\mu},
\label{small_mu_expansion}
\eeq
which is written by taking into account the branch cut at $\Ri = 1$ in the
first squareroot term on the RHS of Eq. (\ref{omega0_sol}).  The sign of the square bracketed term changes sign when $\Ri$ crosses $\Ri_c$.
In the other limit where $\mu \rightarrow \infty$ we find

\beq
\frac{\omega_0^2}{\Omega_0^2} \approx 
\left(1 - \frac{\Ri_c}{\Ri} \right)
\left(1 + i\frac{1}{\nu}\sqrt{\frac{2}{\mu\Ri}}\right) + \order{\frac{1}{\mu}}.
\label{big_mu_expansion}
\eeq
Once again, what stands out is the gross change in character when $\Ri$ passes $\Ri_c$.  
The large $\mu$ limit (high $m$ or small horizontal wavenumbers $k$) shows
that the leading order behavior of $\omega_0$ is that of weakly growing oscillations with 
growth rates $\order{\mu^{-1/2}}$
when $\Ri > \Ri_c$ to very strong $\order 1$ widespread growth when 
$\Ri < \Ri_c$.  While $\Ri = 1$ designates an important boundary indicating
a stark transition in character, we note
that all normal modes show growth for all non-infinite values of $\Ri$.
\par
The limiting form for when the shear is nearly absent recovers at leading order the known oscillatory behavior
of disk intertia-gravity oscillations. 
A series expansion of Eq. (\ref{omega0_sol}) in inverse powers of Ri reveals that

\beq
\frac{\omega_0^2}{\Omega_0^2} =
\sqrt{\mu(2+\mu)} - \mu + \frac{i\sqrt 2}{\nu\sqrt{{\rm Ri}}}
\left(\sqrt{\mu} - \mu\sqrt{\frac{1}{2+\mu}}\right) + \order{\frac{1}{{\rm  Ri}}}
\eeq
If we write $\displaystyle \omega_0 = \omega\sub{00} +  i \omega\sub{i,0} + \cdots$ and 
assume that $ \omega\sub{i,0}$ scales like $\Ri^{-1/2}$, then to leading
order we find

\beq
\frac{\omega\sub{00}}{\Omega_0} = \pm \sqrt{\sqrt{\mu(2+\mu)} - \mu},
\qquad
\frac{\omega\sub{i,0}}{\Omega_0}
 = \frac{1}{\sqrt{2}\nu \omega_{00}}\left(\sqrt{\mu} - \mu\sqrt{\frac{1}{2+\mu}}\right) 
 \label{omega_sol_first_correction}
\eeq
The leading order expression for $\tilde \beta^2$ (denoted by $\beta\sub{00}^2$) is given as

\beq
\frac{\tilde \beta\sub{00}^2}{H_p^2} = 
\frac{1}{\nu\sqrt{2}} \left(1 + \frac{1}{\varepsilon_0}\right)
\sqrt{\left[\mu\left(1+\mu - \sqrt{\mu(2+\mu)}\right)\right]}.
\label{leading_order_beta}
\eeq
We observe that the growth rate $\omega\sub{0,i}$ has a $\mu$ dependence, predicting growth for all values of $\mu$, with a corrsponding maximum value at
$\mu \equiv \mu_{{\rm max}} = 2/3$. 
In this large Ri limit, we define the maximum growth rate 
$\omega\sub{{\rm m}} \equiv \omega\sub i(\mu\sub{{\rm max}}) $
and find that it is given by

\beq
\frac{\omega\sub{{\rm m}}}{\Omega_0}  = 
\frac{\sqrt{1 + \frac{1}{\varepsilon_0}}
}{2\sqrt{2}(2m+ 1)} \left(\frac{\delta v\sub{00}}{2\Omega_0 H_p}\right)
=\frac{1/\sqrt{\Ri}}{2\sqrt{2}(2m+ 1)}.
\eeq
We can similarly identify the fastest growing radial wavenumber based on the definition of $\mu$ 
found in Eq. (\ref{mu_definition}). 
We define a corresponding radial wavelength for the fastest growing mode
($\lambda\sub{{\rm m}}$), related to $k_{{\rm m}} = k(\mu=\mu\sub{{\rm max}})$ via
$k\sub{{\rm m}} \equiv 2\pi/\lambda\sub{{\rm m}}$.  This leads to

\beq
\displaystyle
\frac{\lambda\sub{{\rm m}}}{H_p} = \frac{2\pi/\sqrt{3}}{m+ \sfrac{1}{2}}\sqrt{1 + \frac{1}{\varepsilon_0}}.
\label{fastest_growing_estimates}
\eeq
Finally the vertical scale scale characterizing variations in the parabolic cylinder function is set by $\beta$ and we
estimate its value at $k\sub{{\rm m}}$ based on $\beta$'s 
leading order behavior found in Eq. (\ref{leading_order_beta}).  We therefore say
$\tilde\beta\sub{{\rm max}}$
where 

\beq
\tilde\beta\sub{{\rm m}} \approx \tilde\beta\sub{00}\Big(\mu=\mu\sub{{\rm m}}\Big) = 
\left(\frac{1}{6\sqrt{3}}\right)^{1/4}\left[\frac{\varepsilon_0 + 1}{\varepsilon_0(m+{1}/{2})}\right]^{1/2} H_p.
\label{fastest_growing_beta}
\eeq

\par
In order to help generate an analytic solution to discuss the physical content of the simulations we have developed in the text, we have chosen a vertical profile for $v$ that extends parabolically to infinity. However we know that in the mean time-instantaneous profiles found in section \ref{1d_dust_settling_solutions} approach constant profiles in that limit, but show a parabolic profile in the region around the midplane.  Furthermore, since the function ${\cal D}_m(z/\beta)$ show Gaussian decay on a length scale defined by $\beta$, it is reasonable to suppose that these analytical solutions are representative of the more realistic
circumstances found section \ref{1d_dust_settling_solutions} so long as the decay scale is less than the particle scale height, i.e., $\beta < H_p$.  Additionally, the first few zeros of 
${\cal D}_m(\zeta)$ for given values of $m$, occur for values of $\zeta = \zeta_{0i} < 1$.  This feature ensures that the modes have multiple nodes across the vertical extent.  Based
on this reasoning,
we expect that the fastest growing of the analytic modes we have developed here are
representative so long as $\tilde\beta\sub{{\rm m}} < 1$.  Thus by Eq. (\ref{fastest_growing_beta}) we
require

\beq
\left(\frac{1}{6\sqrt{3}}\right)^{1/4}\left(\frac{\varepsilon_0 + 1}{\varepsilon_0(m+{1}/{2})}\right)^{1/2} \le 1,
\eeq
and this expression places the following constraint on $m$,

\beq
m + \frac{1}{2} >\left(\frac{1}{6\sqrt{3}}\right)^{1/2}\left( \frac{1+\varepsilon_0}{\varepsilon_0}\right),
\eeq
in order to plausibly
apply these approximate -- though analytically derived -- solutions
to interpret the flow transitions considered in this study. \par 
Let $\pm\zeta\sub{m,i}$ denote the $i^{{\rm th}}$  zero of ${\cal D}_m(\zeta)$.  For each order $m$ there are $m$ such zeros which are also those of the
the Hermite polynomials ${\cal H}_m(\zeta)$.  The values of these are: $\zeta\sub{2,\pm 1} = \pm \sqrt{2}/2 $,
$\zeta\sub{3,i=\{0,\pm 1\}} \approx 0,\pm 1.22474 $, $\zeta\sub{4,i=\{\pm 1,\pm 2\}} \approx \pm 0.52464, \pm 1.65068$, 
 $\zeta\sub{4,i=\{0,\pm 1,\pm 2\}} \approx 0, \pm 0.95857, \pm 2.02018 $ and so on \citep{Abramowitz_Stegun_1972}.  Because of the 
 vertical scaling associated with the fastest growing mode, $\beta_{m}$ defined in Eq. (\ref{fastest_growing_beta}), we have therefore the distance between successive nodes for solution index $m$ to be
 
 \beq
 \Delta z\sub{m,i} = \tilde\beta\sub{\rm m} \Big(\zeta\sub{m,i}
 - \zeta\sub{m,i-1}\Big).
 \eeq
  Due to symmetry considerations it is enough to quantify $ \Delta z\sub{m,i}$ by restricting consideration to values of $1 \le i \le m/2$ for $m$ even, and $1 \le i \le (m-1)/2$  for
 $m$ odd.

\section{Hyperdiffusion and hyperviscosity scheme in PENCIL code}\label{sec:hyperdiffusion}

The flux of viscous momentum can be cast to be proportional to the rate of strain tensor, which can be written as 

\begin{equation}\label{eqn:sij}
    S_{mn}=\frac{1}{2}\left(\frac{\partial u_{m}}{\partial x_{n}}+\frac{\partial u_{n}}{\partial x_{m}}-\frac{1}{3}\delta_{mn}\nabla\cdot \bf{\bug}\right),
\end{equation}

\noindent which, in the incompressible limit, along with constant dynamical viscosity $\mu=\nu\rho$, translates to a viscous acceleration as $\nu \nabla^2\bug$, where $\nu$ is the kinematic viscosity. Note that, in Eq. (\ref{eqn:sij}), $p$ and $q$ are used as dummy indices in order to denote the gas velocity components. The viscous acceleration $f_{{\text {visc}}}$, in its most general form can be written as 

\begin{equation}
    f_{{\text {visc}}}=\frac{1}{\rho}\nabla\cdot \left[2\rho\nu\bm{S}\right].
\end{equation}

\noindent With the PENCIL code, however, we are using a higher order (sixth) hyper-dissipation, and hence, the rate-of-strain tensor gets replaced by a higher-order version as below:

\begin{equation}
    f_{{\text {visc}}}^{({\text {hyper})}}=\frac{1}{\rho}\nabla\cdot\left[2\rho\nu_n{\bm {S}}^{(n)}\right].
\end{equation}

\noindent Here, $n=3$ which corresponds to the sixth order hyperdissipation. With this, a simple hyperviscosity is applied in the following form:

\begin{equation}\label{eqn:hyperviscosity}
    f_{{\text {visc}}}=\begin{cases}
      \nu_3\nabla^6\bug , & \text{if $\mu$ = constant;}\\
      \nu_3\left[\nabla^6\bug + 2 \bm{S}^{(3)}\cdot \nabla \ln \rho_g\right], & \text{if $\nu$ = constant.}
    \end{cases}
\end{equation}

 The spectral range over which the hyperviscosity operates is small, and its used only as a high frequency filter in the numerical setup. Moreover, given its artificial nature, the strict requirement of momentum conservation is dropped \Citep[][also see PENCIL code manual]{Lyra_etal_2017}. It is important to note that the simplified expression from Eq. (\ref{eqn:hyperviscosity}) can be written as the divergence of a rate-of-strain tensor as 
 
 \begin{equation}
    S_{mn}^{(3)}=\frac{\partial^5 u_m }{\partial x_n^5}.
\end{equation}

\noindent For more details on the hyperdissipation scheme used in the code, see the PENCIL CODE manual and \citet{Lyra_etal_2017}.

\end{document}